\documentclass[12pt]{article}
\usepackage{amsmath,amsfonts}
\usepackage{hyperref}
\usepackage{graphicx}
\usepackage{xcolor}
\usepackage[numbers,sort&compress]{natbib}

\usepackage{amsthm}
\usepackage{amssymb}
\usepackage[matrix,arrow]{xy}
\usepackage{epsfig}
\usepackage{slashed}
\usepackage{sidecap}
\usepackage{braket}

\makeatletter\@addtoreset{equation}{section}\makeatother

\newcommand{\al}{\alpha}
\newcommand{\be}{\beta}
\newcommand{\te}{\theta}

\newcommand{\zb}{\bar{z}}

\newcommand{\onov}[1]{\frac{1}{#1}}
\newcommand{\mat}[1]{\left(\begin{matrix} #1 \end{matrix}\right)}

\newcommand{\lag}{\mathcal{L}}

\newcommand{\beq}{\begin{equation}}
\newcommand{\eeq}{\end{equation}}
\newcommand{\bea}{\begin{eqnarray}}
\newcommand{\eea}{\end{eqnarray}}

\newcommand{\vev}[1]{{\left< {#1} \right>}}

\newcommand{\eq}[2][ ]{\begin{equation}\label{#1}{\begin{split}#2\end{split}}\end{equation}}
\newcommand{\eql}[2]{\begin{equation}\label{#1}{\begin{split}#2\end{split}}\end{equation}}

\newcommand{\hm}{\hat\mu}
\newcommand{\ha}{\hat a}
\newcommand{\hx}{\hat{\xi}}

\newcommand{\cD}{{\mathcal D}}

\newcommand{\cN}{{\mathcal N}}

\newcommand{\mm}{\mathbb{M}}

\renewcommand{\title}[1]{\vbox{\center\LARGE{#1}}\vspace{5mm}}
\renewcommand{\author}[1]{\vbox{\center\large#1}\vspace{5mm}}

\begin{document}

\bibliographystyle{utphys}
	
\begin{titlepage}
		\vspace{8mm}
	\begin{center}
		\title{\bf Vortex-strings in $\cN=2$ quiver$\times$ U(1) theories}
		\vspace{8mm}
		{\large \bf Avner Karasik}\footnote{avner.karasik@weizmann.ac.il}\\\vspace{2mm}{\large\textit{ Department of Particle Physics and Astrophysics,\\Weizmann Institute of Science, Rehovot 76100, Israel}}
	\end{center}

	\vspace{5mm}
\begin{abstract}
We study $\onov{2}$-BPS vortex-strings in four dimensional $\cN=2$ supersymmetric quiver theories with gauge group $SU(N)^n\times U(1)$. The matter content of the quiver can be represented by what we call a tetris diagram, which simplifies the analysis of the Higgs vacua and the corresponding strings. We classify the vacua of these theories in the presence of a Fayet-Iliopoulos term, and study strings above fully-Higgsed vacua. The strings are studied using classical zero modes analysis, supersymmetric localization and, in some cases, also S-duality. We analyze the conditions for bulk-string decoupling at low energies. When the conditions are satisfied, the low energy theory living on the string's worldsheet is some 2d $\cN=(2,2)$ supersymmetric non-linear sigma model. We analyze the conditions for weak$\to$weak 2d-4d map of parameters, and identify the worldsheet theory in all the cases where the map is weak$\to$weak. For some $SU(2)$ quivers, S-duality can be used to map weakly coupled worldsheet theories to strongly coupled ones. In these cases, we are able to identify the worldsheet theories also when the 2d-4d map of parameters is weak$\to$strong. 

\end{abstract}

\end{titlepage}

\tableofcontents
\section{Introduction and Summary}
In this work we study the worldsheet theories of $\onov{2}$-BPS vortex strings (strings from now on) configurations in four-dimensional $\cN=2$ supersymmetric $SU(N)^n\times U(1)$ gauge theories. These theories are related to the well-studied $SU(N)^n$ quiver theories, by gauging some $U(1)$ flavour symmetry and adding to it a Fayet-Iliopoulos (FI) term. The study of strings in these theories allows us to understand better interesting physical phenomena such as bulk-string decoupling at low energies and weak$\to$ strong mapping of parameters from the four-dimensional theory to the two-dimensional worldsheet theory. In addition, for $SU(2)$ quivers these strings have interesting transformation rules under S-duality that relates strings in theories with different $U(1)$ gauged, and strings in linear quivers to strings in generalized quivers. Some of these properties already appeared in \cite{Gerchkovitz:2017kyi,Gerchkovitz:2017ljt} for the simpler case where the gauge group is $SU(N)\times U(1)$.

 The matter content consists of $N$ fundamental hypermultiplets of $SU(N)_1$, $N$ fundamental hypermultiplets of $SU(N)_n$ and $n-1$ bi-fundamental hypermultiplets of $SU(N)_i\times SU(N)_{i+1}$ for $i=1,...,n-1$. Under the $U(1)$, every hypermultiplet is assigned with a charge $c_i\in\mathbb{Z}$. We will denote the two scalars of the hypermultiplets by $q$ and $\tilde{q}$. When adding an FI term associated with this $U(1)$, some of the hypermultiplets scalars must get non-trivial vacuum expectation value (VEV) $\vev{q_a}=v$ where the index $a$ here labels the scalars that get VEV. The vacuum equations are solved by giving VEV to an $SU(N)^n$ invariant operator, charged under the $U(1)$. The sign of its $U(1)$ charge should be the same as the sign of the FI parameter. This theory supports stable strings. The way to construct a string is to change the boundary conditions for the scalars getting VEV to $\lim_{r\to\infty}\vev{q_a}=ve^{ik_a\phi}$ where $r,\phi$ are the polar coordinates on the plane transverse to the string, and $\{k_a\}$ a set of non-negative or non-positive integers. The string is labelled by the choice of vacuum and the total winding number $K=\sum_a k_a$. The minimal tension configurations within a topological sector $K$, are $\onov{2}$-BPS and preserve $\cN=(2,2)$ supersymmetry on the string worldsheet. These strings are generalizations of the well studied strings in $U(N)$ theories.  For a partial list of references, see \cite{Auzzi:2003fs,Hanany:2003hp,Hanany:2004ea,Shifman:2004dr, Shifman:2006kd,Eto:2005yh, Auzzi:2005gr, Eto:2009zz,  Ferretti:2007rp, Dorey:1999zk, Eto:2007yv, Eto:2006cx}, and the reviews \cite{Shifman:2007ce,Tong:2005un,Tong:2008qd,Eto:2006pg}.  

An important tool we will use in order to study the worldsheet theories of these strings is supersymmetric localization. We can write the partition function of the four-dimensional theory on a squashed sphere using the results of \cite{Pestun:2007rz,Hama:2012bg}. For some range of parameters, we can rewrite the partition function as a sum over Higgs (and mixed) branch contributions. Out of this sum, one can identify the string contributions \cite{Chen:2015fta,Pan:2015hza,Fujimori:2015zaa,Gerchkovitz:2017ljt}. As was explained in \cite{Gerchkovitz:2017kyi}, in some cases the low energy theory factorizes into a product of the four dimensional vacuum theory and the two dimensional worldsheet theory. Correspondingly, the string contribution factorizes into a product of the four-dimensional vacuum partition function and the worldsheet partition function on a two-sphere. In these cases, we can compare the worldsheet partition function to known results of $S^2$ $\cN=(2,2)$ supersymmetric partition functions \cite{Doroud:2012xw,Benini:2012ui}. This comparison of the partition functions allows a highly non-trivial check of any suggestion for the worldsheet theory.
See also \cite{Gomis:2016ljm,Pan:2016fbl} where similar methods were used in the context of surface defects. 

In \cite{Gerchkovitz:2017ljt}, a condition on the U(1) charges was given such that the map of parameters from the 4d theory to the 2d theory is weak to weak. The classical analysis of the zero modes is useful only when the worldsheet theory is weakly coupled. Supersymmetric localization gives exact results for the partition function. However, when the map of parameters is weak to strong, the expression for the partition function we derive is expanded around some strongly coupled point. On the other hand, the expressions available in the literature for $S^2$ partition functions are expanded around the weakly coupled points. This makes the task of identifying the theory very hard in these cases. In this work, we identify the worldsheet theories in all the cases where the map of parameters is weak$\to$weak. For $SU(2)$ quivers, we identify strongly coupled worldsheet theories that are related to weakly coupled ones via S-duality. 

The outline of this paper is as follows. In section \ref{secvacua} we analyze the vacua of $SU(N)^n\times U(1)$ theories in the presence of an FI term. Section \ref{secstrings} is devoted to general properties of the string and its zero modes. We also present the conditions on U(1) charges $\{c_i\}$ that lead to bulk-string decoupling at low energies, and to weak$\to$weak map of parameters. In sections \ref{sun2} and \ref{su2M} we study strings in $SU(N)^2\times U(1)$ theories and $SU(2)^M\times U(1)$ theories respectively, and give ansatzes for the worldsheet theories based on semiclassical analysis. In section \ref{Sdual} we use S-duality properties of the four-dimensional $SU(2)^M\times U(1)$ theories in order to study strongly coupled strings. In particular, S-duality relates linear SU(2) quivers to generalized SU(2) quivers, which allows us to study also strongly coupled strings on generalized quivers. In section \ref{localization} we go back to all the strings studied in the previous sections and study their worldsheet theory using supersymmetric localization. We extract their worldsheet partition function from the four-dimensional partition function and show agreement with our ansatzes. Some technical computations appear in the appendix.

	\section{Vacua analysis and tetris diagrams}
\label{secvacua}
\begin{SCfigure}
	\vspace{10pt}
	\caption{\small{This figure shows the tetris diagram for $SU(2)^3$ theory. Every box (one square) represents one component of a hypermultiplet. The diagram should be read from left-up to right-down. The first two columns represent two fundamentals of $SU(2)_1$. The $2\times 2$ block to their right represents one bifundamental of $SU(2)_1\times SU(2)_2$. The $2\times 2$ block below it, represents one bifundamental of $SU(2)_2\times SU(2)_3$. Finally, the last two columns represent two fundamentals of $SU(2)_3$. }}\label{basictetris}
	\includegraphics[width=0.5\textwidth, height=6cm]{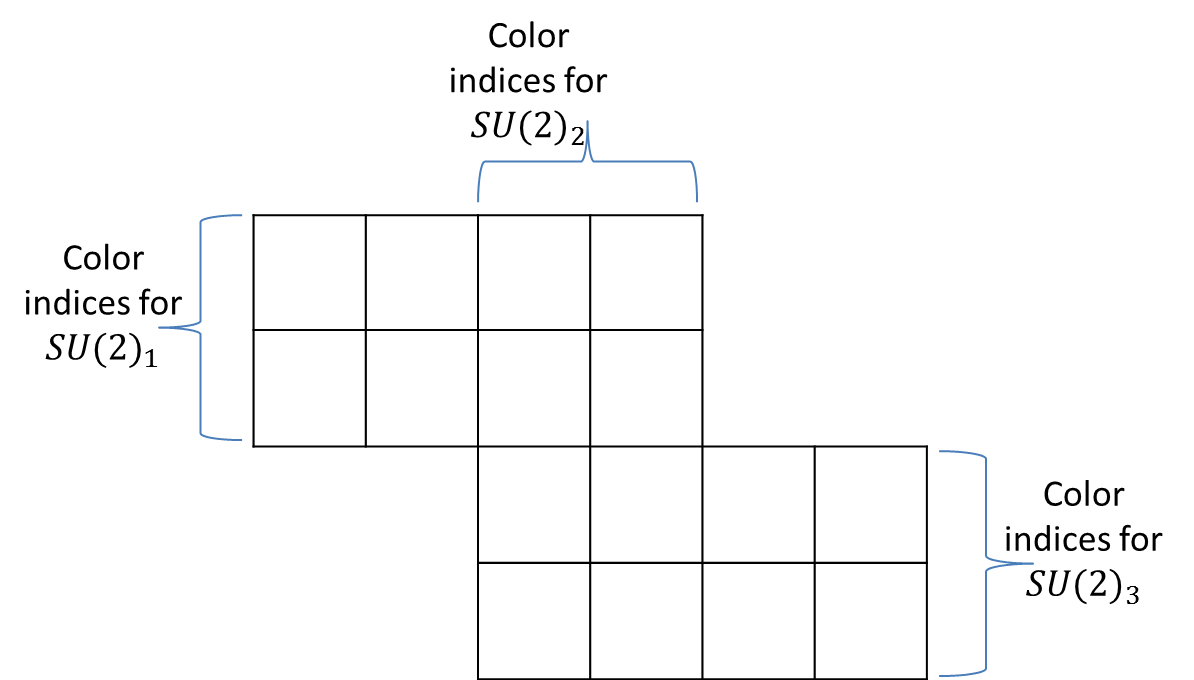}
	\vspace{10pt}
\end{SCfigure}

\begin{figure}
	\vspace{10pt}
         	\includegraphics[width=0.9\textwidth, height=12cm]{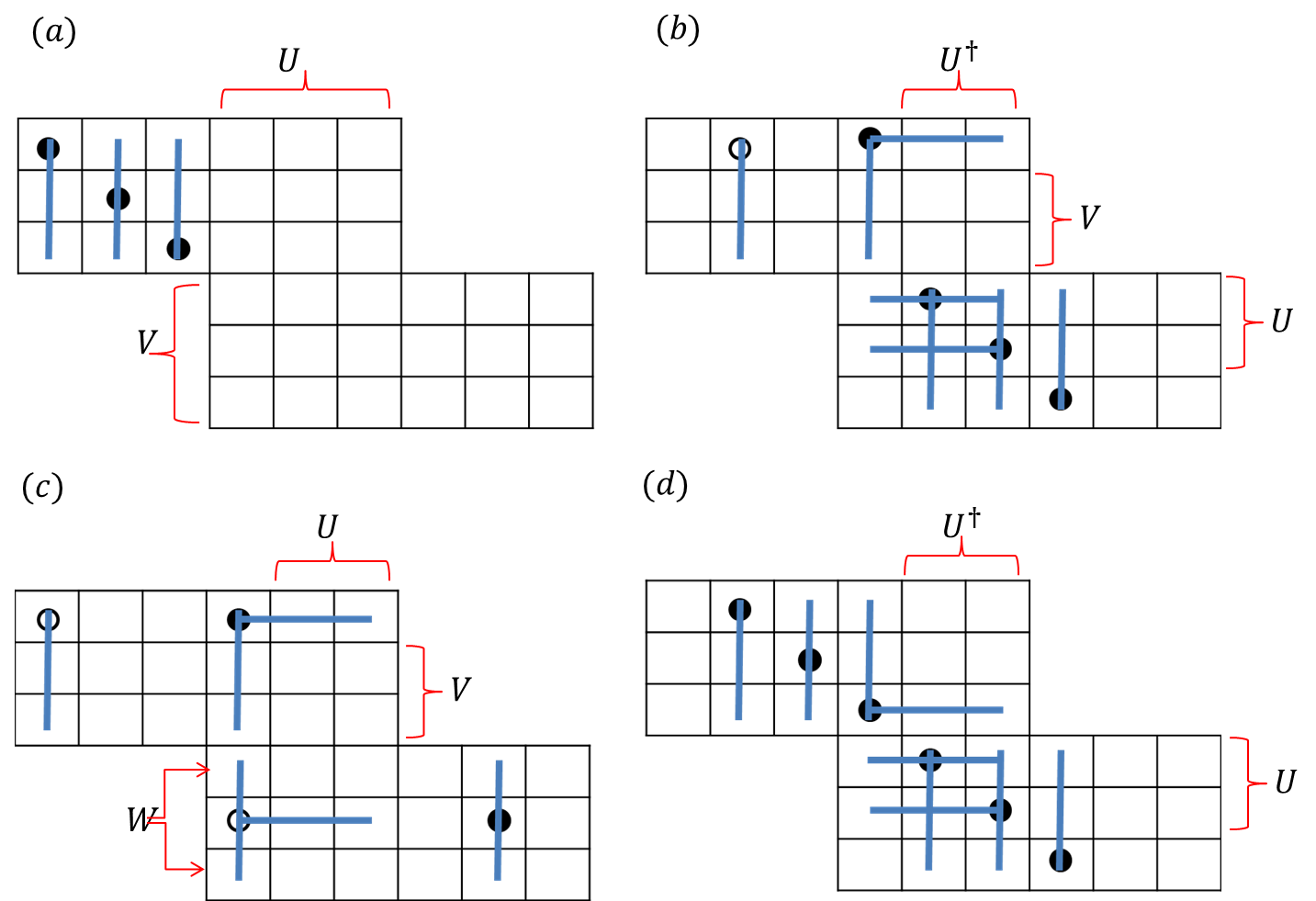}
  \caption{\small{This figure shows four possible vacua for a theory with $G=SU(3)^3$. Full and empty dots represent $q$ and $\tilde{q}$ getting VEV respectively. The lines represent hypermultiplets which are swallawed by gauge multiplets via the Higgs mechanism. In (a) there is a baryon for the first $SU(3)$ and nothing for the rest. In (b) there is a meson for the first $SU(3)$ and baryons for the two other $SU(3)$s. In (c) there is a meson for every $SU(3)$ and in (d) a baryon for every $SU(3)$ factor. Up to the discrete $\mathbb{Z}_{|c|}$ factor, the residual gauge symmetries in these four vacua are: (a) $SU(3)^2$, (b) $SU(2)^2$, (c) $SU(2)^3$, (d) $SU(2)$. The way the residual gauge transformations act on the different indices is shown explicitly on the diagrams.  }}\label{4examples}
	\vspace{10pt}
\end{figure}

In this section we describe what we call a tetris diagram which is a picturial way to represent the matter content of a quiver theory, and use it to classify the vacua in the presence of an FI parameter. Our starting point is the four dimensional $\cN=2$ superconformal quiver theory with gauge group $G=SU(N)^n\equiv SU(N)_1\times SU(N)_2\times...\times SU(N)_n$. The matter content consists of $N$ fundamental hypermultiplets of $SU(N)_1$, $N$ fundamental hypermultiplets of $SU(N)_n$ and $n-1$ bi-fundamental hypermultiplets of $SU(N)_i\times SU(N)_{i+1}$ for $i=1,...,n-1$. The hypermultiplets can be represented by a diagram made out of $n+1$ blocks, each one contains $N^2$ boxes arranged in an $N\times N$ matrix. See figure \ref{basictetris} for an example. We modify the theory first by introducing small and generic masses for the hypermultiplets. In addition, we gauge some global $U(1)$ and add an FI term associated with this $U(1)$. This $U(1)$ can be labelled by asigning an independent U(1) charge $c_i\in\mathbb{Z}$ to every hypermultiplet. Due to the FI parameter, some of the hypermultiplets scalars must get non-trivial vacuum expectation value. The vacuum equations are solved by giving VEV to an $SU(N)^n$ invariant operator, charged under the $U(1)$. The sign of its $U(1)$ charge should be the same as the sign of the FI parameter. These vacua can be represented nicely on the tetris diagram. We can denote a $q$ getting VEV by a black dot and a $\tilde{q}$ getting VEV by a white (empty) dot. We need to give VEV to several scalars such that for every $SU(N)$ factor we have a baryon, a meson or nothing getting VEV. See figure \ref{4examples} for some examples. 

In a given vacuum, the gauge symmetry is broken down to $H\subset G\times U(1)$. $H$ can be read easily from the tetris diagram using the following simple rules:
\begin{itemize}
\item The number of dots equals the reduction in the rank. \eq{\text{Rank}\left(G\times U(1)\right)-\text{Rank}(H)=\text{\# of dots}\ .}
\item If one draws a line (lines) from every dot in the directions of the color indices, all the boxes with line on them represent hypermultiplets which are combined with gauge multiplets into massive W-boson multiplets. Therefore, the number of broken generators equals the number of boxes with line on them.
\item $H$ always contains a discrete $\mathbb{Z}_{|C|}$ factor, where $C$ is the $U(1)$ charge of the operator getting VEV. 
\end{itemize}

\begin{SCfigure}
	\vspace{10pt}
  \caption{\small{A simple example for a seperable vacuum in the $G=SU(3)^2$ theory. The operator getting VEV can be written as a product of two $G$-invariant operators: a baryon of $SU(3)_1$ and a baryon of $SU(3)_2$.  }}\label{seperable}
         	\includegraphics[width=0.3\textwidth, height=4cm]{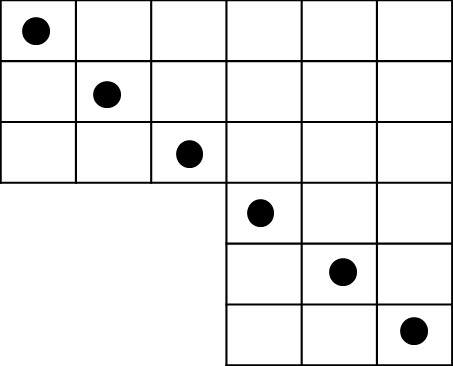}

	\vspace{10pt}
\end{SCfigure}

These rules are correct only for non-seperable vacua. By seperable we mean that the operator getting VEV can be written as a product of more than one $G$-invariant operators. The seperable vacua are excluded from the following reason. The mass term of a hypermultiplet scalar $q$ looks schematically like $(\mu-\sum a)^2|q|^2$ where $\mu$ is the bare mass of $q$ and the sum over $a$ is the sum over the Cartan components of the gauge multiplet scalars in the relevant representations (The off-diagonal elements of $a$ are taken to be zero). If $q$ gets a VEV, this term must vanish and therefore we must also introduce a VEV for the gauge multiplet scalars such that $\sum a=\mu$ in the vacuum. In a seperable vacuum, the VEV of $a^{u(1)}$, the $U(1)$ scalar, can be extracted from every $G$-invariant operator independently. For generic masses, these values will not coincide and therefore this solution is forbidden. As a simple example, consider the $G=SU(3)^2$ theory with the 2-baryonic vacuum illustrated in figure \ref{seperable}. In this example, the operator getting VEV is a product of two $G$-invariant operators. The vanishing of the mass terms gives six equations. By summing over the first three equations, one can extract the value of the U(1) scalar \eq{a^{u(1)}=\frac{\sum_{i=1}^3 \mu_i}{\sum_{i=1}^3 c_i}\ ,} where $\mu_i,\ c_i$ with $i=1,2,3$ are the masses and U(1) charges of the first three hypermultiplets. Similarly, by summing over the last three equations, one can find \eq{a^{u(1)}=\frac{\sum_{i=5}^7 \mu_i}{\sum_{i=5}^7 c_i}\ ,} where now $\mu_i,\ c_i$ with $i=5,6,7$ are the masses and U(1) charges of the last three hypermultiplets. For generic masses $\mu_i$, these values don't coincide and therefore this configuration doesn't solve the vacuum equations. On the same way, every seperable vacuum is excluded once generic values for the hypermultiplets masses $\mu_i$ are turned on. 

 Consider $G=SU(N)^n$ with general $N,n$. We will list all the possible types of non-seperable vacua and the residual gauge symmetry in each of these vacua.

\begin{figure}
	  	\vspace{10pt}
	  	\includegraphics[width=0.9\textwidth, height=6cm]{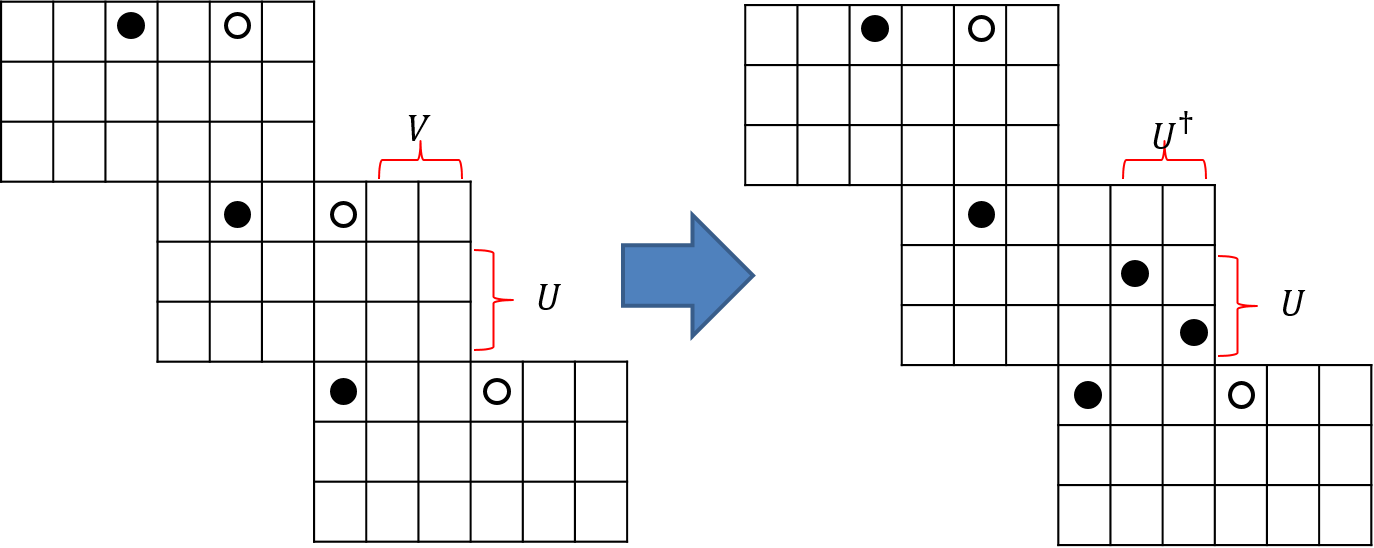}
	  	\caption{\small{This figure shows the contamination of a mesonic chain on the left to a mixed chain containing two baryons, on the right. It is also shown how two independent $SU(N-1)$ symmetries of the mesonic chain, labelled by the matrices $U,V$ are broken to the diagonal combination $V=U^\dagger$ due to the baryons.  }}\label{contamination}
	  	\vspace{10pt}
	  \end{figure}

\begin{itemize}
	\item Single block vacua: The simplest vacua are single block vacua. In these vacua all the dots are in the same block. There are two types of single block vacua:
	\begin{enumerate}
		\item Single block (anti-) baryon: In these vacua there are $N$ dots and $N^2$ lined boxes. The residual gauge symmetry in this case is $H=SU(N)^{n-1}\times \mathbb{Z}_{|C|}$.
\item Single block meson: In these vacua there is one full and one empty dot in the first block or in the last block. There are 2N lined boxes. The residual gauge symmetry is $H=SU(N-1)\times SU(N)^{n-1}\times \mathbb{Z}_{|C|}$.
	\end{enumerate}
	The rest of the vacua contain dots in all the blocks.
	\item Mesonic chain: These vacua are similar to the one presented in \ref{4examples} (c). In these vacua there is a meson for every $SU(N)$ factor. There is one dot in every block which reduces the rank of the gauge group by $n+1$. There are $2nN-n+1$ broken generators, which means that the residual gauge symmetry has rank $n(N-2)$ and dimension $n((N-1)^2-1)$. The residual gauge group is
	\eq{H=(SU(N-1))^{n}\times \mathbb{Z}_{|C|}\ .}
	\item (anti-) Baryonic chain: These vacua are similar to the one presented in \ref{4examples} (d). In these vacua there is a baryon for every $SU(N)$ factor. We can classify the vacua according to the number of dots in the first block, denoted by $m$ with $1\leq m\leq N-1$. With this choice, the number of dots in all the odd blocks is $m$ while the number of dots in all the even blocks is $N-m$. We will divide the analysis to two cases:
	\begin{enumerate}
		\item odd $n$: If $n$, the number of $SU(N)$ factors, is odd, there is an even number of blocks. The number of dots is always $\frac{N(n+1)}{2}$, regardless of $m$. The number of broken generators is $N^2+\frac{n-1}{2}(N^2-2m^2+2mN)$. This implies that the residual gauge symmetry has rank $\frac{n-1}{2}(N-2)$
		and dimension $\frac{n-1}{2}(N^2+(N-m)^2-2)$. The residual gauge symmetry is
		\eq{H=\left(SU(m)\times SU(N-m)\right)^{\frac{n-1}{2}}\times \mathbb{Z}_{|C|}\ .}
		\item even $n$: In this case, there is an odd number of blocks. The number of dots is $\frac{Nn}{2}+m$ and the number of broken generators is $m^2+\frac{n}{2}(N^2-2m^2+2Nm)$. This implies that the residual gauge symmetry has rank $\frac{nN}{2}-n-m+1$ and dimension $\frac{n}{2}((N-m)^2-1)+(\frac{n}{2}-1)(m^2-1)$. The residual gauge symmetry in this case is
		\eq{H=(SU(m))^{\frac{n}{2}-1}\times (SU(N-m))^{\frac{n}{2}}\times \mathbb{Z}_{|C|}\ .}
			\end{enumerate}
	\item Mixed chains: In these vacua there is a mixture of mesonic and baryonic chains. See figures \ref{4examples} (b) and \ref{contamination} for examples. Notice that the block of the exchange from mesonic to baryonic chains contains only one dot. Therefore, intermediate baryonic chains must contain even number of baryons and are characterized by $m=1$. If the vacuum starts (ends) with a baryonic chain, it may contain odd number of baryons and then it has $N-1$ dots in the first (last) block. The easiest way to analyse the residual gauge symmetry is to characterize the mixed vacua by the number of blocks with one dot, denoted by $l$. The residual gauge symmetry is then
	\eq{H=(SU(N-1))^{l-1}\times \mathbb{Z}_{|C|}\ .} 
	This can be simply understood in the following way. Start from a mesonic chain which is a special case of the mixed vacua with $l=n+1$. One can contaminate the chain with baryons by replacing a block with 1 dot by a block with $N-1$ dots, where two "contaminated" blocks cannot be one next to the other. It is easy to see that every contamination of this type, breaks one $SU(N-1)$ factor. See figure \ref{contamination}. Notice that the minimal value for $l$ is $\frac{n}{2}+1$ for even $n$ and $\frac{n+1}{2}$ for odd $n$ which are barynoic chains.
	  \end{itemize}

An immediate result of this analysis is that fully Higgsed vacua exist only for $N=2$ with arbitrary $n$, or $n\leq 2$ with arbitrary $N$.
In the next sections we will study vortex-strings above the fully Higgsed vacua of $G=SU(2)^n$ and $G=SU(N)^2$.

\section{Strings: Generalities and classical analysis}
\label{secstrings}
In the next sections we will generalize the analysis made in \cite{Gerchkovitz:2017ljt} and study the low energy theories living on the strings worldsheet in cases where the original gauge group is $SU(N)^n\times U(1)$. We will do it only for the cases where the vacuum is fully Higgsed. In this section we will go over some of the main steps in the way. Our starting point will be a baryonic chain vacuum of an $SU(N)^n\times U(1)$ theory illustrated by some tetris diagram. The diagram contains $nN+1-n$ dots describing $nN+1-n$ hypermultiplets scalars getting VEV $\vev{q_i}=v\ ,\ i=1,...,nN+1-n$. One also need to give VEV to the $nN+1-n$ Cartan gauge multiplet scalars $a_I$ in order to eliminate the mass terms of $q_i$. At energies much smaller than the mass of the W-bosons $m_W$, the vacuum excitations consists of $N^2+n-1$ light hypermultiplets with masses which are some linear combinations of the hypermultiplets bare masses (The exact values will be specified later). These masses are taken to be much smaller than $m_W$. Excitations with these masses are considered as light and dynamical, while excitations with masses $\sim m_W$ are considered as heavy and frozen. In order to construct strings, one needs to modify the VEV of the scalars by changing the boundary conditions to $\lim_{r\to\infty} q_i=ve^{ik_i\phi}$ where $r,\phi$ are the polar coordinates on the plane transverse to the string and $\{k_i\}$ is a set of non-negative integers.\footnote{Equivalently, we can take $\{k_i\}$ to be non-positive integers. Then the string will carry negative flux but all the analysis will be exactly the same.} The VEVs of $a_I$ are left untouched, but one needs to introduce VEVs to the Cartan gauge fields $A_\phi$ such that the kinetic terms $|\cD_\mu q_i|^2$ vanish at $r\to\infty$. The $U(1)$ flux carried by such a configuration is
\eq{\Phi=\lim_{r\to\infty}\int d\phi A^{u(1)}_\phi=\frac{2\pi K}{C}}
where $K=\sum k_i$, and $C=\sum c_i$ is the charge of the operator getting VEV. 
Minimal tension configurations within a topological sector labelled by $K$ satisfy a set of BPS equations and preserve $\cN=(2,2)$ supersymmetry on the string worldsheet. Configurations with the same $K$ but different $\{k_i\}$ are connected by magnetic monopoles \cite{Tong:2003pz} and are part of the same worldsheet theory. Different $\{k_i\}$ correspond to different vacua of the worldsheet sigma model, and the monopoles correspond to worldsheet kinks that connect the different vacua \cite{Dorey:1999zk}. The worldsheet theory also inherits two vector-like $U(1)_R$ symmetries. The first one is a combination of the four-dimensional $U(1)_R\subset SU(2)_R$ and gauge transformations preserved by the string. The second is a combination of rotation on the plane transverse to the string and gauge transformations. The exact massless zero modes of the string are only the positions of the cores of the string and their superpartners. However, we will treat light modes with masses of order of the hypermultiplets masses as approximate zero modes and include them in our zero modes analysis. There are three types of light modes. We will focus on the bosonic modes where their fermionic partners can be found using supersymmetry.
\begin{enumerate}
\item Size modes: These are modes that come from excitations of the light scalars $q$ and $\tilde{q}$. They can be found in the following way. Among the BPS equations there are the equations 
\eql{Bogomolnyi}{\left(\cD_1+i\cD_2\right)q=\left(\cD_1+i\cD_2\right)\tilde{q}=0\ .}
These equations together with the boundary conditions $\lim_{r\to\infty}q_i=ve^{ik_i\phi}$ implies that $q_i$ has $k_i$ zeros at positions $\vec{r}_{l_i}$. Close to the zeros it behaves as $q_i\sim z-z_{l_i}$, where $z=re^{i\phi}$.  Given the boundary conditions, one can ask whether equations \eqref{Bogomolnyi} allow non-trivial solutions for the light scalars. These solutions are reffered to as size modes. The number of modes equals the number of independent solutions. The number and functional behaviour of the size modes depend highly on the U(1) charges.

\item Off-diagonal modes: Off-diagonal modes are related to gapless excitations of the massive W-bosons. Their number is independent of the U(1) charges. Every pair of swallawed scalars $q_{ij}$ and $q_{ji}$ give rise to $k_i+k_j$ complex modes. Evidences from localization and S-duality for the existance of these modes were given in \cite{Gerchkovitz:2017ljt}.
\item Center of mass modes: These modes parametrize the positions of the string cores on the $(x^1,x^2)$ plane. There are $K$ cores and therefore $2K$ real zero modes. These are the only exact zero modes of the string once the hypermultiplets masses are included.
\end{enumerate}
As was already emphasized in \cite{Gerchkovitz:2017kyi, Gerchkovitz:2017ljt}, the size modes are responsible for two major properties of the low energy effective theory in the string background. We will summarize these two properties here for convenivence.
\subsection{Bulk-string decoupling}
The first property is related to the question whether the string light modes and the bulk light modes decouple at low energies such that the effective action can be written as a sum of two decoupled actions
\eq{S_{eff}=S_{bulk}+S_{string}\ .}
The answer to this question depends on the asymptotic $r$-dependence of the size modes. Modes that decay like $\onov{r^\be}$ with $\be\geq 1$ decouple from the bulk modes at low energies while long range modes that decay like $\onov{r^\be}$ with $0<\be<1$ stay coupled to the bulk modes even at low energies. The demand that there are no long range modes coincides with the condition for no non-trivial Aharonov-Bohm phases for particles in the spectrum encircling the string.

\subsection{Weak to weak mapping}
Starting from a weakly coupled four-dimensional theory, the two-dimensional worldsheet theory can be either weakly coupled or strongly coupled. In other words, the 2d-4d map of parameters can map the weakly coupled regime of the 4d theory to weakly coupled or to strongly coupled regimes of the worldsheet theory. This property depends on the F-term constraints of the four-dimensional theory, that take the form
\eq{\sum_{i}c_i\tilde{q}_iq_i=\sum_{i}\tilde{q}_iT_{R_i}^\al q_i=0\ ,}
where the sum over $i$ is the sum over flavors, $T^\al_{R_i}$ is an SU(N) generator in the representation of $q_i$ and the color index is suppressed. If for every flavor, there are only $q$ zero modes or $\tilde{q}$ zero modes but not both, all the F-terms vanish identically without imposing any constraints on the worldsheet. In this case the map of parameters will be weak to weak. If, on the other hand, there exists a flavor for which there are both $q$ and $\tilde{q}$ zero modes, the F-term constraints act non-trivially on the worldsheet and as a result, the worldsheet theory will be strongly coupled.

\section{Strings in $SU(N)^2\times U(1)$ theories}
\label{sun2}
In this section we will study the $SU(N)^2$ quiver in which we gauge some $U(1)$ flavor symmetry.
The matter content of this theory consists $N$ fundamentals of the first $SU(N)$, $N$ fundamentals of the second $SU(N)$ and one bi-fundamental. Therefore, this theory is parameterized by $2N+1$ charges and masses. We will label the rows of the tetris diagram by $a,b=1,...,2N$ and the columns by $a,b=0,...,2N-1$. In the fully Higgsed vacua, $2N-1$ scalars get VEV. We will take them to be $q_{aa}$ with $\vev{q_{aa}}=v$ for $a=1,...,2N-1$. We will also denote the masses and U(1) charges by $\mu_a,\ c_a$ with $a=0,...,2N$. 
As mentioned above, the gauge symmetry is broken in this vacuum to $\mathbb{Z}_C$ where $C=\sum_{a=1}^{2N-1} c_a$ is the U(1) charge of the operator getting VEV.\footnote{Without loss of generality, we take the FI parameter and the charge $C$ to be positive.} The spectrum of strings above this vacuum is given by $\pi_1\left(SU(N)^2\times U(1)/\mathbb{Z}_C\right)$ which allows fractional magnetic fluxes quantized as $\Phi_{u(1)}^K=\frac{2\pi K}{C}$ with $K\in\mathbb{Z}$.
The string solution is obtained by changing the boundary conditions of the scalars to \eql{bc}{\lim_{r\to\infty}q_{aa}=ve^{ik_a\phi}\ ,} where $k_a$ are non-negative integers. 
For finite tension configurations, $|\cD_\mu q|^2$ must decay at $r\to\infty$ faster than $r^{-2}$. This implies that the Cartan components of the gauge fields $A_\phi$ must be turned on. In particular, it is straight forward to show that the U(1) magnetic flux carried by the string is  
\eq{\Phi_{u(1)}=\lim_{r\to\infty}\int d\phi A'_\phi=\frac{2\pi K}{C}\ ,} where $A'$ is the U(1) gauge field, in agreement with the allowed spectrum.

We will be interested in computing the mass and the two U(1) R-charges of every hypermultiplet around the string solution. These three quantities are computed in a similar way.

\begin{itemize}
	\item Mass: Taking all the off-diagonal elements of the adjoint scalars to zero, the mass of the scalars $q_{ab}$ is $M_{ab}$ with
\eql{mass}{M_{ab}=\begin{cases}
		\mu_b+c_b a'+a_{a}^1\ &\text{for}\ a\leq N\ ,\ b< N\\
		\mu_a+c_a a'+a_{b-N+1}^2\ &\text{for}\ a>N\ ,\ b\geq N\\
		\mu_N+c_N a'+a_{a}^1+a_{b-N+1}^2\ &\text{for}\  a\leq N\ ,\ b\geq N\\
	\end{cases}}
 
where $a',\ a^{1,2}_a$ denote the Cartan elements of the adjoint scalars of the $U(1), SU(N)_{1,2}$ gauge multiplets with $\sum_{a=1}^{N}a^{1,2}_a=0$.
\item $R^{(R)}$-charge: This is a two dimensional vectorlike U(1) R-charge which is a combination of the four-dimensional $U(1)_R\subset SU(2)_R$ and gauge transformations preserved by the string solution. Under a general combination of $U(1)_R\subset SU(2)_R$ and Cartan gauge transformation, the scalar $q_{ab}$ tranform as $q_{ab}\rightarrow e^{i\omega_{ab}}q_{ab}$ with
\eql{RR}{\omega_{ab}=\begin{cases}
		\omega_R+c_b\omega'+\omega_{a}^1\ &\text{for}\ a\leq N\ ,\ b< N\\
		\omega_R+c_a\omega'+\omega_{b-N+1}^2\ &\text{for}\ a>N\ ,\ b\geq N\\
		\omega_R+c_N\omega'+\omega_{a}^1+\omega_{b-N+1}^2\ &\text{for}\  a\leq N\ ,\ b\geq N\\
	\end{cases}}
	where $\omega_R,\ \omega'$ are the $U(1)_R$ and U(1) gauge symmetry parameters respectively, and $\omega^{1,2}_a$ are the Cartan parameters of $SU(N)_{1,2}$ transformations satisfying $\sum_{a=1}^{N}\omega^{1,2}_a=0$.
\item $R^{(J)}$-charge: This is a two dimensional vectorlike U(1) R-charge which is a combination of rotation and gauge transformations. Rotation is broken by the string due to the explicit $\phi$ dependence of the string solution. If all the cores of the string coincide at the same point, there is a combination of rotation and gauge transformation that leaves the string solution invariant. Under a general combination of rotation $\delta\phi= 2\phi_0$\footnote{The factor of 2 was chosen such that $\phi_0$ is the symmetry parameter of the appropriately normalized two-dimensional $R$-symmetry.} and Cartan gauge transformation, the scalars $q_{ab}$ transform as $q_{ab}\rightarrow e^{i\phi_{ab}}q_{ab}$ with
\eql{RJ}{\phi_{ab}=\begin{cases}
		2\phi_0k_a\delta_{ab}+c_b\omega'+\omega_{a}^1\ &\text{for}\ a\leq N\ ,\ b< N\\
		2\phi_0k_a\delta_{ab}+c_a\omega'+\omega_{b-N+1}^2\ &\text{for}\ a>N\ ,\ b\geq N\\
		2\phi_0k_a\delta_{ab}+c_N\omega'+\omega_{a}^1+\omega_{b-N+1}^2\ &\text{for}\  a\leq N\ ,\ b\geq N\\
	\end{cases}}
\end{itemize}
In the string vacuum, we need to demand that \eql{zerofordiagonal}{M_{aa}=\omega_{aa}=\phi_{aa}=0\ .}
These are $2N-1$ equations for the $2N-1$ Cartan parameters. Plugging the values of the Cartan generators back gives us the mass and R-charges of the scalars $q_{ab}$. It will be convenient to package the three quantities into 
\eq{\hat{I}=\left\{\mu_I-\frac{\mu}{C}c_I\ ,\ 1-\frac{(2N-1)c_I}{C}\ ,\ \frac{2Kc_I}{C}-2k_I\right\}\ ,\ I=0,...,2N\ ,}
with \eq{C\equiv \sum_{a=1}^{2N-1}c_a\ ,\ \mu\equiv \sum_{a=1}^{2N-1}\mu_a\ ,\ K\equiv \sum_{a=1}^{2N-1}k_a\ ,\ k_0=k_{2N}=0\ .}
Notice that $\sum_{I=1}^{2N-1}\hat{I}=0$ is satisfied.

\begin{figure}
	\vspace{10pt}
         	\includegraphics[width=0.6\textwidth, height=8cm]{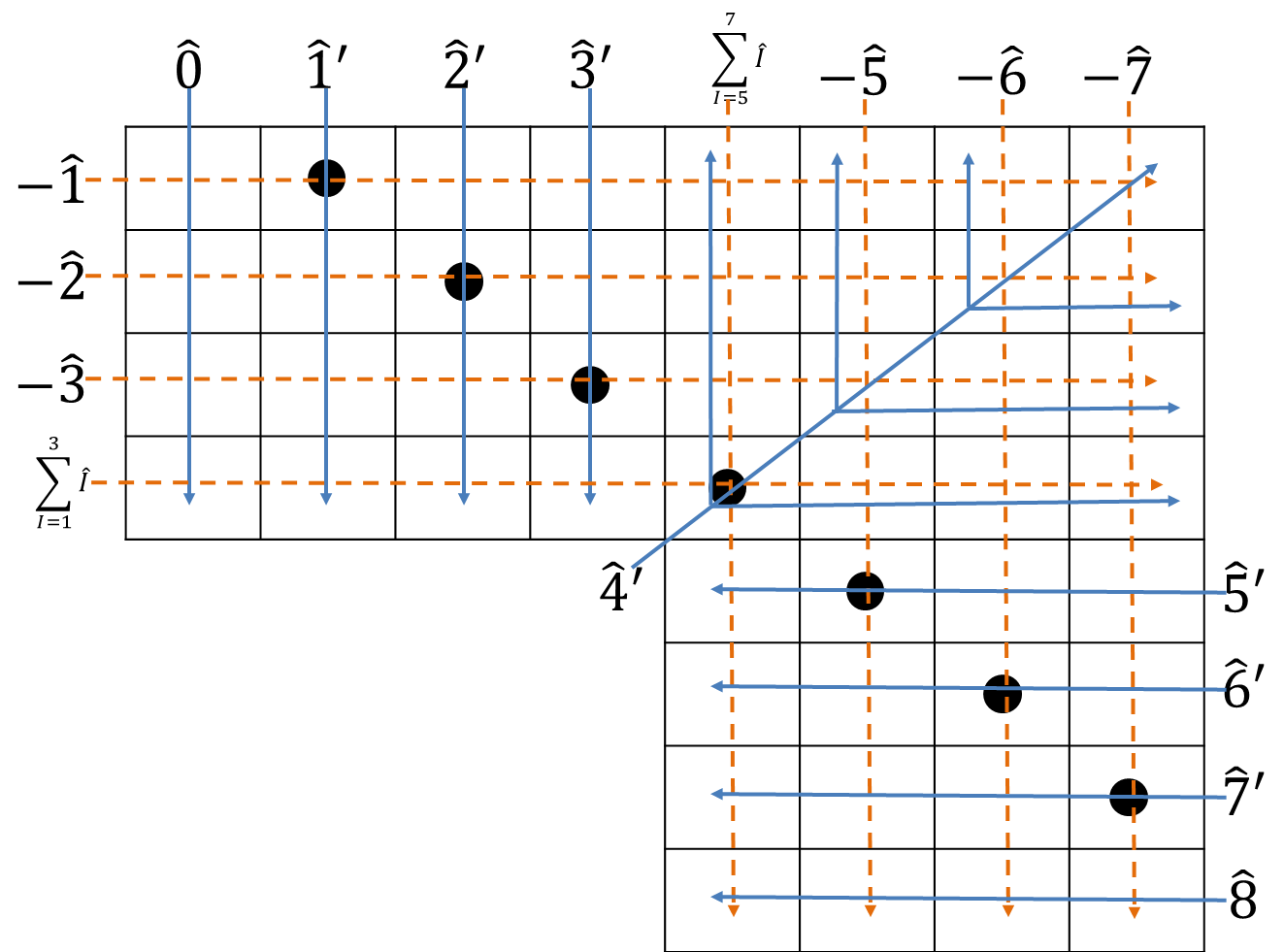}
  \caption{\small{This figure represents a picturial way to solve equation \eqref{zerofordiagonal} and find the mass and R-charges of all the hypermultiplets. The value of the U(1) element $a',\ \omega'$ in equations \eqref{mass},\eqref{RR},\eqref{RJ} is easy to find by summing over the $2N-1$ diagonal equations. This results in $\left\{-\frac{\mu}{C},\ -\frac{(2N-1)\omega_R}{C},\ -\frac{2K\phi_0}{C}\right\}$. The mass and $R$-charges of every hypermultiplet after plugging in the value of the U(1) element are given by $\hat{I}$ on the tail of the blue solid arrow crossing the corresponding box. The dashed orange arrows represent the value subtracted by the $SU(N)^2$ gauge transformation needed in order to satisfy \eqref{zerofordiagonal}. At the end, $\hat{I}_{ab}$ is given by the sum of the $\hat{I}$s that appear on the tail of the arrows crossing the corresponding block. Here $\hat{I}'=\hat{I}+\{0,0,2k_I\}$ except for diagonal elements for which $\hat{I}'=\hat{I}$. For example, there are 3 arrows crossing the box $q_{2,5}$. The sum of them gives $\hat{I}_{2,5}=\hat{4}'-\hat{5}-\hat{2}$. }}\label{Imultiindex}
	\vspace{10pt}
\end{figure}

Every box labelled by the row and column indices $1\leq A\leq 2N\ ,\ 0\leq B\leq 2N$ is parametrized by \eq{\hat{I}_{AB}=\left\{\text{mass}\ ,\ R^{(R)}\ ,\ R^{(J)}\right\}\ ,} with
\eql{multiindexeq}{\hat{I}_{AB}=\begin{cases}\hat{B}-\hat{A}+\{0,0,2k_B\}\ &\text{for}\ 1\leq A\leq N-1\ ,\ 0\leq B\leq N-1\\
                            \hat{B}+\sum_{C=1}^{N-1}\hat{C}+\{0,0,2k_B\}\ &\text{for}\ A=N\ ,\ 0\leq B\leq N-1\\
                            \hat{A}-\hat{B}+\{0,0,2k_A\}\ &\text{for}\ N+1\leq A\leq 2N-1\ \ N\leq B\leq 2N-1\\
                            \hat{A}+\sum_{C=N+1}^{2N-1}\hat{C}+\{0,0,2k_A\}\ &\text{for}\ A=2N\ ,\ N\leq B\leq 2N-1\\
                            \hat{N}-\hat{A}-\hat{B}+\{0,0,2k_N\}\ &\text{for}\ 1\leq A\leq N-1\ ,\ N+1\leq B\leq 2N-1\\
                            -\hat{A}-\sum_{C=1}^{N-1}\hat{C}+\{0,0,2k_N\}\ &\text{for}\ 1\leq A\leq N-1\ ,\ B=N\\
                            -\hat{B}-\sum_{C=N+1}^{2N-1}\hat{C}+\{0,0,2k_N\}\ &\text{for}\ A=N\ ,\ N+1\leq B\leq 2N-1\\\end{cases}}
For example, for $1\leq A\leq N-1\ ,\ 0\leq B\leq N-1$ the mass and R-charges are  \eq{\left\{\text{mass},R^{(R)},R^{(J)}\right\}=\left\{\mu_B-\mu_A+\frac{\mu}{C}(c_A-c_B)\ ,\ \frac{(2N-1)(c_A-c_B)}{C}\ ,\ 2k_A-\frac{2K(c_A-c_B)}{C} \right\}\ .}
A more illustrative representation of equation \eqref{multiindexeq} appears in figure \ref{Imultiindex}.

As explained around equation \eqref{Bogomolnyi}, the size modes are excitations of the light hypermultiplets. These are the hypermultiplets that sit in the first column $q_{a,0}$ with $a=1,...,N$, in the last row $q_{2N,b}$ with $b=N,...,2N-1$, and the upper right $(N-1)\times(N-1)$ block $q_{a,b}$ with $a=1,...,N-1$ and $b=N+1,...,2N-1$.
The size modes are the solutions to equations \eqref{Bogomolnyi} given the boundary conditions \eqref{bc}.\eqref{Bogomolnyi} can be written as
\eql{dlog}{\bar{\partial}\log(q_{ab})=\begin{cases}ic_bA'+iA_{a}^1\ &\text{for}\ a\leq N\ ,\ b< N\\
       ic_aA'+iA_{b-N+1}^2\ &\text{for}\ a>N\ ,\ b\geq N\\
       ic_NA'+iA_{a}^1+iA_{b-N+1}^2\ &\text{for}\  a\leq N\ ,\ b\geq N\\
\end{cases}}
where we emphasize the similarity to equations \eqref{mass}\eqref{RR}\eqref{RJ}.
Solutions to these equations are given by
\eq{q_{i,0}&=\left(\prod_{b}q_{b,b}\right)^{\frac{c_0-c_i}{C}}q_{i,i}\,f_{i,0}(z)\ ,\ i=1,...,N-1\\
q_{N,0}&=\left(\prod_{b}q_{b,b}\right)^{\frac{c_0+\sum_{j=1}^{N-1}c_j}{C}}\prod_{i=1}^{N-1}q^{-1}_{i,i}\,f_{N,0}(z)\ ,\\
q_{2N, i}&=\left(\prod_{b}q_{b,b}\right)^{\frac{c_{2N}-c_i}{C}}q_{i,i}\,f_{2N,i}(z)\ ,\ i=N+1,...,2N-1\\
q_{2N,N}&=\left(\prod_{b}q_{b,b}\right)^{\frac{c_{2N}+\sum_{j=N+1}^{2N-1}c_j}{C}}\prod_{i=N+11}^{2N-1}q^{-1}_{i,i}\,f_{2N,N}(z)\ ,\\
q_{i,j}&=\left(\prod_{b}q_{b,b}\right)^{\frac{c_N-c_i-c_j}{C}}q_{i,i}\,q_{j,j}\,f_{i,j}(z)\ ,\ i=1,...,N-1\ ,\ j=N+1,...,2N-1}
and on the same way

\eq{\tilde{q}_{i,0}&=\left(\prod_{b}q_{b,b}\right)^{-\frac{c_0-c_i}{C}}q_{i,i}^{-1}\,\tilde{f}_{i,0}(z)\ ,\ i=1,...,N-1\\
	\tilde{q}_{N,0}&=\left(\prod_{b}q_{b,b}\right)^{-\frac{c_0+\sum_{j=1}^{N-1}c_j}{C}}\prod_{i=1}^{N-1}q_{i,i}\,\tilde{f}_{N,0}(z)\ ,\\
	\tilde{q}_{2N, i}&=\left(\prod_{b}q_{b,b}\right)^{-\frac{c_{2N}-c_i}{C}}q^{-1}_{i,i}\,\tilde{f}_{2N,i}(z)\ ,\ i=N+1,...,2N-1\\
	\tilde{q}_{2N,N}&=\left(\prod_{b}q_{b,b}\right)^{-\frac{c_{2N}+\sum_{j=N+1}^{2N-1}c_j}{C}}\prod_{i=N+11}^{2N-1}q_{i,i}\,\tilde{f}_{2N,N}(z)\ ,\\
	\tilde{q}_{i,j}&=\left(\prod_{b}q_{b,b}\right)^{-\frac{c_N-c_i-c_j}{C}}q^{-1}_{i,i}\,q^{-1}_{j,j}\,\tilde{f}_{i,j}(z)\ ,\ i=1,...,N-1\ ,\ j=N+1,...,2N-1\ .}
The functions $f(z),\ \tilde{f}(z)$ are general functions independent of $\zb$ that come from integrating over $\zb$ in \eqref{dlog}. Given the boundary conditions $\lim_{r\to\infty} q_{aa}=ve^{ik_a\phi}$, $q_{aa}$ has $k_a$ zeros on the $x^1-x^2$ plane. The functions  $f(z),\ \tilde{f}(z)$ should be restricted such that the scalars $q_{ab},\ \tilde{q}_{ab}$ vanish at $r\to\infty$ and are regular everywhere. For simplicity, the solution is written here for the case where all the zeros of $q_{aa}$ coincide at $r=0$. 
\eql{f}{f_{i,0}(z)&=\sum_n z^{n-\frac{(c_0-c_i)K}{C}-k_i}\rho_{i,0}^{(n)}\ ,\ 0\leq n< \frac{(c_0-c_i)K}{C}+k_i\ ,\\
f_{N,0}(z)&=\sum_n z^{n-\frac{(c_0+\sum_{j=1}^{N-1}c_j)K}{C}+\sum_{j=1}^{N-1}k_j}\rho_{N,0}^{(n)}\ ,\ 0\leq n< \frac{(c_0+\sum_{j=1}^{N-1}c_j)K}{C}-\sum_{j=1}^{N-1}k_j\ ,\\
f_{2N,i}(z)&=\sum_n z^{n-\frac{(c_{2N}-c_i)K}{C}-k_i}\rho_{2N,i}^{(n)}\ ,\ 0\leq n< \frac{(c_{2N}-c_i)K}{C}+k_i\ ,\\
f_{2N,N}(z)&=\sum_n z^{n-\frac{(c_{2N}+\sum_{j=N+1}^{2N-1}c_j)K}{C}+\sum_{j=N+1}^{2N-1}k_j}\rho_{2N,N}^{(n)}\ ,\ 0\leq n< \frac{(c_{2N}+\sum_{j=N+1}^{2N-1}c_j)K}{C}-\sum_{j=N+1}^{2N-1}k_j\ ,\\
f_{i,j}(z)&=\sum_n z^{n-\frac{(c_N-c_i-c_j)K}{C}-k_i-k_j}\rho_{i,j}^{(n)}\ ,\ 0\leq n< \frac{(c_N-c_i-c_j)K}{C}+k_i+k_j\ ,}

and
\eql{ftilde}{\tilde{f}_{i,0}(z)&=\sum_n z^{n+\frac{(c_0-c_i)K}{C}+k_i}\tilde{\rho}_{i,0}^{(n)}\ ,\ 0\leq n< -\frac{(c_0-c_i)K}{C}-k_i\ ,\\
\tilde{f}_{N,0}(z)&=\sum_n z^{n+\frac{(c_0+\sum_{j=1}^{N-1}c_j)K}{C}-\sum_{j=1}^{N-1}k_j}\tilde{\rho}_{N,0}^{(n)}\ ,\ 0\leq n< -\frac{(c_0+\sum_{j=1}^{N-1}c_j)K}{C}+\sum_{j=1}^{N-1}k_j\ ,\\
\tilde{f}_{2N,i}(z)&=\sum_n z^{n+\frac{(c_{2N}-c_i)K}{C}+k_i}\tilde{\rho}_{2N,i}^{(n)}\ ,\ 0\leq n< -\frac{(c_{2N}-c_i)K}{C}-k_i\ ,\\
\tilde{f}_{2N,N}(z)&=\sum_n z^{n+\frac{(c_{2N}+\sum_{j=N+1}^{2N-1}c_j)K}{C}-\sum_{j=N+1}^{2N-1}k_j}\tilde{\rho}_{2N,N}^{(n)}\ ,\ 0\leq n< -\frac{(c_{2N}+\sum_{j=N+1}^{2N-1}c_j)K}{C}+\sum_{j=N+1}^{2N-1}k_j\ ,\\
\tilde{f}_{i,j}(z)&=\sum_n z^{n+\frac{(c_N-c_i-c_j)K}{C}+k_i+k_j}\tilde{\rho}_{i,j}^{(n)}\ ,\ 0\leq n< -\frac{(c_N-c_i-c_j)K}{C}-k_i-k_j\ .}
The parameters $\rho,\ \tilde{\rho}$ are arbitrary complex numbers which parametrize the size modes. For general zeros, the solution is modified as in equation (4.28) of \citep{Gerchkovitz:2017kyi}. However, the number of zero modes and the asymptotic behaviour will not be affected.
The conditions for bulk-string decoupling and for the weak to weak mapping can be read directly from equations \eqref{f},\eqref{ftilde}.

\underline{\textbf{Bulk-string decoupling}}: Decoupling happens if there are no long range size modes that decay slower than $\onov{r}$. The conditions for this are
	\eql{deccondition}{\frac{(c_0-c_i)K}{C}\ ,\ \frac{(c_0+\sum_{j=1}^{N-1}c_j)K}{C}\ ,\ \frac{(c_{2N}-c_{N+i})K}{C}\ ,\ \frac{(c_{2N}+\sum_{j=N+1}^{2N-1}c_j)K}{C}\ ,\ \frac{(c_N-c_i-c_{N+i'})K}{C}\in \mathbb{Z} }
	for every $i,i'=1,...,N-1$.
	This condition also coincides with the condition for no non-trivial Aharonov-Bohm phases.

 \underline{\textbf{Weak to weak mapping}}: The mapping of parameters from the four dimensional theory to the two dimensional worldsheet theory is weak to weak if all the F-term constraints are satisfied trivially. This happens if for every flavor, there are only $q$ or only $\tilde{q}$ modes, but not both. in addition to the size modes, there are also the off-diagonal modes as explained above. In particular, the scalars $q_{iN},\ q_{Nj}$ with $i=1,...,N-1$ and $j=N+1,...,2N-1$ give rise to the off-diagonal modes and therefore we must forbid $\tilde{q}$ modes for the entire bi-fundamental. It means that $\tilde{q}_{ij}=0$ which leads to the condition \eql{weakcondition}{\frac{(c_N-c_i-c_j)K}{C}\geq 0\ ,\ i=1,...,N-1\ ,\ j=N+1,...,2N-1\ .} From the first column and the last row, we get four possibilities for weak to weak mapping:
 \begin{enumerate}
\item $c_0\geq c_i \forall 1\leq i\leq N-1\ ,\ \sum_{j=0}^{N-1}c_j\geq C\ ,\ c_{2N}\geq c_i \forall N+1\leq i\leq 2N-1\ ,\ \sum_{j=N+1}^{2N}c_j\geq C$.
\item $c_0\geq c_i \forall 1\leq i\leq N-1\ ,\ \sum_{j=0}^{N-1}c_j\geq C\ ,\ c_{2N}<c_i \forall N+1\leq i\leq 2N-1\ ,\ \sum_{j=N+1}^{2N}c_j< C$.
\item $c_0< c_i \forall 1\leq i\leq N-1\ ,\ \sum_{j=0}^{N-1}c_j< C\ ,\ c_{2N}\geq c_i \forall N+1\leq i\leq 2N-1\ ,\ \sum_{j=N+1}^{2N}c_j\geq C$.
\item $c_0< c_i \forall 1\leq i\leq N-1\ ,\ \sum_{j=0}^{N-1}c_j< C\ ,\ c_{2N}< c_i \forall N+1\leq i\leq 2N-1\ ,\ \sum_{j=N+1}^{2N}c_j< C$.
\end{enumerate}
In the cases where the two conditions are satisfied, i.e. the bulk and the string decouple at low energies, and all the F-term constraints vanish identically, we give an ansatz for the low energy worldsheet theory with topological charge $K=1$. 
We will show that the given ansatz is consistent both with the classical zero modes analysis and with results obtained from localization.

\subsection{No $\tilde{q}$ case}
\label{notildecase}
\begin{SCfigure}
	\vspace{10pt}
         	\includegraphics[width=0.6\textwidth, height=8cm]{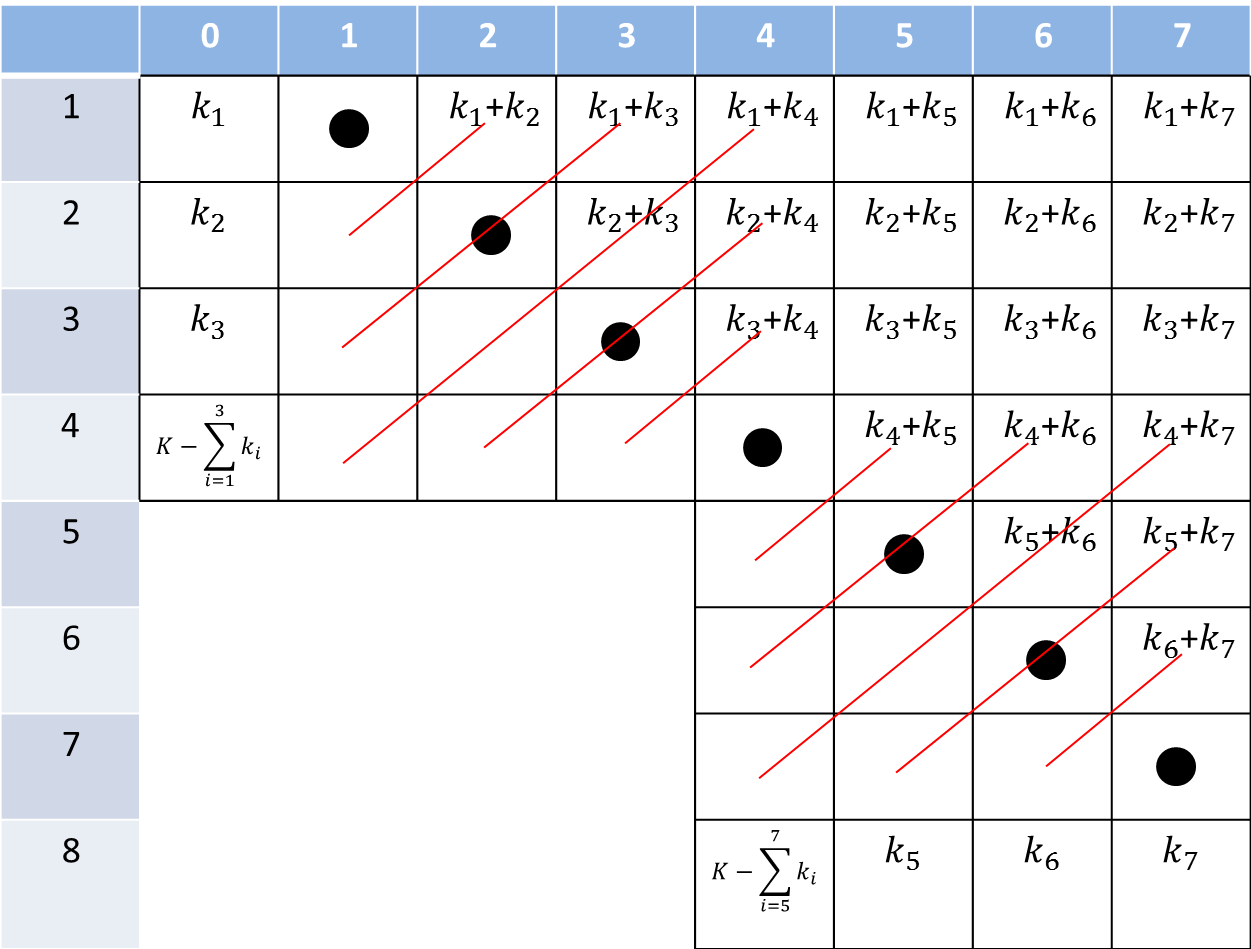}\label{chargedmodes}
  \caption{\small{This figure shows the number of intercating modes from every hypermultiplet in the case described in \ref{notildecase}. Every pair of scalars $q_{ab},\ q_{ba}$ give rise to $k_a+k_b$ off-diagonal modes. Their boxes are connected by red lines and the number of modes is written only in one of the boxes. The size modes appear in the boxes without red line on them. }}
	\vspace{10pt}
\end{SCfigure}

In this section we will describe the worldsheet theory in the case where there are no $\tilde{q}$ excitations. This happens when 
\eq{c_0\geq c_i\ \forall\ 1\leq i\leq N-1\ ,\ \sum_{j=0}^{N-1}c_j\geq C\ ,\ c_{2N}\geq c_i \forall N+1\leq i\leq 2N-1\ ,\ \sum_{j=N+1}^{2N}c_j\geq C\ .}
In addition, we also assume that the other conditions required for bulk-string decoupling and weak to weak mapping \eqref{deccondition} and \eqref{weakcondition} are satisfied.
The size modes are given by the parameters $\rho$ of equation \eqref{f}. It is usefull to distinguish between two types of size modes. The size modes that exist for every choice of the partition $\{k_a\}$ give rise to decoupled chiral fields on the worldsheet. 
\begin{itemize}
\item $\rho_{i,0}^{(n)}$ with $n=k_i,...,\frac{(c_0-c_i)K}{C}+k_i-1$.
\item $\rho_{N,0}^{(n)}$ with $n=K-\sum_{j=1}^{N-1}k_j,...,\frac{K\sum_{j=0}^{N-1}c_j}{C}-\sum_{j=1}^{N-1}k_j-1$.
\item $\rho_{2N,i}^{(n)}$ with $n=k_i,...,\frac{c_{2N}-c_i)K}{C}+k_i-1$.
\item $\rho_{2N,N}^{(n)}$ with $n=K-\sum_{j=N+1}^{2N-1}k_j,...,\frac{K\sum_{j=N+1}^{2N}c_j}{C}-\sum_{j=N+1}^{2N-1}k_j-1$.
\item $\rho_{i,j}^{(n)}$ with $n=k_i+k_j,...,\frac{(c_N-c_i-c_j)K}{C}+k_i+k_j-1$.
\end{itemize}
Stripping off the decoupled modes and ignoring the center of mass modes, we are left with the interacting size modes and off-diagonal modes as can be seen in the tetris diagram \ref{chargedmodes}.

There are $2NK$ complex interacting modes. The fact that the number of modes is independent of the partition $\{k_a\}$ is a sign for weak$\to$weak mapping. Our ansatz is that the $K=1$ worldsheet theory in this case is given by the low energy limit of a two-dimensional $\cN=(2,2)$ $U(1)\times U(1)$ gauged linear sigma model (GLSM) with two complexified FI parameters
\eq{t_{1,2}=\tau_{1,2}\ ,}
where $\tau_a=\frac{\te_a}{2\pi}+\frac{4\pi i}{g_a^2}$ is the 4d complexified gauge coupling, and $t_a=\frac{\te_a^{(2d)}}{2\pi}+i\xi_a$ is the complexified 2d FI parameter.
In addition, the theory contains the following chiral fields
\begin{itemize}
	\item $X$, parametrizing the center of mass modes.
	\item $\psi_{I}^\pm$ with $I=0,...,N$, parametrizing the interacting size modes and off-diagonal modes.
	\item  $\eta_{i,j,r}$ with $i=1,...,N-1,\ j=N+1,...,2N-1,\ r=1,...,\frac{(c_N-c_i-c_j)}{C}$.
	\item $\eta_{0,j,r}$ with $j=1,...,N-1,\ r=1,...,\frac{(c_0-c_i)}{C}$.
	\item $\eta_{0,N,r}$ with $r=1,...,\frac{\sum_{j=0}^{N-1}c_j}{C}-1$.
	\item $\eta_{2N,j,r}$ with $j=N+1,...,2N-1,\ r=1,...,\frac{(c_{2N}-c_i)}{C}$.
	\item $\eta_{2N,N,r}$ with $r=1,...,\frac{\sum_{j=N+1}^{2N}c_j}{C}-1$.
\end{itemize}
The $\eta$ fields parametrize the size modes that exist for every partition $\{k_a\}$. The quantum numbers of these fields are given in table \ref{tablenotilde}.

\begin{table}[h]
	\caption{\small{The spectrum on the worldsheet in cases where the string admits no $\tilde{q}$ excitations.}}\label{tablenotilde}
	\begin{center}
		\begin{tabular}{|c|c|c|c|c|}
			\hline
			Field&$U(1)\times U(1)$ &Twisted Mass&$R^{(R)}$&$R^{(J)}$\\
			$X$&(0,0)&0&0&2\\
			$\psi_0^-$   & (-1,0)              &   $\mu'_0$        &  $1-\frac{(2N-1)c_0}{C}$    &      $\frac{2c_0}{C}$            \\
			$\psi_N^-$   & (0,-1)              &   $\mu'_{2N}$        &  $1-\frac{(2N-1)c_{2N}}{C}$      &      $\frac{2c_{2N}}{C}$          \\
			$\psi_0^+$   & (1,0)              &   $\sum_{j=1}^{N-1}\mu'_j$       &  $\sum_{j=1}^{N-1}\left(1-\frac{(2N-1)c_j}{C}\right)$        &      $\sum_{j=1}^{N-1}\frac{2c_j}{C}$         \\$\psi_N^+$   & (0,1)              &   $\sum_{j=N+1}^{2N-1}\mu'_j$       &  $\sum_{j=N+1}^{2N-1}\left(1-\frac{(2N-1)c_j}{C}\right)$     &      $\sum_{j=N+1}^{2N-1}\frac{2c_j}{C}$            \\
			$\psi_J^+$&(1,-1)&$-\mu'_J-\sum_{i=N+1}^{2N-1}\mu'_i $&  $\frac{2N-1}{C}\left(c_J+\sum_{i=N+1}^{2N-1}c_i\right)-N$ & $\frac{2}{C}\left(\sum_{i=1}^{N}c_i-c_J\right)$ \\
			$\psi_J^-$&(-1,1)&$-\mu'_{N+J}-\sum_{i=1}^{N-1}\mu'_i $&  $\frac{2N-1}{C}\left(c_{N+J}+\sum_{i=1}^{N-1}c_i\right)-N$ & $\frac{2}{C}\left(\sum_{i=N}^{2N-1}c_i-c_{N+J}\right)$ \\
			
			$\eta_{i,j,r}$& (0,0) &$\mu'_N-\mu'_i-\mu'_j$&  $\frac{(2N-1)(c_i+c_j-c_N)}{C}-1$ & $2r$ \\
			$\eta_{0,i,r}$&(0,0)&   $\mu'_0-\mu'_i$     &    $\frac{(2N-1)(c_i-c_0)}{C} $ &   $2r$       \\
			$\eta_{0,N,r}$&(0,0)&  $\sum_{j=0}^{N-1}\mu'_j$   &    $N-\frac{2N-1}{C}\sum_{j=0}^{N-1}c_j$ &  $2r$  \\
			$\eta_{2N,j,r}$&(0,0)&    $\mu'_{2N}-\mu'_{j}$     &    $\frac{(2N-1)(c_j-c_{2N})}{C} $  &   $2r$     \\
			$\eta_{2N,N,r}$&(0,0)&  $\sum_{j=N+1}^{2N}\mu'_j$  &    $N-\frac{2N-1}{C}\sum_{j=N+1}^{2N}c_j$  &  $2r$  \\
			\hline
		\end{tabular}
	\end{center}
\end{table}
\subsubsection{Comparison with the classical spectrum}
\label{classical}
In this section we will show that the ansatz for the worldsheet theory agrees with classical zero modes analysis. We will start from the decoupled sector. It is straight forward to see that the quantum numbers of the $\eta$ fields coincide with the quantum numbers of the decoupled size modes. These are the $\rho^{(n)}$s of equation \eqref{f} with $n=n_{max}+1-r$.
Now we will move on to the charged sector.
 The worldsheet theory has $2N-1$ vacua that correspond to the $2N-1$ choices of $\{k_a\}$ with $K=1$. Due to the twisted masses, only two chiral fields can get non-trivial VEV. This is allowed thanks to the two gauge multiplets scalars $\sigma_{1,2}$ as the mass terms for the charged fields are
\eq{\lag_{\text{mass}}=&(\sigma_1+\mu'_0)^2|\psi_0^-|^2+\left(\sigma_2+\mu'_{2N}\right)^2|\psi_N^-|^2+\left(\sigma_1-\sum_{j=1}^{N-1}\mu'_j\right)^2|\psi_0^+|^2+\left(\sigma_2-\sum_{j=N+1}^{2N-1}\mu'_j\right)^2|\psi_N^+|^2\\
&+\sum_{J=1}^{N-1}\left(\sigma_1-\sigma_2+\mu'_J+\sum_{i=N+1}^{2N-1}\mu'_i\right)^2|\psi_J^+|^2+\sum_{J=1}^{N-1}\left(\sigma_2-\sigma_1+\mu'_{N+J}+\sum_{i=1}^{N-1}\mu'_i\right)^2|\psi_J^-|^2 \ .} 
The chiral fields that get VEV must satisfy the D-term equations
\eql{WSDterm}{&|\psi_0^+|^2+\sum_{J=1}^{N-1}|\psi_J^+|^2-|\psi_0^-|^2-\sum_{J=1}^{N-1}|\psi_J^-|^2=\zeta_1\ ,\\
&|\psi_N^+|^2+\sum_{J=1}^{N-1}|\psi_J^-|^2-|\psi_N^-|^2-\sum_{J=1}^{N-1}|\psi_J^+|^2=\zeta_2\ .}
The $2N-1$ vacua and the corresponding $\{k_a\}$ partitions are given by
\eql{WSvacua}{k_a=\delta_{a,J}\quad\Rightarrow\quad &\psi_N^+=\zeta_1+\zeta_2\ ,\ \psi_J^+=\zeta_1\ ,\ \sigma_1=-\mu'_J \ ,\ \sigma_2=\sum_{j=N+1}^{2N-1}\mu'_j\ ,\\ k_a=\delta_{a,N}\quad\Rightarrow \quad &\psi_0^+=\zeta_1\ ,\ \psi_N^+=\zeta_2\ ,\ \sigma_1=\sum_{j=1}^{N-1}\mu'_j\ ,\ \sigma_2=\sum_{j=N+1}^{2N-1}\mu'_j\ ,\\
        k_a=\delta_{a,N+J}\quad\Rightarrow\quad &\psi_0^+=\zeta_1+\zeta_2\ ,\ \psi_{J}^-=\zeta_2\ ,\ \sigma_1=\sum_{j=1}^{N-1}\mu'_j  \ ,\ \sigma_2=-\mu'_{N+J}\ .\\
}
Lets focus for example on the vacua appearing on the first line of \eqref{WSvacua}. By plugging in the VEV for $\sigma_{1,2}$, we find that the masses of the dynamical fields around the vacuum are given by
\eq{\lag_{\text{mass},J}=&(\mu'_0-\mu'_J)^2|\psi_0^-|^2+\left(\sum_{j=N+1}^{2N}\mu'_j\right)^2|\psi_N^-|^2+\left(\mu'_J+\sum_{j=1}^{N-1}\mu'_j\right)^2|\psi_0^+|^2\\
&+\sum_{I=1,I\neq J}^{N-1}\left(\mu'_J-\mu'_I\right)^2|\psi_I^+|^2+\sum_{I=1}^{N-1}\left(\mu'_J+\mu'_{N+I}-\mu'_N\right)^2|\psi_I^-|^2 \ .} 

Similarly, the R-symmetries preserved by the vacuum are linear combinations of the original R-symmetries with some gauge transformations. This leads to a shift in the charges of the fields. The shift for a field with $U(1)\times U(1)$ charges $(q_1,q_2)$  is
\eq{&\delta R^{(J)}=-q_2\sum_{j=N+1}^{2N-1}\frac{2c_j}{C}-2q_1\left(1-\frac{c_J}{C}\right)\ ,\\
&\delta R^{(R)}=q_2\left(\sum_{j=N+1}^{2N-1}\frac{(2N-1)c_j}{C}-N+1\right)-q_1\left(\frac{(2N-1)c_J}{C}-1\right)\ .}
The equations for these three quantum numbers can be summarized as
\eq{\delta M_{(q_1,q_2)}=-q_1M_{\psi_J^+}-(q_2+q_1)M_{\psi_N^+}\ ,}
where $M$ can be mass or any of the two R-charges.
Straight forward computation shows agreement between the spectrum of the charged fields in the vacuum to the off-diagonal and interacting size modes around the string vacuum $k_a=\delta_{aJ}$, with the identifiaction
\begin{itemize}
\item $\psi_0^-$ with the size mode $\rho_{J,0}^{(0)}$.
\item $\psi_N^-$ with the size mode $\rho_{2N,N}^{(0)}$.
\item $\psi_0^+$ with the off-diagonal mode of the pair $q_{N,J},\ q_{J,N}$.
\item $\psi_{I\neq J}^+$ with the off-diagonal mode of the pair $q_{I,J},\ q_{J,I}$.
\item $\psi_I^-$ with the size modes $\rho_{J,N+I}^{(n)}$.
\end{itemize}

The quantum numbers of the charged fields in the vacuum are exactly given by the $\hat{I}_{AB}$ of equations \eqref{multiindexeq} as summarized in table \ref{tablevacuum}. 
\begin{table}[h]
	\caption{\small{The spectrum of the charged sector around the first vacuum in equation \eqref{WSvacua}.}}\label{tablevacuum}
	\begin{center}
		\begin{tabular}{|c|c|c|c|c|}
			\hline
			Field&Twisted Mass&$R^{(R)}$&$R^{(J)}$\\
			$\psi_0^-$     &   $\mu'_0-\mu'_J$        &  $\frac{(2N-1)(c_J-c_0)}{C}$    &      $2+\frac{2(c_0-c_J)}{C}$            \\
			$\psi_N^-$    &   $\sum_{j=N+1}^{2N}\mu'_{j}$        &  $N-\frac{(2N-1)}{C}\sum_{j=N+1}^{2N}c_j$      &      $\frac{2}{C}\sum_{j=N+1}^{2N}c_j$          \\
			$\psi_0^+$   &   $\mu'_J+\sum_{j=1}^{N-1}\mu'_j$       &  $\sum_{j=1}^{N-1}\left(1-\frac{(2N-1)c_j}{C}\right)+1-\frac{(2N-1)c_J}{C}$        &      $\sum_{j=1}^{N-1}\frac{2c_j}{C}-2+\frac{2c_J}{C}$         \\
			$\psi_{I\neq J}^+$&$\mu'_J-\mu'_I$&  $\frac{(2N-1)(c_I-c_J)}{C}$ & $\frac{2(c_J-c_I)}{C}$ \\
			$\psi_I^-$&$\mu'_J+\mu'_{N+I}-\mu'_N $&  $\frac{2N-1}{C}\left(c_{N+I}+c_J-c_N\right)-1$ & $\frac{2}{C}\left(c_N-c_{N+I}-c_J\right)+2$ \\
			
			\hline
		\end{tabular}
	\end{center}
\end{table}
\subsection{Including $\tilde{q}$ excitations}
\label{withqtilde}
In this section we will describe the worldsheet theory in the cases where some of the hypermultiplets admit $\tilde{q}$ excitations but still the F-term constraints are satisfied trivially. This happens in one of the following three cases
 \begin{enumerate}
\item $c_0\geq c_i \forall 1\leq i\leq N-1\ ,\ \sum_{j=0}^{N-1}c_j\geq C\ ,\ c_{2N}<c_i \forall N+1\leq i\leq 2N-1\ ,\ \sum_{j=N+1}^{2N}c_j< C$.
\item $c_0< c_i \forall 1\leq i\leq N-1\ ,\ \sum_{j=0}^{N-1}c_j< C\ ,\ c_{2N}\geq c_i \forall N+1\leq i\leq 2N-1\ ,\ \sum_{j=N+1}^{2N}c_j\geq C$.
\item $c_0< c_i \forall 1\leq i\leq N-1\ ,\ \sum_{j=0}^{N-1}c_j< C\ ,\ c_{2N}< c_i \forall N+1\leq i\leq 2N-1\ ,\ \sum_{j=N+1}^{2N}c_j< C$.
\end{enumerate}
It is easy to see that almost all the analysis will be the same as in the previous section. The only difference comes from the size modes analysis of the first column or/and the last row. In case 1 in the above list, $\rho_{2N,i}^{(n)}$ and $\rho_{2N,N}^{(n)}$ of \eqref{f} are replaced with $\tilde{\rho}_{2N,i}^{(n)}$ and $\tilde{\rho}_{2N,N}^{(n)}$ of \eqref{ftilde}. In case 2, $\rho_{i,0}^{(n)}$ and $\rho_{N,0}^{(n)}$ of \eqref{f} are replaced with $\tilde{\rho}_{i,0}^{(n)}$ and $\tilde{\rho}_{N,0}^{(n)}$ of \eqref{ftilde}. In case 3, the two replacements should be made. From the worldsheet point of view, these replacements include two changes. One is a trivial change in the spectrum of the decoupled fields $\eta\to\tilde{\eta}$. The second change includes adding neutral fields and couple them to the charged fields via a superpotential. Consider for example the first case in the list. The spectrum is the same as in table \ref{tablenotilde} with the following changes: Replace $\eta_{2N,j,r},\ \eta_{2N,N,r}$ with the decoupled fields $\tilde{\eta}_{2N,j,r}$ with $j=N+1,...,2N-1\ ,\ r=1,...,\frac{(c_j-c_{2N})K}{C}-K$  and $\tilde{\eta}_{2N,N,r}$ with $r=1,...,-\frac{K}{C}\sum_{j=N+1}^{2N}c_j$. Their quantum numbers appear in table \ref{tabledecoupledtilde}.
\begin{table}[h]
	\caption{\small{The decoupled fields that come from $\tilde{q}$ size mode in the first case of \ref{withqtilde}.}}\label{tabledecoupledtilde}
	\begin{center}
		\begin{tabular}{|c|c|c|c|}
			\hline
			Field&Twisted Mass&$R^{(R)}$&$R^{(J)}$\\
			$\tilde{\eta}_{2N,j,r}$&$\mu'_{j}-\mu'_{2N}$& $2+\frac{(2N-1)(c_{2N}-c_j)}{C}$& $2r$\\
                                $\tilde{\eta}_{2N,N,r}$&$-\sum_{i=N+1}^{2N}\mu'_i$& $2-N+\frac{(2N-1)}{C}\sum_{i=N+1}^{2N}c_i$& $2r$\\
			\hline
		\end{tabular}
	\end{center}
\end{table}
These fields represent the size modes $\tilde{\rho}_{2N,j}^{(n)}$ and $\tilde{\rho}_{2N,N}^{(n)}$ with $n=n_{max}+1-r$.
The second change invloves adding neutral chiral fields $\chi_{2N}$ and $\chi_i$ with $i=N+1,...,2N-1$ together with the superpotential 
\eq{W=\al_{2N}\chi_{2N}\psi_N^+\psi_N^-+\al_J\chi_{N+J}\psi_0^+\psi_J^-\psi_N^-\ ,}
where $\al_{2N},\ \al_J$ are some non-zero coefficients that cannot be fixed by our analysis. The quantum numbers of the $\chi$ fields are fixed from the superpotential. We will show that due to the superpotential, the field $\psi_N^-$ is fixed to be zero on the target space. From the superpotential, we get (among others) the constraints 
\eq{\psi_N^+\psi_N^-=0\ ,\ \psi_0^+\psi_J^-\psi_N^-=0\ \forall\ 1\leq J\leq N-1\ .}
In order to satisfy the D-term equations \eqref{WSDterm}, we must have $\psi_N^+\neq 0$ or $\psi_0^+\psi_J^-\neq 0$ for some $1\leq J\leq N-1$, which means that $\psi_N^-=0$ on every point on the targetspace. The other non-trivial equation we get from the superpotential is
\eq{\frac{\partial W}{\partial \psi_N^-}=\al_{2N}\psi_N^+\chi_{2N}+\sum_{J=1}^{N-1}\al_J\psi_0^+\psi_J^-\chi_{N+J}=0\ .}
On the vacua \eqref{WSvacua}, one of the $\chi$s vanishes and we are left with $N-1$ $\chi$s. They represent the interacting size modes $\tilde{\rho}_{2N,N}^{(n)},\ \tilde{\rho}_{2N,j}^{(n_j)}$ with $n=0,...,\sum_{i=N+1}^{2N-1}k_i$ and $n_j=0,...,1-k_j$. There are $N-1$ such modes at any vacuum \eqref{WSvacua}, and there is an exact agreement between the quantum numbers of the $\chi$s and the $\tilde{\rho}$s. As an example consider the vacuum described on the first line of \eqref{WSvacua}. At this vacuum $\psi_N^+\neq 0$ and therefore, $\chi_{2N}=0$. Similarly, in this vacuum $k_a=\delta_{a,J}$ for some $J=1,...,N-1$. This implies that there are no $\tilde{\rho}_{2N,N}^{(n)}$ modes and 1 mode for every $\tilde{\rho}_{2N,j}^{(0)}$.
Due to the obvious symmetry between the first column and the last row of the tetris diagram, the other cases in the list presented at the beginning of \ref{withqtilde} will be exactly the same.

\section{Strings in $SU(2)^M\times U(1)$ theories}
\label{su2M}
\begin{SCfigure}
		\vspace{10pt}
		\includegraphics[width=0.65\textwidth, height=8cm]{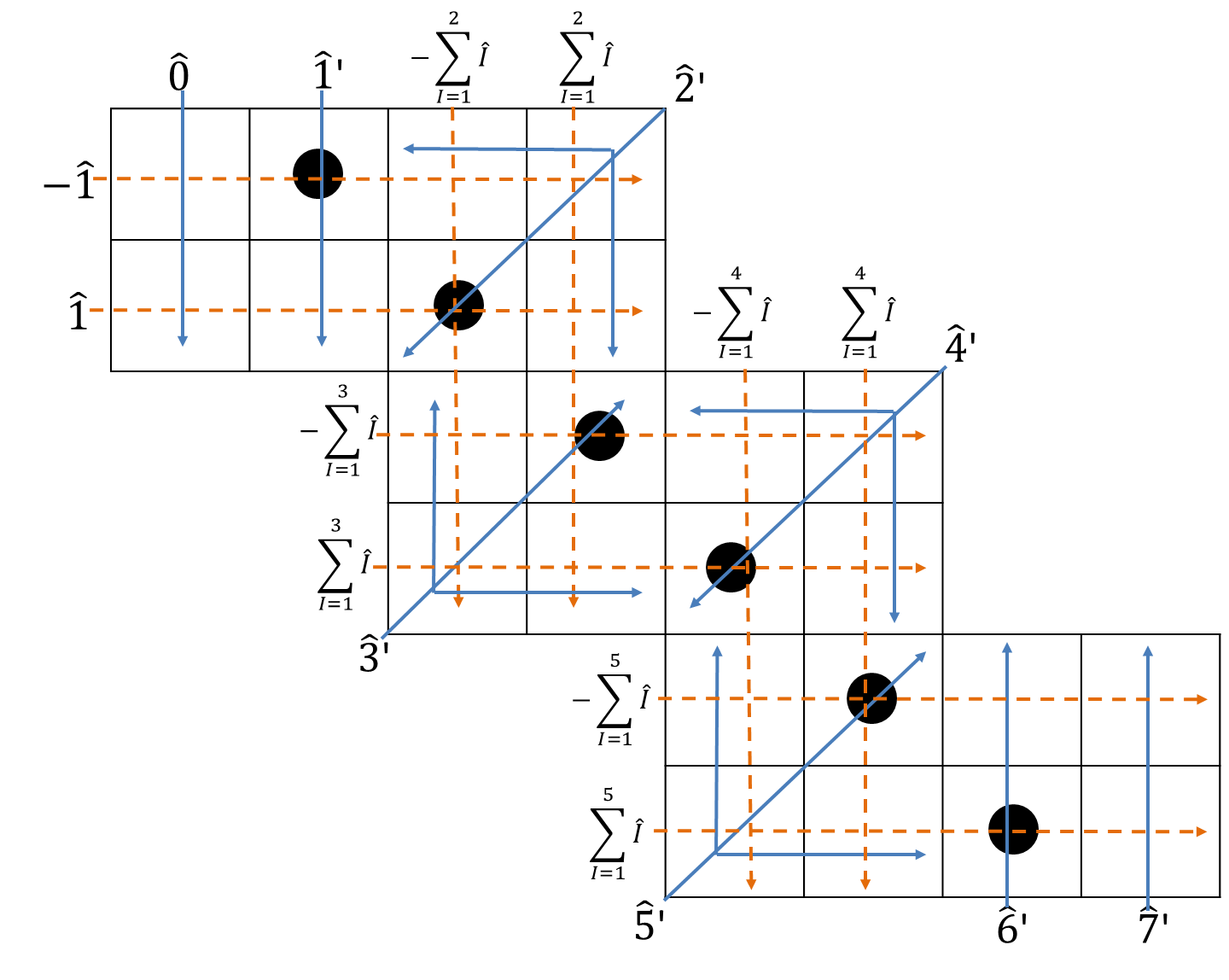}
		\caption{\small{Similar to figure \ref{Imultiindex}, this figure represents the masses and R-charges of all the hypermultiplets. $\hat{I}_{ab}=\{\text{Mass},\ R^{(R)},\ R^{(J)}\}$ of every hypermultiplet is given by the sum of the $\hat{I}$s that appear on the tail of the arrows crossing the corresponding block where $\hat{I}=\{\mu'_I\ ,\ 1-\frac{(M+1)c_I}{C}\ ,\ \frac{2Kc_I}{C}-2k_I\}$ and $\hat{I}'=\hat{I}+\{0,0,2k_I\}$. Notice that $\sum_{I=1}^{M+1}\hat{I}=0$. See figure  \ref{Imultiindex} for a more detailed explanation.}}\label{su2quiver}
		\vspace{10pt}
	\end{SCfigure}

In this section we will study the $SU(2)^M$ quiver theories in which we gauge some U(1) flavor symmetry. We will focus on the fully Higgsed vacua in which $M+1$ scalars get VEV. The tetris diagram contains $M+1$ $2\times 2$ blocks. Each one of the $M-1$ inner blocks represent one bifundamental and therefore the entire block should be accompannied with one mass and one U(1) charge. Each of the two outer blocks represent two fundamentals and therefore it is accompannied with two masses and two U(1) charges. There are $M+3$ masses $\{\mu_i\}$ and charges $\{c_i\}$ which are labelled by the index $i,j=0,...,M+2$. without loss of generality, we will give VEV to the boxes that sit on the same diagonal. We will denote the boxes by two indices, where the first row is denoted by 1 and the first column is denoted by 0, such that the dots are located in $q_{aa}$. See figure \ref{su2quiver}. The gauge symmetry is broken in this vacuum to $\mathbb{Z}_C$ where $C=\sum_{a=1}^{M+1}c_a$. The spectrum of strings above this vacuum is given by $\pi_1\left(SU(2)^M\times U(1)/\mathbb{Z}_C\right)$ which allows fractional magnetic fluxes quantized as $\Phi_{u(1)}^K=\frac{2\pi K}{C}$ with $K\in\mathbb{Z}$. Repeating the same procedure as in the previous section, we construct the string by changing the boundary conditions to
\eql{ka}{\lim_{r\to\infty} q_{aa}=ve^{ik_a\phi}\ ,} where $k_a$ are non-negative integers. The asymptotic value of the gauge field can be easily computed from the demand that the tension is finite. The U(1) flux carried by this string is
\eq{\Phi_{u(1)}=\lim_{r\to\infty}\int d\phi A'_\phi=\frac{2\pi K}{C}\ , K=\sum_{a=1}^{M+1} k_a} in agreement with the allowed spectrum.
As before, we will compute the mass and two $U(1)$ R-charges of every hypermultiplet around the string solution. 
\begin{itemize}
\item Mass: The VEV of the adjoint scalars should be chosen such that the mass terms of $q_{aa}$ vanish. Denoting by $a'$ the U(1) scalar and by $a_I$ the Cartan scalar of the I'th SU(2) gauge group, the following equations must be satisfied in the vacuum
\eq{&c_1a'+a_1=\mu_1\ ,\ c_{M+1}a'-a_M=\mu_{M+1}\ ,\\
	&c_Ia'+a_I-a_{I-1}=\mu_I\ ,\ 2\leq I\leq M\ .}
These equations are solved by
\eq{a'=\mu/C\ ,\ a_I=\sum_{J=1}^I\mu_J'\ ,\ \mu\equiv \sum_{I=1}^{M+1}\mu_I\ ,\ \mu_I'\equiv\mu_I-\frac{\mu}{C}c_I\ .}
\item $R^{(R)}$-charge: the vacuum preserves a $U(1)$ R-charge which is the original $U(1)_R\in SU(2)_R$ accompannied by a Cartan gauge transformation that keeps the vacuum invariant. This R-symmetry transformation should satisfy
\eq{&c_1\omega'+\omega_1=\al\ ,\ c_{M+1}\omega'-\omega_M=\al\ ,\\
	&c_I\omega'+\omega_I-\omega_{I-1}=\al\ ,\ 2\leq I\leq M}
where $\omega',\ \omega_I$ are the gauge parameters related to the U(1) and the Cartan of the I'th SU(2) respectivly, and $\al$ is the $U(1)_R\in SU(2)_R$ parameter. These equations are solved by \eq{\omega'=(M+1)\al/C\ ,\ \omega_I=\sum_{J=1}^I\left(1-\frac{(M+1)c_J}{C}\right)\al\ .}
\item $R^{(J)}$-charge: The string solution where all the cores of the string coincide, preserves a combination of rotation and gauge transformations. This symmetry transformation should satisfy
\eq{&2\phi_0 k_1+c_1\omega'+\omega_1=0 \ ,\ 2\phi_0k_{M+1}+c_{M+1}\omega'-\omega_M=0\ ,\\
&2\phi_0 k_I+c_I\omega'+\omega_I-\omega_{I-1}=0\ ,\ 2\leq I\leq M\ .}
These equations are solved by
\eq{\omega'=-\frac{2K\phi_0}{C}\ ,\ \omega_I=\sum_{J=1}^I\left(\frac{2Kc_J}{C}-2k_J\right)\phi_0\ .}
\end{itemize}
Using the previous computations, we can write the mass and R-charges of the hypermultiplets around the string. The results are represented on the tetris diagram \ref{su2quiver}. 
Our next step is to derive the spectrum of size modes which are solutions to equations \eqref{Bogomolnyi} with the boundary conditions \eqref{ka}. This results in\footnote{As in equations \eqref{f}, \eqref{ftilde}, we write the solution for the case where all the zeros of the string coincide at the origin to simplify expressions.}
\eql{bisizemodes}{q_{a\pm 1,a\mp 1}&=\left(\prod_b q_{bb}\right)^{2c_a/C}q_{aa}^{-1}f_{a}(z)\ ,\ f_a(z)=\sum_n\frac{\rho^{(n)}_{a}}{z^{2Kc_a/C-k_a-n}}\ ,\ 0\leq n<\frac{2Kc_a}{C}-k_a\\
\tilde{q}_{a\pm 1,a\mp 1}&=\left(\prod_b q_{bb}\right)^{-2c_a/C}q_{aa}\tilde{f}(z)\ ,\ \tilde{f}_a(z)=\sum_n\frac{\tilde{\rho}^{(n)}_{a}}{z^{k_a-2Kc_a/C-n}}\ ,\ 0\leq n<k_a-\frac{2Kc_a}{C}}
for $2\leq a\leq M$. These are excitations of the light hypermultiplets inside every bifundamental.\footnote{The $\pm$ signs in \eqref{bisizemodes} are chosen such that the box $q_{a\pm1,a\mp1}$ is inside the tetris diagram. It means that we take $-$ sign for even $a$ and $+$ sign for odd $a$.} 
The other light hypermultiplets are the ones in the first column denoted by $q_{10},\ q_{20}$ and the ones in the last row (column) which are denoted by $q_{M+2,M},\ q_{M+2,M+1}$ ($q_{M,M+2},\ q_{M+1,M+2}$). 
The size modes analysis for these hypermultiplets results in
\eql{moresizemodes}{&q_{10}=\left(\prod_b q_{bb}\right)^{(c_0-c_1)/C}q_{11}f_{10}(z)\ ,\ f_{10}(z)=\sum_{n}\frac{\rho^{(n)}_{1,0}}{z^{\frac{(c_0-c_1)K}{C}+k_1-n}}\ ,\ 0\leq n<\frac{(c_0-c_1)K}{C}+k_1\\
&q_{20}=\left(\prod_b q_{bb}\right)^{(c_0+c_1)/C}q_{11}^{-1}f_{20}(z)\ ,\ f_{20}(z)=\sum_{n}\frac{\rho^{(n)}_{2,0}}{z^{\frac{(c_0+c_1)K}{C}-k_1-n}}\ ,\ 0\leq n<\frac{(c_0+c_1)K}{C}-k_1\\
&\tilde{q}_{10}=\left(\prod_b q_{bb}\right)^{(c_1-c_0)/C}q_{11}^{-1}\tilde{f}_{10}(z)\ ,\ \tilde{f}_{10}(z)=\sum_{n}\frac{\tilde{\rho}^{(n)}_{1,0}}{z^{\frac{(c_1-c_0)K}{C}-k_1-n}}\ ,\ 0\leq n<\frac{(c_1-c_0)K}{C}-k_1\\
&\tilde{q}_{20}=\left(\prod_b q_{bb}\right)^{-(c_0+c_1)/C}q_{11}\tilde{f}_{20}(z)\ ,\ \tilde{f}_{20}(z)=\sum_{n}\frac{\tilde{\rho}^{(n)}_{2,0}}{z^{\frac{-(c_0+c_1)K}{C}+k_1-n}}\ ,\ 0\leq n<\frac{-(c_0+c_1)K}{C}+k_1\ ,}
and exactly the same for the last row/column with the replacement of $0\to M+2\ ,\ 1\to M+1$.
The conditions for bulk-string decoupling and the for weak to weak mapping can be read directly from equations \eqref{bisizemodes},\eqref{moresizemodes}.

\underline{\textbf{Bulk-string decoupling}}: Decoupling happens if there are no long range size modes that decay slower than $\onov{r}$. The conditions for this are
	\eql{deccondition2}{\frac{(c_0\pm c_1)K}{C}\ ,\ \frac{(c_{M+2}\pm c_{M+1})K}{C}\ ,\ \frac{2Kc_a}{C}\in \mathbb{Z}\ \forall\ 2\leq a\leq M\ .}
		This condition also coincides with the condition for no non-trivial Aharonov-Bohm phases.

 \underline{\textbf{Weak to weak mapping}}: The mapping of parameters from the four dimensional theory to the two worldsheet theory is weak$\to$ weak if all the F-term constraints are satisfied trivially. This happens if for every flavor, there are only $q$ or only $\tilde{q}$ modes, but not both. in addition to the size modes, there are also the off-diagonal modes as explained above. These forbid $\tilde{q}$ modes for all bi-fundamental size modes \eqref{bisizemodes}. It means that $\tilde{q}_{a\pm1,a\mp1}=0$  for every $2\leq a\leq M$, which leads to the condition \eql{weakcondition2}{2c_a\geq C\ \forall\ 2\leq a\leq M\ .} From the first column, we get that if $c_0\geq c_1$, then $c_0+c_1\geq C$ and vice versa. Similarly, from the last row/column we get that if $c_{M+2}\geq c_{M+1}$, then $c_{M+2}+c_{M+1}\geq C$ and vice versa. We will assume that all these conditions are satisfied, with $c_0\geq c_1$ and $c_{M+2}\geq c_{M+1}$. The other cases can be dealt similarly with $q\leftrightarrow \tilde{q}$.\footnote{This is special for SU(2) quivers because the fundamental representation of SU(2) is pseudo-real. In SU(N) theories with $N>2$, the different cases should be studied independently as done in the previous section.}
We will give an ansatz for the low energy worldsheet theory with topological charge $K=1$, and show that the given ansatz is consistent both with the classical zero modes analysis and with results obtained from localization.
\subsection{$K=1$ worldsheet theory}
\label{secsu2ansatz}
\begin{SCfigure}
	\vspace{10pt}
         	\includegraphics[width=0.6\textwidth, height=8cm]{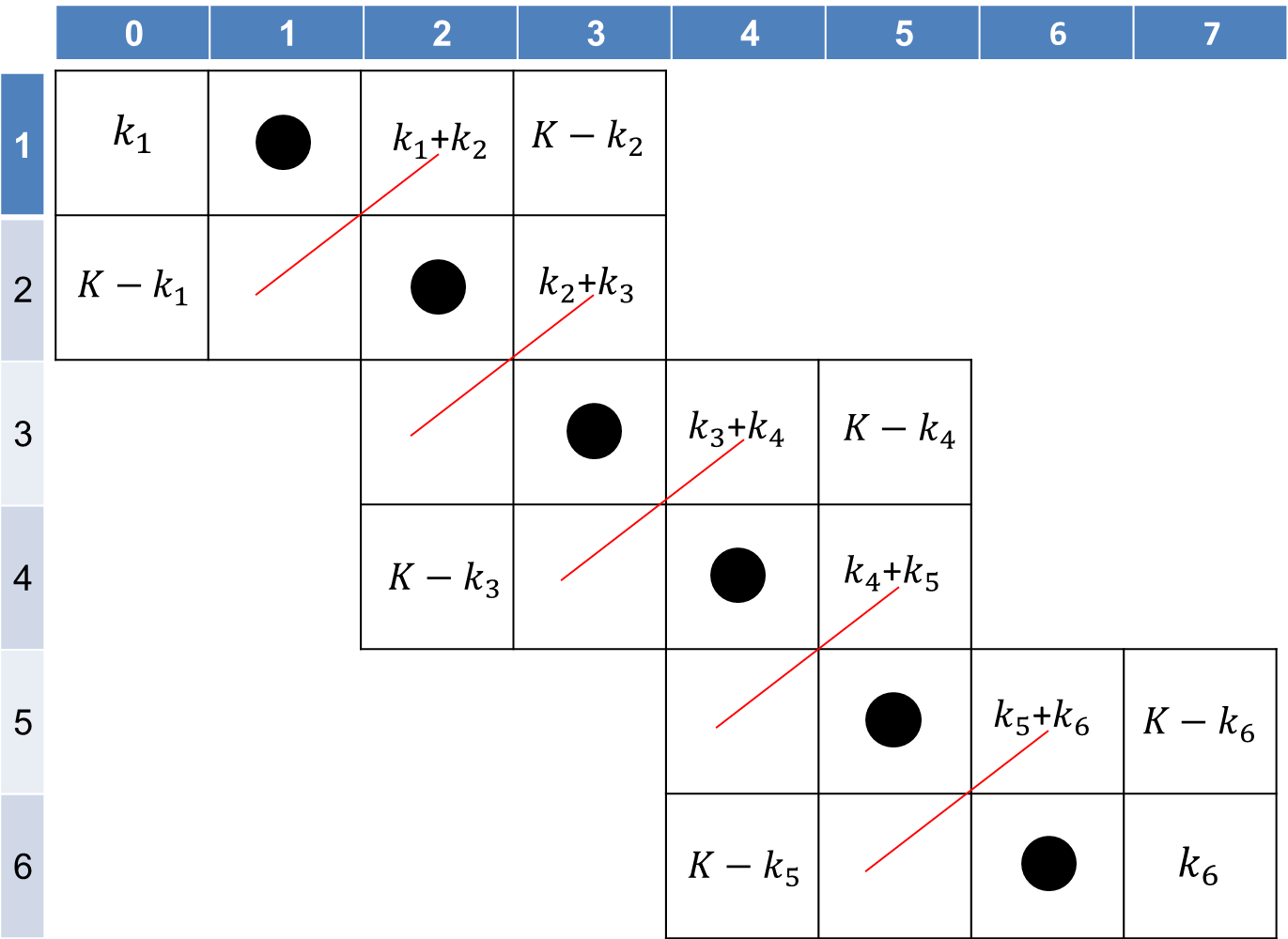}\label{chargedsu2}
  \caption{\small{This figure shows the number of interacting modes from every hypermultiplet in the case described in \ref{secsu2ansatz}. Every pair of scalars $q_{ab},\ q_{ba}$ give rise to $k_a+k_b$ off-diagonal modes. Their boxes are connected by red lines and the number of modes is written only in one of the boxes. The size modes appear in the boxes without red line on them. Simple counting results in $(M+2)K$ modes.}}
	\vspace{10pt}
\end{SCfigure}

In this section we will describe the worldsheet theory for the $K=1$ string in the case where \eqref{deccondition2}, \eqref{weakcondition2} are satisfied and \eql{weaknotilde}{c_0\geq c_1\ ,\ c_{M+2}\geq c_{M+1}\ ,\ c_0+c_1\geq C\ ,\ c_{M+2}+c_{M+1}\geq C\ .}
We will keep for now $K$ and restrict to $K=1$ later on. In this case, there are no $\tilde{q}$ excitations. The size modes are given by the parameters $\rho$ in \eqref{bisizemodes} and \eqref{moresizemodes}. We can identify the size modes that exist for every choice of the partition. They give rise to decoupled chiral fields on the worldsheet. These size modes are
\begin{itemize}
\item $\rho_a^{(n)}$ with $n=K-k_a\ ,\ ...\ ,\ \frac{2Kc_a}{C}-k_a-1$.
\item $\rho_{1,0}^{(n)}$ with $n=k_1\ ,\ ...\ ,\ \frac{(c_0-c_1)K}{C}+k_1-1$.
\item $\rho_{2,0}^{(n)}$ with $n=K-k_1\ ,\ ...\ ,\ \frac{(c_0+c_1)K}{C}-k_1-1$.
\item $\rho_{M+1,M+2}^{(n)}$ with $n=k_{M+1}\ ,\ ...\ ,\ \frac{(c_{M+2}-c_{M+1})K}{C}+k_{M+1}-1$.
\item $\rho_{M,M+2}^{(n)}$ with $n=K-k_{M+1}\ ,\ ...\ ,\ \frac{(c_{M+2}+c_{M+1})K}{C}-k_{M+1}-1$.
\end{itemize}
Stripping off the decoupled modes and ignoring the center of mass modes, we are left with the interacting size modes and off-diagonal modes as can be seen in the tetris diagram \ref{chargedsu2}.
Our ansatz for the worldsheet theory in the $K=1$ case is the low energy limit of a two-dimensional $\cN=(2,2)$ $U(1)^M$ GLSM with $M$ complexified FI parameters
\eq{t_{a}=\tau_a\ \forall\ 1\leq a\leq M\ ,} and the following chiral fields:
\begin{itemize}
\item $X$, parametrizing the center of mass modes.
\item $\psi_I^\pm$ with $I=1,...,M+1$, parametrizing the interacting size modes and the off-diagonal modes.
\item $\eta_{i,r}$ with $i=2,...,M$ and $r=1,...,\frac{2c_i}{C}-1$.
\item $\eta_{1,0,r}$ with $r=1,...,\frac{c_0-c_1}{C}$.
\item $\eta_{2,0,r}$ with $r=1,...,\frac{c_0+c_1}{C}-1$.
\item $\eta_{M+1,M+2,r}$ with $r=1,...,\frac{c_{M+2}-c_{M+1}}{C}$.
\item $\eta_{M,M+2,r}$ with $r=1,...,\frac{c_{M+2}+c_{M+1}}{C}-1$.
\end{itemize}
Again, the $\eta$ fields represent the decoupled size modes.
The quantum numbers of these fields are given in table \ref{su2spectrum}.
\begin{table}[h]
	
	\caption{\small{The spectrum on the worldsheet for strings described in \ref{secsu2ansatz}   }}\label{su2spectrum}
	\begin{center}
		\begin{tabular}{|c|c|c|c|c|}
			\hline
			Field&$U(1)^M$&Twisted Mass&$R^{(R)}$&$R^{(J)}$\\
			$X$        &Neutral          &0       &0&    2         \\
			$\psi_1^-$  &(-1,0...,0)     & $\mu'_0$&$1-\frac{(M+1)c_0}{C}$  &$\frac{2c_0}{C}$\\
                                $\psi_{1}^+$&(1,0...,0)  &$\mu'_1$&$1-\frac{(M+1)c_1}{C}$  &$\frac{2c_1}{C}$\\
                                $\psi_{M+1}^-$&(0,...,0,1)&$\mu'_{M+1}$&$1-\frac{(M+1)c_{M+1}}{C}$  &$\frac{2c_{M+1}}{C}$\\
			$\psi_{M+1}^+$&(0...,0,-1)&$\mu'_{M+2}$&$1-\frac{(M+1)c_{M+2}}{C}$  &$\frac{2c_{M+2}}{C}$\\
                                $\psi_{I\neq 1,M+1}^\pm$&$(0,,...,0,\overbrace{\mp1}^{I-1},\overbrace{\pm1}^{I},0,...,0)$&$\mu'_I$&$1-\frac{(M+1)c_I}{C}$  &$\frac{2c_I}{C}$\\
			$\eta_{1,0,r}$&Neutral& $\mu'_{0}-\mu'_1$  & $\frac{(M+1)(c_1-c_0)}{C}$&2r\\
		    $\eta_{2,0,r}$&Neutral& $\mu'_{0}+\mu'_1$  & $2-\frac{(M+1)(c_0+c_1)}{C}$&2r\\
		    $\eta_{M+1,M+2,r}$&Neutral& $\mu'_{M+2}-\mu'_{M+1}$  & $\frac{(M+1)(c_{M+1}-c_{M+2})}{C}$&2r\\
		    $\eta_{M,M+2,r}$&Neutral& $\mu'_{M+2}+\mu'_{M+1}$  & $2-\frac{(M+1)(c_{M+1}+c_{M+2})}{C}$&2r\\
		    $\eta_{i,r}$&Neutral& $2\mu'_{i}$  & $2-\frac{2(M+1)c_i}{C}$&2r\\
		    \hline
		\end{tabular}
	\end{center}
\end{table}
\subsection{Comparison with the classical spectrum}
It is straight forward to check that the spectrum of $\eta$ fields matches exactly the spectrum of decoupled size modes $\rho$ with $n=n_{\text{max}}+1-r$. We will show that the charged  sector also agrees with the expectations. The GLSM has $M+1$ vacua corresponding to the $M+1$ partitions $\{k_a\}$ in the following way
\eql{su2vacua}{k_a=\delta_{a,b}\Leftrightarrow \vev{\psi_I^+},\ \vev{\psi_J^-}\neq 0\ ,\ \sigma_I=\sum_{I'=1}^I\mu'_{I'}\ ,\ \sigma_J=\sum_{J'=J}^M\mu'_{J'+1}\text{ for }1\leq I<b<J\leq M+1\ ,}
where $\sigma_I$ is the gauge multiplet scalar of the I'th U(1). 
As explained in \ref{classical}, the mass and $R$-charges are shifted in the vacuum due to the VEV of the fields. For example, in the case of $k_a=\delta_{a,M+1}$, the obtained spectrum around the vacuum is given by table \ref{k1vacuum}. Straight forward computation shows agreement between the spectrum of the charged fields in the vacuum to the off-diagonal and interacting size modes around the string vacuum $k_a=\delta_{a,M+1}$, with the identification
\begin{itemize}
\item $\psi_1^-$ with the size mode $\rho_{2,0}^{(0)}$.
\item $\psi_{M+1}^-$ with the off-diagonal mode of the pair $q_{M,M+1},\ q_{M+1,M}$.
\item $\psi_{M+1}^+$ with the size mode $\rho_{M+1,M+2}^{(0)}$.
\item $\psi_{I\neq 1,M+1}^-$ with the size modes $\rho^{(0)}_{I}$.
\end{itemize}
Exactly on the same way, the matching of the spectra holds also when expanded around the other vacua. 
\begin{table}[h]
	
	\caption{\small{The spectrum of the charged sector around the first vacuum $k_a=\delta_{a,M+1}$ of equation \eqref{su2vacua}.}}\label{k1vacuum}
	\begin{center}
		\begin{tabular}{|c|c|c|c|c|}
			\hline
			Field&Twisted Mass&$R^{(R)}$&$R^{(J)}$\\
			$\psi_1^-$  & $\mu'_0+\mu'_1$&$2-\frac{(M+1)(c_0+c_1)}{C}$  &$\frac{2(c_0+c_1)}{C}$\\
                                $\psi_{M+1}^-$&$2\mu'_{M+1}$&$2-\frac{2(M+1)c_{M+1}}{C}$  &$\frac{4c_{M+1}}{C}-2$\\
			$\psi_{M+1}^+$&$\mu'_{M+2}-\mu'_{M+1}$&$\frac{(M+1)(c_{M+1}-c_{M+2})}{C}$  &$2+\frac{2(c_{M+2}-c_{M+1})}{C}$\\
                                $\psi_{I\neq 1,M+1}^-$&$2\mu'_I$&$2-\frac{2(M+1)c_I}{C}$  &$\frac{4c_I}{C}$\\
		    \hline
		\end{tabular}
	\end{center}
\end{table}

\section{S-duality for SU(2) quivers}
\label{Sdual}
S-duality properties of $\cN=2$ superconformal SU(2) quivers were studied for example in \cite{Seiberg:1994aj,Argyres:1999fc,Gaiotto:2009we}. In the case of one SU(2) gauge group with four fundamental hypermultiplets, the theory enjoys a classical SO(8) global symmetry that acts on the eight half hypermultiplets (or, if you like, the eight $\cN=1$ chiral multiplets). SO(8) has an $S_3$ outer automorphism group. The theory is invariant under the outer automorphism of SO(8) accompanied by $S_3$ transformations of the gauge coupling $q\equiv e^{2\pi i\tau}$.\footnote{We use here what is known in the literature as $\tau_{uv}$ that transforms non-trivially only under an $S_3$ subgroup of  the $SL(2,\mathbb{Z})$ that acts on $\tau_{IR}$. For a discussion about the differences between the two, see for example section 9.2 in \cite{Tachikawa:2013kta}.} The S-duality group $S_3$ has six elements which are generated by two generators. We will denote them by $\mathcal{S}$ and $\mathcal{T}'$.\footnote{$\mathcal{T}'$ is the same as $\mathcal{S}\mathcal{T}\mathcal{S}$ in the conventions of \cite{Gerchkovitz:2017ljt}. We decided to specify here the action of $\mathcal{T}'$ instead of $\mathcal{T}$ since we work with it explicitly in \ref{tautominustau}.} They act on the four $SO(8)$ Cartan generators as
\eql{S}{M_{12}&\to\onov{2}\left(M_{12}+M_{34}+M_{56}+M_{78}\right)\\
M_{34}&\to\onov{2}\left(M_{12}+M_{34}-M_{56}-M_{78}\right)\\
M_{56}&\to\onov{2}\left(M_{12}-M_{34}+M_{56}-M_{78}\right)\\
M_{78}&\to\onov{2}\left(M_{12}-M_{34}-M_{56}+M_{78}\right)\ ,}for $\mathcal{S}$ and
\eql{Tprime}{M_{12}&\to\onov{2}\left(M_{12}+M_{34}+M_{56}-M_{78}\right)\\
M_{34}&\to\onov{2}\left(M_{12}+M_{34}-M_{56}+M_{78}\right)\\
M_{56}&\to\onov{2}\left(M_{12}-M_{34}+M_{56}+M_{78}\right)\\
M_{78}&\to\onov{2}\left(-M_{12}+M_{34}+M_{56}+M_{78}\right)\ ,}
for $\mathcal{T}'$. 
The corresponding transformations of the gauge coupling $\tau$ are
\eq{\mathcal{S}: e^{2\pi i\tau}\rightarrow 1-e^{2\pi i\tau}\quad ,\quad \mathcal{T}': \tau\rightarrow-\tau\ .} Turning on masses for the hypermultiplets and/or gauging some U(1), break explicitly the SO(8) global symmetry. Now, instead of relating the theory to itself, the S-duality transformations relate between two theories that differ by their masses and U(1) charges. The masses and U(1) charges transform on the same way as the Cartan generators. This fact was used in \cite{Gerchkovitz:2017ljt} to find the worldsheet theory of strings in some cases where the 4d-2d map is weak$\to$strong. This is done in the following way. Consider two different U(1)s with charges $\{c_i\}$ and $\{c'_i\}$ related by SO(8) outer automorphism. It means that there exists some mapping $\tau\to\tau'(\tau)$ such that a theory with U(1) charges $\{c_i\}$ and SU(2) gauge coupling $\tau$ is equivalent to a theory with U(1) charges $\{c'_i\}$ and SU(2) gauge coupling $\tau'$. The same story holds also for the worldsheet theories. Lets say that in the first theory, there is a string such that the FI parameter of its worldsheet theory is given by $t=f(\tau)$ where $f(\tau)$ is some function. Under S-duality, the string is mapped to a string in the second theory. The worldsheet theory of the dual string will be the same up to the map of parameters, which is $t=f(\tau'(\tau))$. As an example, we can start from a known weakly coupled worldsheet theory for which $t=\tau$ and act with the $\mathcal{S}$ transformation that takes \eql{Softau}{e^{2\pi i\tau}\rightarrow e^{2\pi i\tau'}=1-e^{2\pi i\tau}\ .}
The dual worldsheet theory will be the same theory but with \eq{t=\onov{2\pi i}\log\left(1-e^{2\pi i\tau}\right)=O\left(e^{-S_{\text{inst}}}\right)\ ,}
which is of course strongly coupled.
We would like to apply this method on SU(2) quivers. Every SU(2) factor is coupled to four fundamental hypermultiplets. However, there are some difficulties coming from the coupling of these hypermultiplets to other SU(2)s. The exact S-duality group involves a complicated transformation of all the gauge couplings simultanously \cite{Gaiotto:2009we} \eq{\tau_i\rightarrow \tau'_i\left(\{\tau_j\}\right)\ .}
For simplicity, we will use an approximate S-duality that acts only on one gauge group, as in the case of one SU(2) coupled to four fundamental hypermultiplets.
The approximate S-duality is broken due to the other gauge couplings, however, we still expect it to be applicable in the limit where the gauge couplings of the adjacent SU(2)s are much smaller than the gauge coupling of the discussed SU(2). 
In the next section, we will consider two weakly coupled worldsheet theories related by S-duality and show that the correct mapping between the two as predicted from the approximate S-duality is achieved only in the limit described above. Later on, in \ref{tautolog}, \ref{genquivers}, we will use the S-duality transformation \eqref{Softau} to find strongly coupled worldsheet theories of some (generalized) quivers. 
\subsection{$\mathcal{T}'$ transformation}
\label{tautominustau}
Consider two different $SU(2)^M\times U(1)$ theories with U(1) charges related by $\mathcal{T}'$ tranformation \eqref{Tprime} of one of the SU(2) factors. Lets consider first the case where the SU(2) is in the middle of the quiver, i.e. coupled to two bifundamentals. We can take it to be $SU(2)_L$ for some $2\leq L\leq M-1$. In order to understand the action of S-duality, it will be usefull to follow the Cartans. The two bifundamentals charged under $SU(2)_L$ have masses $\mu_{L},\ \mu_{L+1}$ and charges $c_L,\ c_{L+1}$. In addition, they are charged under $SU(2)_{L-1}$ and $SU(2)_{L+1}$ respectively. Their Cartan generators, denoted by $\al_{L\mp1}$, act with an opposite phase on the two halves of every bifundamental. The  SO(8) automorphism acts on the four charges \eql{Cartan}{\left\{c_L+\al_{L-1},\ c_L-\al_{L-1},\ c_{L+1}+\al_{L+1},\ c_{L+1}-\al_{L+1}\right\}\ .} 
Under $\mathcal{T}'$ which takes $\tau_{L}\to-\tau_L$, the two Cartan generators $\al_{L\pm1}$ are interchanged $\al_{L-1}\leftrightarrow \al_{L+1}$. This is equivalent to interchanging the two bifundamentals. The meaning is that in the limit of $Im(\tau_L)\ll Im(\tau_{L\pm 1})$, the worldsheet theory should be invariant under 
\eql{flip}{\tau_L\to-\tau_L\ ,\  \mu_L\leftrightarrow \mu_{L+1}\ ,\ c_{L}\leftrightarrow  c_{L+1}\ .}
Lets see that this is indeed true for the worldsheet theory described in \ref{secsu2ansatz}. The conditions \eqref{deccondition2}, \eqref{weakcondition2}, \eqref{weaknotilde} are invariant under \eqref{flip}. The decoupled sector, which is made out of the $\eta$ fields of table \ref{su2spectrum} is also invariant under \eqref{flip}. The non-trivial part comes from the charged sector of table \ref{su2spectrum}. It will be useful to see what happens to the D-term constraints. Before the transformation, the relevant D-terms were
\eql{originalDterms}{|\psi^+_{L-1}|^2+|\psi_{L}^-|^2-|\psi_{L-1}^-|^2-|\psi_{L}^+|^2&=\xi_{L-1}\  ,\\  |\psi_{L+1}^-|^2+|\psi_{L}^+|^2-|\psi_{L+1}^+|^2-|\psi_{L}^-|^2&=\xi_L\ ,\\ |\psi_{L+1}^+|^2+|\psi^-_{L+2}|^2-|\psi_{L+1}^-|^2-|\psi^+_{L+2}|^2&=\xi_{L+1}\ .}
The transformation \eqref{flip} takes $\xi_L\to-\xi_L$ and exchanges the masses and R-charges of $\psi^{\pm}_{L}$ and $\psi^{\pm}_{L+1}$. We can relabel $\psi_L^{\pm}\leftrightarrow \psi_{L+1}^{\pm}$ and write the new D-term constraints
\eq{|\psi^+_{L-1}|^2+|\psi_{L+1}^-|^2-|\psi_{L-1}^-|^2-|\psi_{L+1}^+|^2&=\xi_{L-1}\  ,\\  |\psi_{L}^-|^2+|\psi_{L+1}^+|^2-|\psi_{L}^+|^2-|\psi_{L+1}^-|^2&=-\xi_L\ ,\\ |\psi_{L}^+|^2+|\psi^-_{L+2}|^2-|\psi_{L}^-|^2-|\psi^+_{L+2}|^2&=\xi_{L+1}\ .}
These three equations can be written as
\eq{|\psi^+_{L-1}|^2+|\psi_L^-|^2-|\psi_{L-1}^-|^2-|\psi_L^+|^2&=\xi_{L-1}-\xi_L\  ,\\  |\psi_{L}^+|^2+|\psi_{L+1}^-|^2-|\psi_{L}^-|^2-|\psi_{L+1}^+|^2&=\xi_L\ ,\\ |\psi_{L+1}^+|^2+|\psi^-_{L+2}|^2-|\psi_{L+1}^-|^2-|\psi^+_{L+2}|^2&=\xi_{L+1}-\xi_L\ .}
In the $Im(\tau_L)\ll Im(\tau_{L\pm 1})\Leftrightarrow \xi_L\ll \xi_{L\pm1}$ limit, these equations are the same as equations \eqref{originalDterms} which is what we expect to get from S-duality.

Now we will consider the case where the transformed gauge group is on one of the edges. We will take it to be $SU(2)_1$. The case of $SU(2)_M$ is equivalent. The relevant hypermultiplets are two $SU(2)_1$ fundamentals whose masses and charges are denoted by $\mu_{0,1},\ c_{0,1}$ and one $SU(2)_1\times SU(2)_2$ bifundamental, whose mass and charge are denoted by $\mu_2,\ c_2$. $\mathcal{T}'$ now acts on the four charges  \eql{Cartanedge}{\left\{c_0,\ c_1,\ c_{2}+\al_{2},\ c_{2}-\al_{2}\right\}\ .} This transformation takes
\eql{flipedge}{\tau_1\to-\tau_1\ ,\ c_0\to \onov{2}(c_0-c_1)+c_2\ ,\ c_1\to\onov{2}(c_1-c_0)+c_2\ ,\ c_2\to \onov{2}(c_0+c_1)}
and similarly for the masses. As before,  the conditions \eqref{deccondition2}, \eqref{weakcondition2}, \eqref{weaknotilde} are invariant under \eqref{flipedge}. The decoupled sector, which is made out of the $\eta$ fields of table \ref{su2spectrum} is also invariant under \eqref{flipedge}. The relevant charged part of the theory contains 6 fields and two U(1)s with the D-terms
\eq{|\psi_1^+|^2+|\psi_{2}^-|^2-|\psi_1^-|^2-|\psi_{2}^+|^2=\xi_1\  ,\  |\psi_{3}^-|^2+|\psi_{2}^+|^2-|\psi_{3}^+|^2-|\psi_{2}^-|^2=\xi_2\ .}
Under $\xi_1\to-\xi_1$ the D-terms become
\eq{|\psi_1^-|^2+|\psi_{2}^+|^2-|\psi_1^+|^2-|\psi_{2}^-|^2=\xi_1 ,\  |\psi_{3}^-|^2+|\psi_1^+|^2-|\psi_{3}^-|^2-|\psi_1^-|^2=\xi_2-\xi_1\ ,}
and the masses and R-charges of the fields are given by the action of \eqref{flipedge} on table \ref{su2spectrum}.
Recall that the R-charges and the  masses themselves are not physical because they can be shifted by gauge transformations and redefinitions of the scalars $\sigma_i$. If we shift $\sigma_1$ by $\onov{2}(\mu'_0-\mu'_1)$ then the new masses are
\eq{m_{\psi_1^\pm}=\mu'_2\ ,\ m_{\psi_2^+}=\mu'_1\ ,\ m_{\psi_2^-}=\mu'_0\ ,} and similarly for the R-charges.
We can rename $\psi_1^-\leftrightarrow \psi_2^-$ and $\psi_1^+\leftrightarrow \psi_2^+$ and in the limit where $\xi_2\gg \xi_1$ we get exactly the same theory before the $\mathcal{T}'$ transformation with the correct spectrum.

\subsection{$\mathcal{S}$ transformation on the edge}
\label{tautolog}
\begin{SCfigure}
		\vspace{10pt}
		\includegraphics[width=0.35\textwidth, height=4.5cm]{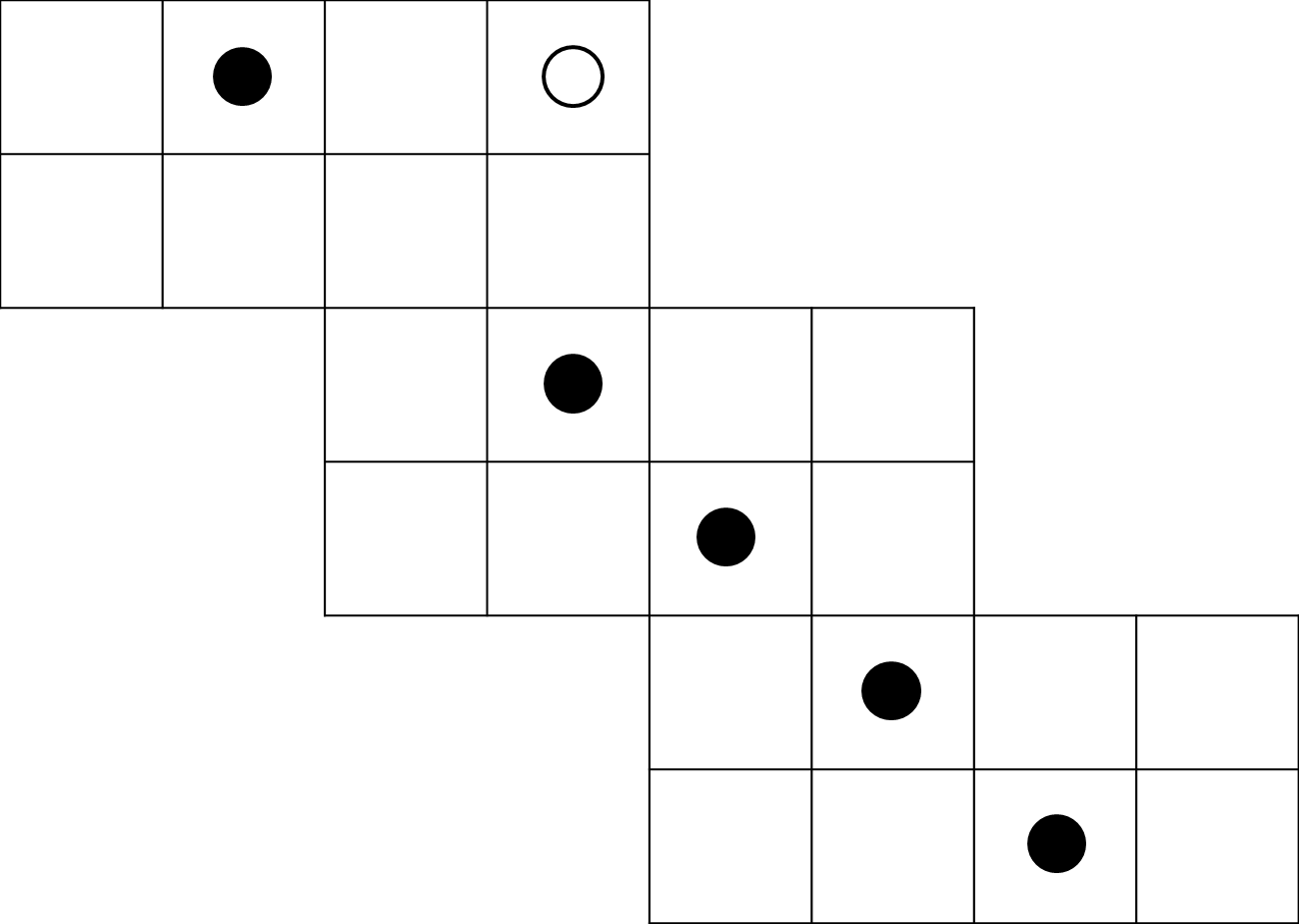}
		\caption{\small{The $\mathcal{S}$-dual vacuum to the baryonic vacuum is the mixed mesonic-baryonic vacuum shown in this figure. $\mathcal{S}$ transformation on the first $SU(2)$ takes the baryonic vacuum to the vacuum presented in this figure. One can check that the two dual vacua carry the same U(1) charge $C'=C$.  }}\label{mixed}
		\vspace{10pt}
	\end{SCfigure}

In this section we consider the action of the $\mathcal{S}$ transformation \eqref{S} on $SU(2)_1$, the first SU(2) factor. The $\mathcal{S}$ transformation acts on \eqref{Cartanedge} as
\eql{SCartanedge}{c_0\to c_0'= \onov{2}(c_0+c_1)-c_2\ ,\ c_1\to c_1'=\onov{2}(c_1+c_0)+c_2\ ,\ c_2\to c_2' =\onov{2}(c_0-c_1)\ .}
The first thing that we observe is that $C'\neq C$, simply because $c'_1+c'_2\neq c_1+c_2$. This means that we are not comparing the correct strings. The string we need to examine is a string above the mixed mesonic-baryonic vacuum illustrated in figure \ref{mixed}. Instead of analysing this string from the beginning, we can use the fact that the fundamental representation of SU(2) is pseudoreal, and therefore we can exchange mesons with baryons, and take the vacuum described in figure \ref{mixed} back to the baryonic vacuum. This transformation effectively takes $c_2'\to-c_2'=\onov{2}(c_1-c_0)$.

The next thing that we observe is that unlike the $\mathcal{T}'$ transformation, the conditions \eqref{weakcondition2}, \eqref{weaknotilde} are not invariant under the $\mathcal{S}$ transformation. This is becaue $2c_2'<C$ and $c_0'<c_1'$ (After we took $c'_2\to-c'_2$).
The size modes counting \eqref{bisizemodes}, \eqref{moresizemodes} now implies that there are size modes excitations of $\tilde{q}_{13}$ and $\tilde{q}_{10}$. Therefore there are non-trivial F-term constraints of the form
\eq{\tilde{q}_{10}q_{20}+\tilde{q}_{13}q_{23}=0\ ,\ \tilde{q}_{13}q_{12}=0\ ,} and as a result the worldsheet theory is strongly coupled.
The approximate S-duality gives us the worldsheet theory only in the limit $\tau_1\ll \tau_2$. The $K=1$ worldsheet theory in this case is given by the low energy limit of an $\cN=(2,2)$ $U(1)^M$ GLSM with M complexified FI parameters
\eq{e^{2\pi it_1}=1-e^{2\pi i\tau_1}\ ,\ t_{a}=\tau_a\ \forall\ 2\leq a\leq M,}
and the following chiral fields 
\begin{itemize}
\item $X$, parametrizing the center of mass modes.
\item $\psi_I^\pm$ with $I=1,...,M+1$, parametrizing the interacting size modes and the off-diagonal modes.
\item $\eta_{i,r}$ with $i=3,...,M$ and $r=1,...,\frac{2c_i}{C}-1$.
\item $\eta_{2,r}$ with $r=1,...,\frac{c_1-c_0}{C}-1$.
\item $\eta_{1,0,r}$ with $r=1,...,-\frac{2c_2}{C}$.
\item $\eta_{2,0,r}$ with $r=1,...,\frac{c_0+c_1}{C}-1$.
\item $\eta_{M+1,M+2,r}$ with $r=1,...,\frac{c_{M+2}-c_{M+1}}{C}$.
\item $\eta_{M,M+2,r}$ with $r=1,...,\frac{c_{M+2}+c_{M+1}}{C}-1$.
\end{itemize}
The quantum numbers of the fields are summarized in table \ref{Sontheedge}.
\begin{table}[h]
	
	\caption{\small{The spectrum on the worldsheet for $\mathcal{S}$-dual strings described in \ref{tautolog}   }}\label{Sontheedge}
	\begin{center}
		\begin{tabular}{|c|c|c|c|c|}
			\hline
			Field&$U(1)^M$&Twisted Mass&$R^{(R)}$&$R^{(J)}$\\
			$X$        &Neutral          &0       &0&    2         \\
			$\psi_1^-$  &(-1,0...,0)     & $\onov{2}(\mu'_0+\mu'_1-2\mu'_2)$&$1+\frac{(M+1)(2c_2-c_0-c_1)}{2C}$  &$\frac{c_0+c_1-2c_2}{C}$\\
                                $\psi_{1}^+$&(1,0...,0)  &$\onov{2}(\mu'_0+\mu'_1+2\mu'_2)$&$1-\frac{(M+1)(c_0+c_1+2c_2)}{2C}$  &$\frac{c_0+c_1+2c_2}{C}$\\
                                $\psi_2^\pm$&$(\mp1,\pm1,0,...)$&$\onov{2}(\mu'_1-\mu'_0)$&$1+\frac{(M+1)(c_0-c_1)}{2C}$&$\frac{c_1-c_0}{C}$\\			
                                $\psi_{M+1}^-$&(0,...,0,1)&$\mu'_{M+1}$&$1-\frac{(M+1)c_{M+1}}{C}$  &$\frac{2c_{M+1}}{C}$\\
			$\psi_{M+1}^+$&(0...,0,-1)&$\mu'_{M+2}$&$1-\frac{(M+1)c_{M+2}}{C}$  &$\frac{2c_{M+2}}{C}$\\
                                $\psi_{I\neq 1,2,M+1}^\pm$&$(0,,...,0,\overbrace{\mp1}^{I-1},\overbrace{\pm1}^{I},0,...,0)$&$\mu'_I$&$1-\frac{(M+1)c_I}{C}$  &$\frac{2c_I}{C}$\\
			$\eta_{1,0,r}$&Neutral& $-2\mu'_{2}$  & $\frac{2(M+1)c_2}{C}$&2r\\
		    $\eta_{2,0,r}$&Neutral& $\mu'_{0}+\mu'_1$  & $2-\frac{(M+1)(c_0+c_1)}{C}$&2r\\
                         $\eta_{2,r}$&Neutral&$\mu'_1-\mu'_0$& $2+\frac{(M+1)(c_0-c_1)}{C}$&2r\\
		    $\eta_{M+1,M+2,r}$&Neutral& $\mu'_{M+2}-\mu'_{M+1}$  & $\frac{(M+1)(c_{M+1}-c_{M+2})}{C}$&2r\\
		    $\eta_{M,M+2,r}$&Neutral& $\mu'_{M+2}+\mu'_{M+1}$  & $2-\frac{(M+1)(c_{M+1}+c_{M+2})}{C}$&2r\\
		    $\eta_{i\geq 3,r}$&Neutral& $2\mu'_{i}$  & $2-\frac{2(M+1)c_i}{C}$&2r\\
		    \hline
		\end{tabular}
	\end{center}
\end{table}

\subsection{$\mathcal{S}$ transformation on the middle: Generalized quivers}
\label{genquivers}
\begin{SCfigure}
	\vspace{10pt}
	\includegraphics[width=0.5\textwidth, height=5cm]{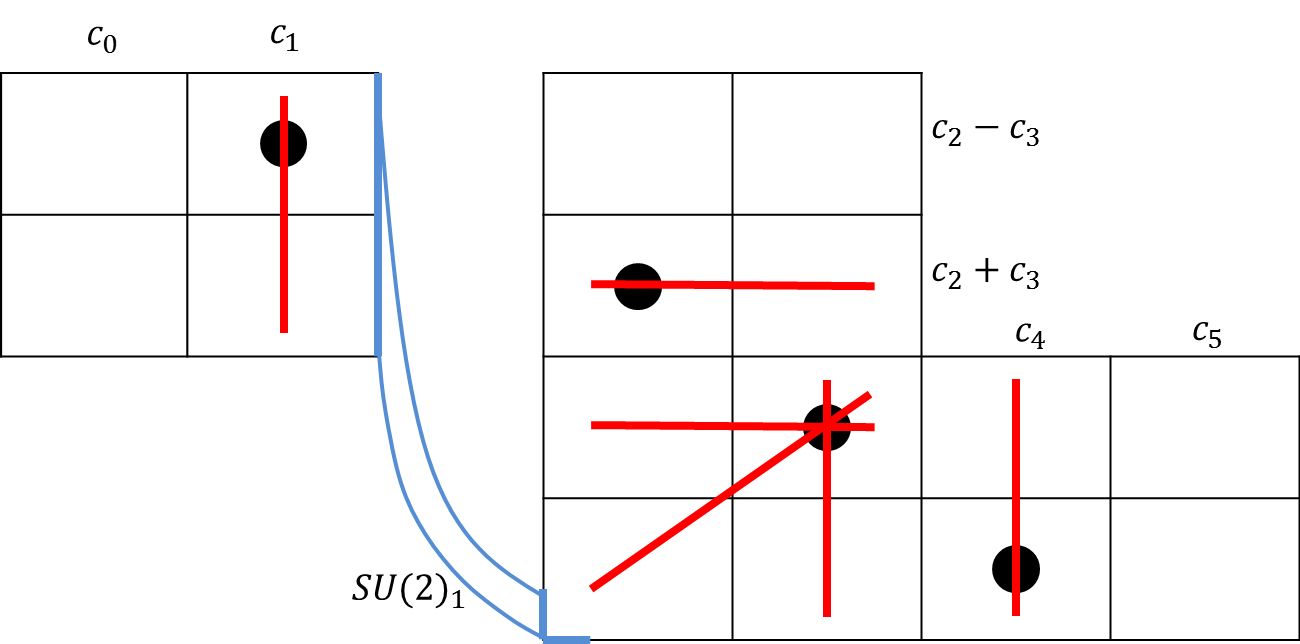}
\caption{\small{The tetris diagram of the $SU(2)^3$ generalized quiver. The coupling of the hypermultiplets to $SU(2)_1$ is represented by the curved blue lines. The charges of the hypermultiplets are written on the diagram. The vacuum presented here is the fully Higgsed vacuum S-dual to the fully Higgsed vacuum of the linear quiver studied above. }}\label{genquiver}
	\vspace{10pt}
\end{SCfigure}

In this section we will study the $\mathcal{S}$ transformation acting on an intermediate $SU(2)_L$ with $2\leq L\leq M-1$. This SU(2) is coupled to 2 bifundamentals. As in section \ref{tautominustau}, the $\mathcal{S}$ transformation acts as \eqref{S} on \eqref{Cartan}, which results in 
\eql{u1Saction}{c_{L}\to c_L\ ,\ \al_{L+1}\to\al_{L+1}\ ,\ \al_{L-1}\leftrightarrow c_{L+1}\ .}
An interesting consequence is that the $\mathcal{S}$ transformation takes the quiver to the so called generalized quiver represented in figure \ref{genquiver}. Lets see exactly how it works. Under $\mathcal{S}$, we see from equation \eqref{u1Saction} that the two gauge groups $SU(2)_{L\pm 1}$ act on the same block. This block is now a trifundamental of the three gauge groups $SU(2)_{L,L\pm1}$. Take as an example $L=2$.
 In order to understand better how $SU(2)_{1}$ acts on the trifundamental hypermultiplet, it will be usefull to look at the scalars \eq{q=\mat{q_{32}&q_{33}\\q_{42}&q_{43}}\ ,\ \tilde{q}=\mat{\tilde{q}_{32}&\tilde{q}_{33}\\\tilde{q}_{42}&\tilde{q}_{43}}\ .} 
Under $SU(2)_2\times SU(2)_3$, they transform as $q\rightarrow U_3qU_2^T\ ,\ \tilde{q}\rightarrow U_3^*\tilde{q}U_2^\dagger$, or equivalently
\eq{q\rightarrow U_3qU_2^T\ ,\ \sigma_2\tilde{q}\sigma_2\rightarrow U_3\sigma_2\tilde{q}\sigma_2U_2^T\ .}
The third gauge group, $SU(2)_{1}$, now acts on $\mat{q & \sigma_2\tilde{q}\sigma_2}^T$ as doublets. Notice that the mixing between $q$ and $\tilde{q}$ implies that the trifundamental field must be massless and U(1) neutral. One of the difficulties in studying strings on generalized quivers is that the worldsheet theory is inherently strongly coupled. In the previous cases, the F-terms couple $q$ with $\tilde{q}$, then one can find simple conditions on the U(1) charges such that the F-terms are satisfied trivially and as a result, the worldsheet theory is weakly coupled. On the other hand, trifundamental F-terms couple also $q$ with $q$ and $\tilde{q}$ with $\tilde{q}$. For example, from the Cartan of $SU(2)_1$ we get the constraint
\eq{q_{43}q_{32}-\tilde{q}_{43}\tilde{q}_{32}+q_{33}q_{42}-\tilde{q}_{33}\tilde{q}_{42}+...=0\ .}
The off-diagonal modes that come from $q_{43},\ q_{32}$ impose non-trivial constraints on the worldsheet. As a result, the worldsheet theory is strongly coupled, regardless of the U(1) charges. We will study generalized quiver strings in the case where they are $\mathcal{S}$-dual to weakly coupled strings. We will write everything explicitly for the $SU(2)^3$ quiver. The generalization to longer quivers is straight forward. The masses and U(1) charges of the first two columns and the last two columns, parametrized by $a=0,1,4,5$, are invariant under the $\mathcal{S}$ transformation and are equal to $\mu_a,\ c_a$. The two $SU(2)_2$ fundamentals have masses and U(1) charges $\mu_2\pm \mu_3\ ,\ c_2\pm c_3$. The trifundamental field is massless and neutral, as explained above. The U(1) charges satisfy \eq{c_0\geq c_1\ ,\ c_5\geq c_4\ ,\ c_0+c_1\geq C\ ,\ c_5+c_4\geq C\ ,\ 2c_2\geq C\ ,\ 2c_3\geq C\ .}
It will be convenient to denote the charges by
\eq{c_{1+}=c_1\ ,\ c_{1-}=c_0\ ,\ c_{2\pm}=c_2\pm c_3\ ,\ c_{3+}=c_4\ ,\ c_{3-}=c_5\ ,}
and similarly for the masses.

Putting all the details together, our ansatz for the worldsheet theory is a $U(1)^3$ GLSM with complexified FI parameters
\eq{t_{1,3}=\tau_{1,3}\ ,\ e^{2\pi it_2}=1-e^{2\pi i\tau_2}\ ,}
and the following chiral fields
\begin{itemize}
\item One neutral chiral field $X$.
\item Six chiral fields, $\psi_{I\pm}\ ,\ I=1,2,3$ with charges $(\pm1,0,0),\ (0,\pm1,0),\ (0,0,\pm1)$.
\item Two chiral fields  $\psi_{4\pm}$ with charges $(\pm1,\pm1,\mp1)$.
\item Neutral fields $\eta_{I,r}^+$ with $I=1,2,3$ and $r=1,...,\frac{c_{I+}+c_{I-}}{C}-1$.
\item Neutral fields $\eta_{I,r}^-$ with $I=1,3$ and $r=1,...,\frac{c_{I-}-c_{I+}}{C}$.
\item Neutral fields $\eta_{2,r}^-$ with $r=1,...,\frac{c_{2+}-c_{2-}}{C}-1$.
\end{itemize}
The quantum numbers of these fields are summarized in table \ref{generalizedtable}
\begin{table}[h]
	
	\caption{\small{The spectrum on the worldsheet for a generalized quiver string}}\label{generalizedtable}
	\begin{center}
		\begin{tabular}{|c|c|c|c|c|}
			\hline
			Field&Twisted Mass&$R^{(R)}$&$R^{(J)}$\\
			$X$                  &0               &      0    &    2        \\
			$\psi_{I\pm}$       & $\mu'_{I\pm}$&$1-\frac{4c_{I\pm}}{C}$&$\frac{2c_{I\pm}}{C}$  \\
			$\psi_{4+}$&$\mu'_{2+}-\mu'_{2-}$&$1-\frac{4(c_{2+}-c_{2-})}{C}$  &$\frac{2(c_{2+}-c_{2-})}{C}$\\
			$\psi_{4-}$&$0$&$1$&$0$  \\
			$\eta_{I,r}^+$& $\mu'_{I+}+\mu'_{I-}$ & $2-\frac{4(c_{I+}+c_{I-})}{C}$ &2r\\
		    $\eta_{I\neq 2,r}^-$&$\mu'_{I-}-\mu'_{I+}$  & $\frac{4(c_{I+}-c_{I-})}{C}$&2r\\
		    $\eta_{2,r}^-$& $\mu'_{2+}-\mu'_{2-}$  & $2-\frac{4(c_{2+}-c_{2-})}{C}$&2r\\

			\hline
		\end{tabular}
	\end{center}

\end{table}

One can easily check that at the limit $g_2\gg g_{1,3}$, the spectrum is mapped to spectrum of table \ref{su2spectrum} under the map \eq{e^{2\pi it_2}\to 1-e^{2\pi i t_2}\  .}

\section{Worldsheet partition functions from supersymmetric localization}
\label{localization}
In this section we will derive the worldsheet $S^2$ partition functions for all the strings discussed in the previous sections. The ideas of this derivation were presented in \cite{Chen:2015fta,Pan:2015hza} and elaborated in \cite{Gerchkovitz:2017ljt}. We will review this method briefly. As a start, we put our four dimensional quiver theory with gauge group $SU(N)^M\times U(1)$ on the four ellipsoid 
\eq{\frac{x_0^2}{r^2}+\frac{x_1^2+x_2^2}{l^2}+\frac{x_3^2+x_4^2}{\tilde{l}^2}\ .} 
The partition function on this manifold was computed in \cite{Pestun:2007rz,Hama:2012bg}. The partition function in this representation is written as an $(MN-M+1)$-dimensional integral over the Coulomb branch coordinates 
\eql{quiverpartition}{Z_{S^4_b}=&\int \prod_{I=1}^M\left(\prod_{a=1}^{N-1} d(w_a\cdot\ha_I)\right)\,d\ha'\, \prod_{I=1}^Me^{-\frac{16\pi^2}{g_I^2}\ha_I\cdot \ha_I-\frac{8\pi^2}{e^2}\,\ha'^2+16i\pi^2\hx \ha'}\\ &\frac{\prod_{I=1}^{M}\prod_{a\neq b}\Upsilon_b\left(iw_a\cdot\ha_I-iw_b\cdot\ha_I\right)}{\prod_{a=1 }^{N}\left[\prod_{i=0}^{N-1}\Upsilon_b\left(iw_a\cdot\ha_1+ic_i\ha'-i\hm_i+\frac{Q}{2}\right)\prod_{i=N+M-1}^{2N+M-2}\Upsilon_b\left(iw_a\cdot\ha_M+ic_i\ha'-i\hm_i+\frac{Q}{2}\right)\right]}\\&\prod_{I=1}^{M-1}\prod_{a,b=1}^{N}\onov{\Upsilon_b\left(iw_a\cdot\ha_I+iw_b\cdot \ha_{I+1}+ic_{N-1+I}\ha'-i\hm_{N-1+I}+\frac{Q}{2}\right)}|Z_{\text{inst}}|^2\;. }
Here $g_I,e$ are the gauge couplings for the $SU(N)_I$ and the $U(1)$ factors respectively. Similarly, $\ha_I,\ \ha'$ are the Coulomb branch parameters of $SU(N)_I$ and the $U(1)$ factors, respectively. $\hx,\ \hm,\ c_i$ are the FI parameter, masses and U(1) charges, $w_a$ are the weights of $SU(N)$ in the fundamental representation and $Q\equiv b+b^{-1}$ with $b^2\equiv l/\tilde{l}$. $\ha,\ \hm,\ \hx$ are dimensionless and measured in units of $\sqrt{l\tilde{l}}$. The exponent in the first line of \eqref{quiverpartition} comes from the classical action evaluated in the saddle points, the numerator in the second line comes from the 1-loop determinant of the vector multiplets, the denominator in the second line comes from the 1-loop determinant of the 2N $SU(N)_1$ and $SU(N)_M$ fundamental hypermultiplets, the denominator in the third line comes from the 1-loop determinant of the bifundamental hypermultiplets, and $Z_{\text{inst}}$ is the Nekrasov instanton partition function \cite{Nekrasov:2002qd} with equivariant parameters $\epsilon_1=l^{-1}$ and $\epsilon_2=\tilde{l}^{-1}$.
$\Upsilon_b(x)$  is a holomorphic function which is defined by the conditions and the shift relation
\eql{upsilon}{\Upsilon_b(x)&=\Upsilon_{b^{-1}}(x)\ ,\
\Upsilon_b(Q/2)&=1\ ,\ \Upsilon_b(x+b)&=\frac{\Gamma(bx)}{\Gamma(1-bx)}b^{1-2bx}\Upsilon_b(x)\ .}
$\Upsilon_b(x)$ has zeros at
\eq{x+mb+nb^{-1}&=0\ ,\ m,n\in\mathbb{N}\ ,\\
Q-x+mb+nb^{-1}&=0\ ,\ m,n\in\mathbb{N}\ .}

For a wide range of the parameters of the theory, one can close the contour of integration over the U(1) Coulomb branch parameter in the complex plane. 
 The partition function is then written as a sum over the residues of the poles that lie inside the contour of integration. The poles come from the zeros of the $\Upsilon_b(x)$ functions that appear in the denominator of \eqref{quiverpartition}. One can try and close the contour of integration also for the other $M(N-1)$ integrals. For some of the terms, it is possible to eliminate all the integrals in this way. Inspired by the analysis of \cite{Gaiotto:2012xa}, these terms are interpreted as Higgs branch saddle points. Indeed, among these terms we can identify the fully Higgsed vacua and strings above these vacua that were studied in the previous sections. For the rest of the terms, it is not possible to close the contour for all the integrals. These terms are identified with mixed Higgs-Coulomb saddle points. The leftover Coulomb branch integrals represent the residual gauge symmetry of these configurations. Between these terms one can find the non-fully Higgsed vacua that were described in section \ref{secvacua}. From now on we will focus only on the fully Higgsed contributions. This sum can be written as
\eq{Z_{\text{fully Higgsed}}=\sum_{\{l_a\}}Z_{\text{vac},\{l_a\}}e^{-8\pi^2Q\hx(MN-M+1)/|C_{\{l_a\}}|}e^{16\pi^2i\hx\hm_{\{l_a\}}/C_{\{l_a\}}}\sum_{K,K'}e^{-16\pi^2\hx\left(Kb+K'b^{-1}\right)/|C_{\{l_a\}}|}Z_{K,K',\{l_a\}}\ ,}
where $Z_{\text{vac},\{l_a\}}$ and $Z_{K,K',\{l_a\}}$ are independent of $\hx$. The sum over $\{l_a\}$ is the sum over fully Higgsed vacua where the choice of $\{l_a\}$ is the choice of $NM-M+1$ hypermultiplets that get VEV, $C_{\{l_a\}}=\sum_{a=1}^{NM-M+1}c_{l_a}$ and $\hm_{\{l_a\}}=\sum_{a=1}^{NM-M+1}\hm_{l_a}$, and $Z_{\text{vac},\{l_a\}}$ is the partition function of the $N^2+M-1$ light hypermultiplets in the vacuum. $K,\ K'$ are the winding numbers of strings that wrap the two-spheres $\frac{x_0^2}{r^2}+\frac{x_1^2+x_2^2}{l^2}=1$ and $\frac{x_0^2}{r^2}+\frac{x_3^2+x_4^2}{\tilde{l}^2}=1$ respectively. We will focus on configurations for which $K'=0$ that correspond to the $\onov{2}$-BPS strings studied above. Configurations for which both $K$ and $K'$ are non-zero are $\onov{4}$-BPS configurations of intersecting strings. The function $Z_{K,\{l_a\}}\equiv Z_{K,0,\{l_a\}}$ contains the information about the string dynamics. If the factorization conditions hold and the string is decoupled from the bulk at low energies, $Z_{K,\{l_a\}}$ is identifed with the $S^2$ partition function of the string worldsheet theory. In these cases, one can use the identities \eqref{upsilon} to transform all the $\Upsilon_b(x)$ functions to $\Gamma$-matrices, which are the "building blocks" of $S^2$ partition functions. The weak$\to$weak condition can also be seen from the form of $Z_{K,\{l_a\}}$. When the weak$\to$weak conditions are satisifed, $Z_{K,\{l_a\}}$ has a fixed number of $\Gamma$ functions, independent of the partition $\{k_a\}$. This is a property of $S^2$ partition functions expanded around the weakly coupled point. When the weak$\to$weak conditions are not satisifed, different terms in $Z_{K,\{l_a\}}$ contain different number of $\Gamma$ functions. This can happen for $S^2$ partition functions expanded around some strongly coupled point.\footnote{See section (6.2.3) in \cite{Gerchkovitz:2017ljt} and section \ref{Sedgesection} for examples.}
Once $Z_{K,\{l_a\}}$ is found, we want to identify it with the $S^2$ partition function of the string worldsheet theory under some 2d-4d map of parameters.
 In the next sections, we will compute $Z_{K,\{l_a\}}$ for all the strings studied in sections \ref{sun2}, \ref{su2M} and show that the results agree with our suggestions for their worldsheet theories. 
\subsection{$SU(N)^2\times U(1)$}
In this section we will extract the worldsheet $S^2$ partition functions for the strings studied in section \ref{sun2} and show that the results are consistent with the worldsheet theories presented in \ref{sun2}.
The four-ellipsoid partition function for the $SU(N)\times SU(N)\times U(1)$ theory is
\eq{Z_{S^4_b}=&\int \left(\prod_{a=1}^{N-1} d(w_a\cdot\ha_1)\right)\,\left(\prod_{a=1}^{N-1} d(w_a\cdot\ha_2)\right)\,d\ha'\, e^{-\frac{16\pi^2}{g_1^2}\ha_1\cdot \ha_1-\frac{16\pi^2}{g_2^2}\ha_2\cdot \ha_2-\frac{8\pi^2}{e^2}\,\ha'^2+16i\pi^2\hx \ha'}\\ &\frac{\prod_{a\neq b}\Upsilon_b\left(iw_a\cdot\ha_1-iw_b\cdot\ha_1\right)}{\prod_{a=1 }^{N}\prod_{i=0}^{N-1}\Upsilon_b\left(iw_a\cdot\ha_1+ic_i\ha'-i\hm_i+\frac{Q}{2}\right)}\, \frac{\prod_{a\neq b}\Upsilon_b\left(iw_a\cdot\ha_2-iw_b\cdot\ha_2\right)}{\prod_{a=1 }^{N}\prod_{i=N+1}^{2N}\Upsilon_b\left(iw_a\cdot\ha_2+ic_i\ha'-i\hm_i+\frac{Q}{2}\right)}\\&\prod_{a,b=1}^{N}\onov{\Upsilon_b\left(iw_a\cdot\ha_1+iw_b\cdot \ha_2+ic_N\ha'-i\hm_N+\frac{Q}{2}\right)}|Z_{\text{inst}}|^2\;. }
The fully Higgsed vacuum and strings we are interested in are given by collecting the residues from the following poles 
\eq{&iw_a\cdot \ha_1+ic_a\cdot\ha'-i\hm_a+\frac{Q}{2}+k_{a}b=0\ ,\ a=1,...,N-1\\
       &iw_a\cdot \ha_2+ic_{N+a-1}\cdot\ha'-i\hm_{N+a-1}+\frac{Q}{2}+k_{N+a-1}b=0\ ,\ a=2,...,N\\
       &iw_N\cdot \ha_1+iw_1\cdot \ha_2+ic_N\cdot\ha'-i\hm_N+\frac{Q}{2}+k_{N}b=0\ .}
Summation over all the terms with fixed $K=\sum_{a=1}^{2N-1}k_a$ gives the contribution from the string with winding number $K$, and the term with $K=0$ gives the vacuum contribution. 

It will be usefull to define 
\eq{&\mu=\sum_{i=1}^{2N-1}\hm_i\ ,\ \mu'_a=\hm_a-\frac{\mu c_a}{C}\ ,\ k'_a=k_a-\frac{Kc_a}{C}\ ,\ Q'_a=Q\left(1-\frac{(2N-1)c_a}{C}\right)\ ,\\ &M_a=\mu'_a+\frac{iQ'_a}{2},\ M=\frac{\mu}{C}+\frac{i(2N-1)Q}{2C}\ ,\ \mm_a=M_a-\frac{ibKc_a}{C}\ .}
In terms of these variables, the poles are given by
\eq{i\ha'&=iM-\frac{Kb}{C}\\
      iw_a\cdot \ha_1&=i\mm_a-k_ab\ ,\   iw_{a+1}\cdot \ha_2=i\mm_{N+a}-k_{N+a}b\ ,\ a=1,...,N-1\ ,\\
iw_N\cdot\ha_1&=-\sum_{a=1}^{N-1}\left(i\mm_a-k_a\right)\ ,\ iw_1\cdot\ha_2=-\sum_{a=2}^{N}\left(i\mm_{N+a-1}-k_{N+a-1}\right)\ ,}
The contribution from this choice of poles with a fixed $K=\sum_{a=1}^{2N-1} k_a$ is
\eq{Z_K=&\sum_{\{k_a\}}e^{-\frac{8\pi^2}{g_1^2}\left(\sum_{i=1}^{N-1}(\mm_i+ik_ib)^2+(\sum_{i=1}^{N-1}(\mm_i+ik_ib))^2\right)-\frac{8\pi^2}{g_2^2}\left(\sum_{i=N+1}^{2N-1}(\mm_i+ik_ib)^2+(\sum_{i=N+1}^{2N-1}(\mm_i+ik_ib))^2\right)}\\ 
&\frac{\prod_{i=1}^{N-1}\left[\Upsilon_b\left(-\sum_{a=1}^{N-1}\left(i\mm_a-k_ab\right)-i\mm_i+k_ib\right)\Upsilon_b\left(i\mm_i-k_ib+\sum_{a=1}^{N-1}\left(i\mm_a-k_ab\right)\right)\right]}{\prod_{i=0}^{N-1}\left[\Upsilon_b\left(\sum_{a=1}^{N-1}\left(-i\mm_a+k_ab\right)-i\mm_i\right)\prod_{a=1,\ a\neq i }^{N-1}\Upsilon_b\left(i\mm_a-k_ab-i\mm_i\right)\right]}\\ 
&\frac{\prod_{i\neq j=1}^{N-1}\Upsilon_b\left(i(\mm_i-\mm_j)+(k_j-k_i)b\right)}{\prod_{i=1}^{N-1}\Upsilon_b\left(i\mm_i-k_ib-\sum_{a=2}^{N}\left(i\mm_{N+a-1}-k_{N+a-1}b\right)-i\mm_N\right)}\\
&\frac{\prod_{i=N+1}^{2N-1}\left[\Upsilon_b\left(-\sum_{a=N+1}^{2N-1}\left(i\mm_a-k_ab\right)-i\mm_i+k_ib\right)\Upsilon_b\left(i\mm_i-k_ib+\sum_{a=N+1}^{2N-1}\left(i\mm_a-k_ab\right)\right)\right]}{\prod_{i=N+1}^{2N}\left[\Upsilon_b\left(-\sum_{a=N+1}^{2N-1}\left(i\mm_a-k_ab\right)-i\mm_i\right)\prod_{a=N+1,\ a\neq i }^{2N-1}\Upsilon_b\left(i\mm_a-k_ab-i\mm_i\right)\right]}\\ 
&\frac{\prod_{i\neq j=N+1}^{2N-1}\Upsilon_b\left(i(\mm_i-\mm_j)+(k_j-k_i)b\right)}{\prod_{i=N+1}^{2N-1}\Upsilon_b\left(-\sum_{a=1}^{N-1}\left(i\mm_a-k_ab\right)+i\mm_{i}-k_{i}b-i\mm_N\right)}Res\left(\prod_{i=1}^{2N-1}\Upsilon_b(-k_ib)\right)^{-1}\\
&\prod_{i=1}^{N-1}\prod_{j=N+1}^{2N-1}\onov{\Upsilon_b\left(i\mm_i-k_ib+i\mm_{j}-k_{j}b+i\mm_N\right)}|Z_{\text{inst}}|^2e^{\frac{8\pi^2}{e^2C^2}\,\left(iM-Kb\right)^2+\frac{16\pi^2\hx}{C} \left(iM-Kb\right)}\;.}
First, let us evaluate the vacuum contribution by plugging in $K=0$:
\eq{Z_0=&e^{-\frac{8\pi^2}{g_1^2}\left(\sum_{i=1}^{N-1}M_i^2+(\sum_{i=1}^{N-1}M_i)^2\right)-\frac{8\pi^2}{g_2^2}\left(\sum_{i=N+1}^{2N-1}M_i^2+(\sum_{i=N+1}^{2N-1}M_i)^2\right)}Res\left(\prod_{i=1}^{2N-1}\Upsilon_b(0)\right)^{-1}\\ 
&\frac{1}{\Upsilon_b\left(-\sum_{a=1}^{N-1}iM_a-iM_0\right)\prod_{i=1 }^{N-1}\Upsilon_b\left(iM_i-iM_0\right)}\\ 
&\frac{1}{\Upsilon_b\left(-\sum_{a=N+1}^{2N-1}iM_a-iM_{2N}\right)\prod_{i=N+1 }^{2N-1}\Upsilon_b\left(iM_i-iM_{2N}\right)}\\ 
&\prod_{i=1}^{N-1}\prod_{j=N+1}^{2N-1}\onov{\Upsilon_b\left(iM_i+iM_{j}-iM_N\right)}|Z_{\text{inst}}|^2e^{-\frac{8\pi^2M^2}{e^2C^2}+\frac{16\pi^2i\hx M}{C}}\;.}
$Z_0$ describes the $N^2+1$ light hypermultiplets which are dynamical in the vacuum. Their masses and $R^{(R)}$ charges can be read from the imaginary and real parts of the arguments of the $\Upsilon_b$ functions. It is straight forward to check that these are in agreement with the expectations from the classical analysis.

Now we will move on to evaluating $Z_K$. Using the identities
\eql{upsilonidentities}{\frac{\Upsilon_b(x)}{\Upsilon_b(x-nb)}=\prod_{r=1}^{n}\frac{\gamma(bx-rb^2)}{b^{2bx-2rb^2-1}}\ ,\ \frac{\Upsilon_b(x+nb)}{\Upsilon_b(x)}=\prod_{r=0}^{n-1}\frac{\gamma(bx+rb^2)}{b^{2bx+2rb^2-1}}\ ,}
with \eq{\gamma(x)=\frac{\Gamma(x)}{\Gamma(1-x)}\ ,}
 the partition function can be written as

\eq{Z_K=&Res\left(\prod_{i=1}^{2N-1}\Upsilon_b(0)\right)^{-1}e^{\frac{8\pi^2}{e^2C^2}\,\left(iM-Kb\right)^2+\frac{16\pi^2\hx}{C} \left(iM-Kb\right)}\sum_{\{k_a\}}\prod_{i=1}^{2N-1}\prod_{r=1}^{k_i}\frac{\gamma(-rb^2)}{b^{-2rb^2-1}}|Z_{\text{inst}}^{\{k_i\}}|^2\\&e^{-\frac{8\pi^2}{g_1^2}\left(\sum_{i=1}^{N-1}(\mm_i+ik_ib)^2+(\sum_{i=1}^{N-1}(\mm_i+ik_ib))^2\right)-\frac{8\pi^2}{g_2^2}\left(\sum_{i=N+1}^{2N-1}(\mm_i+ik_ib)^2+(\sum_{i=N+1}^{2N-1}(\mm_i+ik_ib))^2\right)}\\ 
&\frac{1}{\Upsilon_b\left(\sum_{j=0}^{N-1}\left(-i\mm_j+k_jb\right)\right)\prod_{j=1}^{N-1}\Upsilon_b\left(i\mm_j-i\mm_0+(k_0-k_j)b\right)}\\
&\onov{\Upsilon_b\left(\sum_{j=N+1}^{2N}\left(-i\mm_j+k_jb\right)\right)\prod_{a=N+1}^{2N-1}\Upsilon_b\left(i\mm_a-k_ab-i\mm_{2N}\right)}\\
&\prod_{i=1}^{N-1}\prod_{j=N+1}^{2N-1}\onov{\Upsilon_b\left(i\mm_i-k_ib+i\mm_{j}-k_{j}b-i\mm_N\right)}\\
&\prod_{i\neq j=1}^{N-1}\prod_{r=1}^{k_i}\frac{\gamma\left(ib(\mm_j-\mm_i)+(k_i-k_j)b^2-rb^2\right)}{b^{2ib(\mm_j-\mm_i)+2(k_i-k_j)b^2-2rb^2-1}}\prod_{i\neq j=N+1}^{2N-1}\prod_{r=1}^{k_i}\frac{\gamma\left(ib(\mm_j-\mm_i)+(k_i-k_j)b^2-rb^2\right)}{b^{2ib(\mm_j-\mm_i)+2(k_i-k_j)b^2-2rb^2-1}}\\
&\prod_{i=1}^{N-1}\prod_{r=1}^{k_i}\frac{\gamma\left(b\sum_{j=1}^{N-1}(-i\mm_j+k_jb)-ib\mm_i+k_ib^2\right)}{b^{2b\sum_{j=1}^{N-1}(-i\mm_j+k_jb)-2ib\mm_i+2k_ib^2-1}}\prod_{i=N+1}^{2N-1}\prod_{r=1}^{k_i}\frac{\gamma\left(b\sum_{j=N+1}^{2N-1}(-i\mm_j+k_jb)-ib\mm_i+k_ib^2\right)}{b^{2b\sum_{j=N+1}^{2N-1}(-i\mm_j+k_jb)-2ib\mm_i+2k_ib^2-1}}\\
&\prod_{r=1}^{k_N}\prod_{i=1}^{N-1}\frac{\gamma\left(ib\mm_i-k_ib^2+\sum_{j=1}^{N-1}(ib\mm_j-k_jb^2)-rb^2\right)}{b^{2ib\mm_i-2k_ib^2+2\sum_{j=1}^{N-1}(ib\mm_j-k_jb^2)-2rb^2-1}}\\
&\prod_{r=1}^{k_N}\prod_{i=N+1}^{2N-1}\frac{\gamma\left(ib\mm_i-k_ib^2+\sum_{j=N+1}^{2N-1}(ib\mm_j-k_jb^2)-rb^2\right)}{b^{2ib\mm_i-2k_ib^2+2\sum_{j=N+1}^{2N-1}(ib\mm_j-k_jb^2)-2rb^2-1}}\ .}
From this form of $Z_K$ we can easily derive the factorization condition
\eq{\frac{K(c_0-c_i)}{C},\ \frac{K\sum_{a=0}^{N-1}c_a}{C},\ \frac{K(c_{2N}-c_j)}{C},\ \frac{K\sum_{a=N+1}^{2N}c_a}{C},\ \frac{K(c_i+c_j-c_N)}{C}\in\mathbb{Z}\ \begin{cases} i=1,...,N-1\\  j=N+1,...,2N-1\end{cases}}
When these conditions are satisfied, we can use \eqref{upsilonidentities} to write $\frac{Z_K}{Z_0}$ as a product of $\gamma$ functions which can be interpreted as some $S^2$ partition function describing the string worldsheet theory.
Indeed, these are the same conditions as the conditions for no fractional size modes as seen from equations \eqref{f}, \eqref{ftilde}.
Besides the factorization conditions, we also have the weak$\to$weak conditions. The worldsheet theory is weakly coupled when the number of $\gamma$-functions doesn't depend on the partition $\{k_i\}$, only on $K$. For this to be satisfied, we demand
\eq{c_N-c_i-c_j\geq 0\ \forall\ 1\leq i\leq N-1\ ,\ N+1\leq j\leq 2N-1\ .}
In addition, we need to demand one of the following four possibilities
\begin{enumerate}
\item $c_0\geq c_i \forall 1\leq i\leq N-1\ ,\ \sum_{j=0}^{N-1}c_j\geq C\ ,\ c_{2N}\geq c_i \forall N+1\leq i\leq 2N-1\ ,\ \sum_{j=N+1}^{2N}c_j\geq C$.
\item $c_0\geq c_i \forall 1\leq i\leq N-1\ ,\ \sum_{j=0}^{N-1}c_j\geq C\ ,\ c_{2N}<c_i \forall N+1\leq i\leq 2N-1\ ,\ \sum_{j=N+1}^{2N}c_j< C$.
\item $c_0< c_i \forall 1\leq i\leq N-1\ ,\ \sum_{j=0}^{N-1}c_j< C\ ,\ c_{2N}\geq c_i \forall N+1\leq i\leq 2N-1\ ,\ \sum_{j=N+1}^{2N}c_j\geq C$.
\item $c_0< c_i \forall 1\leq i\leq N-1\ ,\ \sum_{j=0}^{N-1}c_j< C\ ,\ c_{2N}< c_i \forall N+1\leq i\leq 2N-1\ ,\ \sum_{j=N+1}^{2N}c_j< C$.
\end{enumerate}
These are the same conditions found using the classical zero modes analysis, see equation \eqref{weakcondition} and the list right after. 
Assuming that the conditions are satisfied, the result for $Z_K$ that we get is
\eql{Id}{Z_K=Z_0Z_{\text{overall}}Z_{S^2,K}\ ,\ Z_{S^2,K}=Z_{\text{dec}}^{(K)}Z_{\text{charged}}^{(K)}\ ,}
where $Z_{\text{overall}}$ contains overall factors that are interpreted as regularization ambiguities, see the discussion at the end of section (4) in \cite{Gerchkovitz:2017ljt}. The factors $Z_{\text{dec}}^{(K)}$ and $Z_{\text{charged}}^{(K)}$ are
\eq{Z_{\text{dec}}^{(K)}&=\prod_{i=1}^{N-1}\prod_{j=N+1}^{2N-1}\prod_{r=1}^{\frac{(c_N-c_i-c_j)K}{C}}\gamma\left(ib(M_i+M_j-M_N)-rb^2\right)I_0^{(d)}I_{2N}^{(d)}\ ,\\
I_0^{(d)}(\text{cases 1,2})&=\prod_{r=1}^{\frac{K}{C}\sum_{j=0}^{N-1}c_j-K}\gamma\left(-ib\sum_{j=0}^{N-1}M_j-rb^2\right)\prod_{i=1}^{N-1}\prod_{r=1}^{\frac{(c_0-c_i)K}{C}}\gamma\left(ibM_i-ibM_0-rb^2\right)\ ,\\
I_0^{(d)}(\text{cases 3,4})&=\prod_{r=0}^{-\frac{K}{C}\sum_{j=0}^{N-1}c_j-1}\gamma\left(1+ib\sum_{j=0}^{N-1}M_j-rb^2\right)\prod_{i=1}^{N-1}\prod_{r=0}^{\frac{(c_i-c_0)K}{C}-K-1}\gamma\left(1+ibM_0-ibM_i-rb^2\right)\ ,\\
I_{2N}^{(d)}(\text{cases 1,3})&=\prod_{r=1}^{\frac{K}{C}\sum_{j=N+1}^{2N}c_j-K}\gamma\left(-ib\sum_{j=N+1}^{2N}M_j-rb^2\right)\prod_{i=N+1}^{2N-1}\prod_{r=1}^{\frac{(c_{2N}-c_i)K}{C}}\gamma\left(ibM_{i}-ibM_{2N}-rb^2\right)\ ,\\
I_{2N}^{(d)}(\text{cases 2,4})&=\prod_{r=0}^{-\frac{K}{C}\sum_{j=N+1}^{2N}c_j-1}\gamma\left(1+ib\sum_{j=N+1}^{2N}M_j-rb^2\right)\prod_{i=N+1}^{2N-1}\prod_{r=0}^{\frac{(c_i-c_{2N})K}{C}-K-1}\gamma\left(1+ibM_{2N}-ibM_i-rb^2\right)\ ,}

and 
\eq{&Z_{\text{charged}}^{(K)}=\sum_{\{k_a\}}e^{-\frac{8\pi^2}{g_1^2}\left(\sum_{i=1}^{N-1}(\mm_i+ik_ib)^2+(\sum_{i=1}^{N-1}(\mm_i+ik_ib))^2\right)-\frac{8\pi^2}{g_2^2}\left(\sum_{i=N+1}^{2N-1}(\mm_i+ik_ib)^2+(\sum_{i=N+1}^{2N-1}(\mm_i+ik_ib))^2\right)}I^{(c)}_0I^{(c)}_{2N}\\&\prod_{i=1}^{N-1}\prod_{j=N+1}^{2N-1}\prod_{r=1}^{k_i+k_j}\gamma\left(ib(\mm_i+\mm_j-\mm_N)-rb^2\right)\prod_{i\neq j=1}^{N-1}\prod_{r=1}^{k_i}\gamma\left(ib(\mm_j-\mm_i)+(k_i-k_j)b^2-rb^2\right)\\&\prod_{i\neq j=N+1}^{2N-1}\prod_{r=1}^{k_i}\gamma\left(ib(\mm_j-\mm_i)+(k_i-k_j)b^2-rb^2\right)\prod_{i=1}^{N-1}\prod_{r=1}^{k_i}\gamma\left(b\sum_{j=1}^{N-1}(-i\mm_j+k_jb)-ib\mm_i+k_ib^2-rb^2\right)\\&\prod_{r=1}^{k_N}\prod_{i=1}^{N-1}\gamma\left(ib\mm_i-k_ib^2+\sum_{j=1}^{N-1}(ib\mm_j-k_jb^2)-rb^2\right)\prod_{r=1}^{k_N}\prod_{i=N+1}^{2N-1}\gamma\left(ib\mm_i-k_ib^2+\sum_{j=N+1}^{2N-1}(ib\mm_j-k_jb^2)-rb^2\right)\\&\prod_{i=N+1}^{2N-1}\prod_{r=1}^{k_i}\gamma\left(b\sum_{j=N+1}^{2N-1}(-i\mm_j+k_jb)-ib\mm_i+k_ib^2-rb^2\right)\prod_{i=1}^{2N-1}\prod_{r=1}^{k_i}\gamma(-rb^2)|Z_{\text{inst}}^{\{k_i\}}|^2\ ,}
with
\eql{Ic}{I^{(c)}_0(\text{cases 1,2})=&\prod_{r=1}^{\sum_{j=N}^{2N-1}k_j}\gamma\left(-ib\sum_{j=0}^{N-1}\mm_j+Kb^2-rb^2\right)\prod_{i=1}^{N-1}\prod_{r=1}^{k_i}\gamma\left(ib\mm_i-ib\mm_0-rb^2\right)\ ,\\
I^{(c)}_0(\text{cases 3,4})=&\prod_{r=1}^{\sum_{j=0}^{N-1}k_j}\gamma\left(1+ib\sum_{j=0}^{N-1}\mm_j-(r-1)b^2\right)\prod_{i=1}^{N-1}\prod_{r=0}^{K-k_i-1}\gamma\left(1+ib\mm_0-ib\mm_i+(K-r)b^2\right)\ ,\\
I^{(c)}_{2N}(\text{cases 1,3})=&\prod_{r=1}^{\sum_{j=1}^{N}k_j}\gamma\left(-ib\sum_{j=N+1}^{2N}\mm_j+(K-r)b^2\right)\prod_{i=N+1}^{2N-1}\prod_{r=1}^{k_i}\gamma\left(ib\mm_{i}-ib\mm_{2N}-rb^2\right)\ ,\\
I^{(c)}_{2N}(\text{cases 2,4})=&\prod_{r=1}^{\sum_{j=N+1}^{2N}k_j}\gamma\left(1+ib\sum_{j=N+1}^{2N}\mm_j-(r-1)b^2\right)\prod_{i=N+1}^{2N-1}\prod_{r=0}^{K-k_i-1}\gamma\left(1+ib\mm_{2N}-ib\mm_i+(K-r)b^2\right)\ .}

For $K=1$, up to an overall factor, and ignoring instantons, $Z_{\text{charged}}^{(1)}$ in case(1) takes the form
\eql{sun2partition}{\frac{Z_{\text{charged}}^{(1)}}{\gamma(-b^2)}&=\prod_{i=1}^{N-1}\gamma\left(ib\mm_i+ib\sum_{j=1}^{N-1}\mm_j-b^2\right)\prod_{i=N+1}^{2N-1}\gamma\left(ib\mm_i+ib\sum_{j=N+1}^{2N-1}\mm_j-b^2\right)\\&\gamma\left(-ib\sum_{j=0}^{N-1}\mm_j\right)\gamma\left(-ib\sum_{j=N+1}^{2N}\mm_j\right)\\
+&\sum_{J=1}^{N-1}e^{-\frac{16\pi^2ib}{g_1^2}\left(\mm_J+\sum_{i=1}^{N-1}\mm_i+ib\right)}\prod_{i=N+1}^{2N-1}\gamma\left(ib(\mm_i+\mm_J-\mm_N)-b^2\right)\prod_{j\neq J=1}^{N-1}\gamma\left(ib(\mm_j-\mm_J)\right)\\
	&\gamma\left(-ib\sum_{j=1}^{N-1}\mm_j-ib\mm_J+b^2\right)\gamma\left(ib\mm_J-ib\mm_0-b^2\right)\gamma\left(-ib\sum_{j=N+1}^{2N}\mm_j\right)\\
	&+\sum_{J=N+1}^{2N-1}e^{-\frac{16\pi^2ib}{g_2^2}\left(\mm_J+\sum_{i=N+1}^{2N-1}\mm_i+ib\right)}\prod_{i=1}^{N-1}\gamma\left(ib(\mm_i+\mm_J-\mm_N)-rb^2\right)\prod_{j\neq J=N+1}^{2N-1}\gamma\left(ib(\mm_j-\mm_J)\right)\\&\gamma\left(-ib\sum_{j=N+1}^{2N-1}\mm_j-ib\mm_J+b^2\right)\gamma\left(-ib\sum_{j=0}^{N-1}\mm_j\right)\gamma\left(ib\mm_{J}-ib\mm_{2N}-b^2\right)\ .}

For case 2, it is the same, multiplied by an overall
\eq{\gamma\left(1+ib\sum_{j=N+1}^{2N}\mm_j\right)\prod_{i=N+1}^{2N-1}\gamma\left(1+ib\mm_{2N}-ib\mm_i+b^2\right)\ .}
For case 3, it is the same, multiplied by an overall
\eq{\gamma\left(1+ib\sum_{j=0}^{N-1}\mm_j\right)\prod_{i=1}^{N-1}\gamma\left(1+ib\mm_0-ib\mm_i+b^2\right)\ .}
 For case 4, it is the same, multiplied by an overall
\eq{&\gamma\left(1+ib\sum_{j=N+1}^{2N}\mm_j\right)\prod_{i=N+1}^{2N-1}\gamma\left(1+ib\mm_{2N}-ib\mm_i+b^2\right)\gamma\left(1+ib\sum_{j=0}^{N-1}\mm_j\right)\prod_{i=1}^{N-1}\gamma\left(1+ib\mm_0-ib\mm_i+b^2\right)\ .}
The expressions derived here for $Z_{\text{dec}}^{(K)}$ coincide with the $S^2$ partition function of the decoupled size modes. For $K=1$, these are the $\eta$ and $\tilde{\eta}$ fields of section \ref{notildecase} for case 1 and \ref{withqtilde} for case 2.  $Z_{\text{charged}}^{(1)}$ for case 1 coincides with the $S^2$ partition function of the $U(1)\times U(1)$ GLSM whose matter content consists of the chiral fields $X$ and $\psi^\pm_I$ described in table \ref{tablenotilde}. Similarly, for case 2, $Z_{\text{charged}}^{(1)}$ coincides with the $S^2$ partition function of the $U(1)\times U(1)$ GLSM whose matter content consists of the chiral fields $X$, $\psi^\pm_I$ and $\chi$ described in section \ref{withqtilde}. Cases 3,4 can be treated exactly the same. The computations of the relevant $S^2$ partition functions and the exact matchings are presented in appendix \ref{S2computations}.

\subsection{$SU(2)^n\times U(1)$}

In this section we will extract the worldsheet $S^2$ partition functions for the strings studied in section \ref{su2M} and show that the results are consistent with the worlsheet theories presented in \ref{su2M}. The four-ellipsoid partition function for the $SU(2)^n\times U(1)$ theory is
\eq{
	Z_{S_b^4}=&\int d^{n} \ha_I\, d \ha'\; \prod_I e^{-\frac{16\pi^2}{g_I^2} \ha_I^2}e^{-\frac{4\pi^2}{e^2}\hat a'^2}e^{16i\pi^2\hat{\xi}\hat a'}\prod_I\Upsilon_b\left(2i\ha_I\right)\Upsilon_b\left(-2i\ha_I\right)|Z_{\text{inst}}(\ha,\ha',c_j,\hm_j)|^2 \\&\left(\Upsilon_b\left(i\ha_1+i(c_0\ha'-\hm_0)+\frac{Q}{2}\right)\Upsilon_b\left(-i\ha_1+i(c_0\ha'-\hm_0)+\frac{Q}{2}\right)\right)^{-1}\\&\left(\Upsilon_b\left(i\ha_1+i(c_1\ha'-\hm_1)+\frac{Q}{2}\right)\Upsilon_b\left(-i\ha_1+i(c_1\ha'-\hm_1)+\frac{Q}{2}\right)\right)^{-1}\\&\left(\prod_{j=n+1}^{n+2}\Upsilon_b\left(i\ha_n+i(c_j\ha'-\hm_j)+\frac{Q}{2}\right)\Upsilon_b\left(-i\ha_n+i(c_j\ha'-\hm_j)+\frac{Q}{2}\right)\right)^{-1}\\&\prod_{I=2}^n\left(\Upsilon_b\left(i\ha_I+i\ha_{I-1}+i(c_I\ha'-\hm_I)+\frac{Q}{2}\right)\Upsilon_b\left(-i\ha_I+i\ha_{I-1}+i(c_I\ha'-\hm_I)+\frac{Q}{2}\right)\right)^{-1}\\&\prod_{I=2}^n\left(\Upsilon_b\left(i\ha_I-i\ha_{I-1}+i(c_I\ha'-\hm_I)+\frac{Q}{2}\right)\Upsilon_b\left(-i\ha_I-i\ha_{I-1}+i(c_I\ha'-\hm_I)+\frac{Q}{2}\right)\right)^{-1}\ . }
The poles that give the baryonic chain studied in section \ref{su2M} are given by
\eql{poles}{&i\ha_1+i(c_1\ha'-\hm_1)+\frac{Q}{2}=-k_1b\ ,\ -i\ha_n+i(c_{n+1}\ha'-\hm_{n+1})+\frac{Q}{2}=-k_{n+1}b\\&i\ha_I-i\ha_{I-1}+i(c_I\ha'-\hm_I)+\frac{Q}{2}=-k_Ib\ \forall\ 2\leq I\leq n\ .}
We will denote
\eq{&\hm=\sum_{I=1}^{n+1}\hm_I\ ,\ \mu_I'=\hm_I-\frac{\hm c_I}{C}\ ,\ k_I'=k_I-\frac{Kc_I}{C}\ ,\ r_I'=Q\left(1-\frac{(n+1)c_I}{C}\right)\ ,\ \\&M_I=\mu'_I+\frac{ir'_I}{2}\ ,\ \mm_I=M_I-\frac{ibKc_I}{C}\ ,\ M=\frac{\hm}{C}+\frac{i(n+1)Q}{2C}\ ,\ \sum_{I=1}^{n+1}\mm_I=-iKb\ .}
The residues of the poles \eqref{poles} are given by
\eq{Z_{\{k_I\}}=&b^{(3K-k_{n+1}-k_1)(1+b^2)-b^2\sum_{I=1}^{n+1}k_I^2-b^2(k_1^2+k_{n+1}^2)+4ib\sum_{I=1}^{n+1}k_I\mm_I}\\
	&\prod_I e^{-\frac{16\pi^2}{g_I^2} (\sum_{J=1}^I(\mm_J+ik_Jb))^2}e^{-\frac{4\pi^2}{e^2}(M+iKb/C)^2}e^{16i\pi^2\hat{\xi}(M+iKb/C)}|Z_{\text{inst}}|^2 \text{Res}\left(\Upsilon_b(0)\right)^{-(n+1)}\\&\left(\Upsilon_b\left(i\mm_1-i\mm_0-k_1b\right)\Upsilon_b\left(i\mm_{n+1}-i\mm_{n+2}-k_{n+1}b\right)\right)^{-1}\prod_{I=1}^{n+1}\prod_{m=1}^{k_I}\gamma(-mb^2)\\&\left(\Upsilon_b\left(-i\mm_0-i\mm_1+k_1b\right)\Upsilon_b\left(-i\mm_{n+2}-i\mm_{n+1}+k_{n+1}b\right)\right)^{-1}\prod_{I=2}^{n}\left(\Upsilon_b\left(-2i\mm_I+k_Ib\right)\right)^{-1}\\&\prod_{I=1}^n\prod_{m=1}^{k_I}\gamma\left(-2ib\sum_{J=1}^I(\mm_J+ik_Jb)-mb^2\right)\prod_{I=1}^{n}\prod_{m=1}^{k_{I+1}}\gamma\left(2ib\sum_{J=1}^I(\mm_J+ik_Jb)-mb^2\right)\ .}
The vacuum of the theory is given by plugging in $k_I=0$. This results in
\eq{Z_{0}=&\prod_I e^{-\frac{16\pi^2}{g_I^2} (\sum_{J=1}^IM_J)^2}e^{-\frac{4\pi^2}{e^2}M^2}e^{16i\pi^2\hat{\xi}M}|Z_{\text{inst}}|^2 \text{Res}\left(\Upsilon_b(0)\right)^{-(n+1)}\prod_{I=2}^{n}\left(\Upsilon_b\left(-2iM_I\right)\right)^{-1}\\&\left(\Upsilon_b\left(iM_1-iM_0\right)\Upsilon_b\left(iM_{n+1}-iM_{n+2}\right)\Upsilon_b\left(-iM_0-iM_1\right)\Upsilon_b\left(-iM_{n+2}-iM_{n+1}\right)\right)^{-1}\ .}

This is the partition function of the $n+3$ light hypermultiplets with the correct masses and R-charges. The conditions for factorization can read easily from the arguments of the $\Upsilon_b(x)$ functions, and result in
\eq{\frac{(c_0\pm c_1)K}{C},\ \frac{(c_{n+2}\pm c_{n+1})K}{C},\ \frac{2c_IK}{C}\in\mathbb{Z}\ \forall 2\leq I\leq n\ .}
These are the same conditions derived from classical size modes analysis \eqref{deccondition2}.
\subsubsection{Weakly coupled worldsheet theories}
Assuming the conditions are satisfied, we can write the worldsheet theory partition function. We will start from strings satisfying the conditions \eq{c_0\geq c_1\ ,\ c_0+c_1\geq C\ ,\ c_{n+2}\geq c_{n+1}\ ,\ c_{n+2}+c_{n+1}\geq C\ , 2c_I\geq C\ \forall\ 2\leq I\leq n\ .}
These are the weak$\to$weak conditions analysed in section \ref{su2M}.
Ignoring the instantons and some overall factors, we get
\eq{\frac{Z_{\{k_I\}}}{Z_{\{0\}}}=Z_{\text{charged}}^{(K)}Z_{\text{dec}}^{(K)}\ ,}with
\eq{Z_{\text{dec}}^{(K)}=&\prod_{m=1}^{\frac{(c_0-c_1)K}{C}}\gamma(ibM_1-ibM_0-mb^2)\prod_{m=1}^{\frac{(c_0+c_1)K}{C}-K}\gamma(-ibM_1-ibM_0-mb^2)\prod_{I=2}^{n}\prod_{m=1}^{2c_IK/c-K}\gamma(-2ibM_I-mb^2)\\
	&\prod_{m=1}^{\frac{(c_{n+2}-c_{n+1})K}{C}}\gamma(ibM_{n+1}-ibM_{n+2}-mb^2)\prod_{m=1}^{\frac{(c_{n+2}+c_{n+1})K}{C}-K}\gamma(-ibM_{n+1}-ibM_{n+2}-mb^2)\ ,}
and
\eq{&Z_{\text{charged}}^{(K)}=\prod_I e^{\frac{16\pi^2}{g_I^2}\left(b^2\left(\sum_{J=1}^Ik_J\right)^2-2ib\sum_{J,J'=1}^I\mm_J k_{J'}\right)}\prod_{I=1}^{n+1}\prod_{m=1}^{k_I}\gamma(-mb^2)\prod_{I=1}^{n}\prod_{m=1}^{k_{I+1}}\gamma\left(2ib\sum_{J=1}^I(\mm_J+ik_Jb)-mb^2\right)\\&\prod_{m=1}^{k_1}\gamma\left(ib\mm_1-ib\mm_0-mb^2\right)\prod_{m=1}^{k_{n+1}}\gamma\left(ib\mm_{n+1}-ib\mm_{n+2}-mb^2\right)\\
	&\prod_{m=1}^{K-k_1}\gamma\left(-ib\mm_1-ib\mm_0-mb^2+b^2K\right)\prod_{I=2}^{n}\prod_{m=1}^{K-k_I}\gamma(-2ib\mm_I-mb^2+b^2K)\\&\prod_{m=1}^{K-k_{n+1}}\gamma\left(-ib\mm_{n+1}-ib\mm_{n+2}-mb^2+b^2K\right)\prod_{I=1}^n\prod_{m=1}^{k_I}\gamma\left(-2ib\sum_{J=1}^I(\mm_J+ik_Jb)-mb^2\right)\ .}
For $K=1$ it simplifies to

\eql{ZS2ofsu2n}{&\frac{Z_{\text{charged}}^{(1)}}{\gamma(-b^2)}=\gamma\left(2ib\sum_{J=1}^n\mm_J-b^2\right)\gamma\left(ib\mm_{n+1}-ib\mm_{n+2}-b^2\right)\gamma\left(-ib\mm_1-ib\mm_0\right)\prod_{I=2}^{n}\gamma(-2ib\mm_I)\\
	&+\prod_Ie^{\frac{16\pi^2}{g_I^2}\left(b^2-2ib\sum_{J=1}^I\mm_J\right)}\gamma\left(ib\mm_1-ib\mm_0-b^2\right)\gamma\left(-ib\mm_{n+1}-ib\mm_{n+2}\right)\gamma\left(-2ib\mm_1+b^2\right)\prod_{I=2}^{n}\gamma(-2ib\mm_I)\\
	&+\sum_{l=2}^{n}\prod_{I=l}^{n} e^{\frac{16\pi^2}{g_I^2}\left(b^2-2ib\sum_{J=1}^I\mm_J\right)}\gamma\left(-ib\mm_1-ib\mm_0\right)\gamma\left(-ib\mm_{n+1}-ib\mm_{n+2}\right)\\&\gamma\left(-2ib\sum_{J=1}^l\mm_J+b^2\right)\gamma\left(2ib\sum_{J=1}^{l-1}\mm_J-b^2\right)\prod_{I=2,I\neq l}^{n}\gamma\left(-2ib\mm_I\right)\ .}
The expression derived here for $Z_{\text{dec}}^{(K)}$ coincides with the $S^2$ partition function of the decoupled size modes. For $K=1$, these are the $\eta$ fields of table \ref{su2spectrum}.  $Z_{\text{charged}}^{(1)}$ coincides with the $S^2$ partition function of the $U(1)^n$ GLSM whose matter content consists of the chiral fields $X$ and $\psi^\pm_I$ described in table \ref{su2spectrum}. The computations of the relevant $S^2$ partition functions and the exact matchings are presented in appendix \ref{S2computations}.

\subsubsection{$\mathcal{S}$-dual strings}
\label{Sedgesection}
In this section we will compute the partition function for strongly coupled wordlsheet theories which are $\mathcal{S}$-dual to the strings studied in the previous section. We will start from the case described in section \ref{tautolog}, in which the charges satisfy
\eq{c_0<c_1\ ,\ c_0+c_1\geq C\ ,\ c_{n+2}\geq c_{n+1}\ ,\ c_{n+2}+c_{n+1}\geq C\ ,\ 2c_2<C\ , \ 2c_I\geq C\ \forall\ 3\leq I\leq n\ .}
For Simplicity we will derive the worldsheet partition function in the case of two $SU(2)$ factors. The generalization to higher number of $SU(2)$s is straight forward. For $K=1$, the worldsheet partition function is
\eq{\frac{Z_{\{k_I\}}^{(K=1)}}{Z_{\{0\}}}=Z_{\text{dec}}Z_{\text{charged}}\ ,}
where 
\eq{&Z_{\text{dec}}=\gamma(-b^2)\prod_{m=0}^{\frac{(c_1-c_0)K}{C}-K-1}\gamma(1-ibM_1+ibM_0-mb^2)\prod_{m=1}^{\frac{(c_{4}-c_{3})K}{C}}\gamma(ibM_{3}-ibM_{4}-mb^2)\\
	&\prod_{m=1}^{\frac{(c_0+c_1)K}{C}-K}\gamma(-ibM_1-ibM_0-mb^2)\prod_{m=1}^{\frac{(c_{4}+c_{3})K}{C}-K}\gamma(-ibM_{3}-ibM_{4}-mb^2)\prod_{m=0}^{-2c_2K/C-1}\gamma(1+2ibM_2-mb^2)\ ,}
and
\eql{ZSontheedge}{&Z_{\text{charged}}=\gamma\left(-ib\mm_{3}-ib\mm_{4}\right)\left[\gamma\left(-2ib\sum_{J=1}^2\mm_J+b^2\right)\right.\gamma\left(2ib\mm_1-b^2\right)\gamma\left(-ib\mm_1-ib\mm_0\right)\\&\gamma\left(1-ib\mm_1+ib\mm_0+b^2\right)\gamma\left(1+2ib\mm_2\right)+\left.e^{\frac{16\pi^2}{g_1^2}\left(b^2-2ib\mm_1\right)}\gamma\left(-2ib\mm_1+b^2\right)\right]\\
	&+e^{\frac{16\pi^2}{g_2^2}\left(b^2-2ib\mm_3\right)}\gamma\left(2ib\sum_{J=1}^2\mm_J-b^2\right)\gamma\left(1-ib\mm_1+ib\mm_0+b^2\right)\gamma\left(ib\mm_{3}-ib\mm_{4}-b^2\right)\gamma\left(-ib\mm_1-ib\mm_0\right)\ .}
$Z_{\text{dec}}$ coincides with the $S^2$ partition function of the decoupled modes $X$ and $\eta$ of table \ref{Sontheedge}. $Z_{\text{charged}}$ coincides with the $S^2$ partition function of the $U(1)^2$ GLSM whose matter content consists of the $\psi_{1,2,3}^{\pm}$ of table \ref{Sontheedge}. The computation of the relevant $S^2$ partition function and the exact mapping of parameters appear in appendix \ref{S2computations}.
\subsection{Generalized quiver localization}
\label{Genquiversection}
In this section we will study the $G=SU(2)^3$ theory with two fundamentals for every $SU(2)$ and one trifundamental. We will denote the masses and U(1) charges of the six fundamental fields by $\mu_{Is},\ c_{Is}$ with $I=1,2,3$ and $s=\pm$. The trifundamental field is massless and U(1) invariant. The $S^4$ partition function reads
\eq{
	Z_{S_b^4}=&\int d^{3} \ha_I\, d \ha'\; \prod_I e^{-\frac{16\pi^2}{g_I^2} \ha_I^2}e^{-\frac{4\pi^2}{e^2}\hat a'^2}e^{16i\pi^2\hat{\xi}\hat a'}\prod_I\Upsilon_b\left(2i\ha_I\right)\Upsilon_b\left(-2i\ha_I\right)|Z_{\text{inst}}(\ha,\ha',c_j,\hm_j)|^2 \\&\prod_{I=1}^3\prod_{s=\pm}\left(\Upsilon_b\left(i\ha_I+i(c_{Is}\ha'-\hm_{Is})+\frac{Q}{2}\right)\Upsilon_b\left(-i\ha_I+i(c_{Is}\ha'-\hm_{Is})+\frac{Q}{2}\right)\right)^{-1}\\&\left(\Upsilon_b\left(i\ha_1+i\ha_{2}+i\ha_3+\frac{Q}{2}\right)\Upsilon_b\left(-i\ha_1+i\ha_{2}+i\ha_3+\frac{Q}{2}\right)\right)^{-1}\\&\left(\Upsilon_b\left(i\ha_1-i\ha_{2}+i\ha_3+\frac{Q}{2}\right)\Upsilon_b\left(-i\ha_1-i\ha_{2}+i\ha_3+\frac{Q}{2}\right)\right)^{-1}\ . }
We are interested in the poles
\eq{-i\ha_I+i(c_{I+}\ha'-\hm_{I+})+\frac{Q}{2}=-k_Ib\ ,\ i\sum_{I}\ha_I+\frac{Q}{2}=-k_4b}
which are solved by
\eq{\ha'=\frac{iKb}{C}+\frac{2iQ}{C}+\frac{\mu}{C}\ ,\ \ha_I=-\mm_{I+}-ik_Ib\ ,}
with
\eq{&\mu=\sum_{I}\hm_{I+}\ ,\ \mu_{I\pm}'=\mu_{I\pm}-\frac{\mu c_{I\pm}}{C}\ ,\ M_{I\pm}=\mu'_{I\pm}+\frac{iQ}{2}-\frac{2iQc_{I\pm}}{C}\ ,\ \mm_{I\pm}=M_{I\pm}-\frac{iKbc_{I\pm}}{C}\ .}
The result is
\eq{Z_{K}=&\sum_{\{k_a\}}\prod_I e^{-\frac{16\pi^2}{g_I^2} \left(\mm_{I+}+ik_Ib\right)^2}e^{-\frac{4\pi^2}{e^2C^2}(\mu+iQ+iK)^2}e^{\frac{16i\pi^2\hat{\xi}}{C}(\mu+iQ+iK)}\left|Z_{\text{inst}}^{(k_a)}\right|^2 \prod_{a=1}^4 Res\left(\onov{\Upsilon_b(-k_ab)}\right)\\&\prod_{I=1}^3\left[b^{k_I+k_4+b^2k_I(k_I+1)+b^2k_4(k_4+1)+4ib(k_I-k_4)(\mm_{I+}+ik_Ib)}\prod_{r=1}^{k_I}\gamma\left(-2ib\mm_{I+}+2b^2k_I-rb^2\right)\right]\\&\prod_{I=1}^3\left[\prod_{r=1}^{k_4}\gamma\left(2ib\mm_{I+}-2b^2k_I-rb^2\right)\left(\Upsilon_b\left(-i\mm_{I+}-i\mm_{I-}+k_Ib\right)\Upsilon_b\left(i\mm_{I+}-i\mm_{I-}-k_Ib\right)\right)^{-1}\right]\ . }
Lets start from the vacuum described by taking $K=0$. The result is
\eq{Z_{0}=&\prod_I e^{-\frac{16\pi^2}{g_I^2} M_{I+}^2}e^{-\frac{4\pi^2}{e^2C^2}(\mu+iQ)^2}e^{\frac{16i\pi^2\hat{\xi}}{C}(\mu+iQ)} Res\left(\onov{\Upsilon_b(0)}\right)^4\\&\prod_{I=1}^3\left[\Upsilon_b\left(-iM_{I+}-iM_{I-}\right)\right]^{-1}\prod_{I=1}^3\left[\Upsilon_b\left(iM_{I+}-iM_{I-})\right)\right]^{-1}\ . }
$Z_{0}$ describes the six light hypermultiplets in the vacuum.

For $K=1$, the contribution is
\eq{Z_{1}=&\sum_{J=1}^3\prod_I e^{-\frac{16\pi^2}{g_I^2} \left(\mm_{I+}+ib\delta_{IJ}\right)^2}e^{-\frac{4\pi^2}{e^2C^2}(\mu+iQ+i)^2}e^{\frac{16i\pi^2\hat{\xi}}{C}(\mu+iQ+i)}\left|Z_{\text{inst}}^{(k_J=1)}\right|^2 Res\left(\onov{\Upsilon_b(0)}\right)^4\gamma(-b^2)b^{2+4ib\mm_{J+}}\\&\gamma\left(-2ib\mm_{J+}+b^2\right)\prod_{I=1}^3\left[\Upsilon_b\left(-i\mm_{I+}-i\mm_{I-}+\delta_{IJ}b\right)\right]^{-1}\prod_{I=1}^3\left[\Upsilon_b\left(i\mm_{I+}-i\mm_{I-}-\delta_{IJ}b\right)\right]^{-1}+\\
&+\prod_I e^{-\frac{16\pi^2}{g_I^2} \left(\mm_{I+}\right)^2}e^{-\frac{4\pi^2}{e^2C^2}(\mu+iQ+i)^2}e^{\frac{16i\pi^2\hat{\xi}}{C}(\mu+iQ+i)}\left|Z_{\text{inst}}^{(k_4=1)}\right|^2  Res\left(\onov{\Upsilon_b(0)}\right)^4b^{2b^2+2}\gamma(-b^2)\\&\prod_{I=1}^3\gamma\left(2ib\mm_I-b^2\right)\prod_{I=1}^3\left[\Upsilon_b\left(-i\mm_{I+}-i\mm_{I-}\right)\right]^{-1}\prod_{I=1}^3\left[\Upsilon_b\left(i\mm_{I+}-i\mm_{I-})\right)\right]^{-1}\ . }
$Z_1$ factorizes if $\frac{c_{I+}+c_{I-}}{C}\in \mathbb{Z}\ \forall I=1,2,3$.
Using the identities \eqref{upsilonidentities}, one can write the expression for $\frac{Z_1}{Z_0}$ and see that there is no choice of charges such that the number of $\gamma$ functions is independent of the partition $\{k_a\}$. This means that the generalized quiver string is inherently strongly coupled due to the trifundamental matter. In order to simplify the expressions, we will assume that the $\mathcal{S}$-dual string is weakly coupled such that the dual charges satisfy \eq{2c_{2,3}\geq C\ ,\ c_0\geq c_1\ ,\ c_5\geq c_4\ ,\ c_0+c_1\geq C\ ,\ c_5+c_4\geq C\ . }
The charges of the generalized quiver are related to these charges via the transformation
\eql{Sgeneralized}{c_{1-}=c_0\ ,\ c_{1+}=c_1\ ,\ c_{3-}=c_5\ ,\ c_{3+}=c_4\ ,\ c_{2\pm}=c_2\pm c_3\ .}
Therefore they satisfy
\eql{genconditions}{c_{1,3-}\geq c_{1,3+}\ ,\ c_{1,3-}+c_{1,3+}\geq C\ ,\ c_{2+}\pm c_{2-}\geq C\ .}
For charges satisfying \eqref{genconditions}, we can write
\eq{Z_1=Z_0Z_{\text{decoupled}}Z_{\text{charged}}\ ,}
where
\eq{&Z_{\text{decoupled}}=\gamma(-b^2)\prod_{I=1}^3\prod_{r=1}^{\frac{c_{I+}+c_{I-}}{C}-1}\gamma\left(-ibM_{I+}-ibM'_{I-}-rb^2\right)\\
&\prod_{r=0}^{\frac{c_{2+}-c_{2-}}{C}-2}\gamma\left(1-ibM_{2+}+ibM_{2-}-rb^2\right)\prod_{I=1,3}\prod_{r=1}^{\frac{c_{I-}-c_{I+}}{C}}\gamma\left(ibM_{I+}-ibM_{I+}-rb^2\right)\ ,}
and 
\eql{partitionfunctiongen}{&Z_{\text{charged}}=e^{-\frac{16\pi^2}{g_1^2} \left(2ib\mm_{1+}-b^2\right)}\gamma\left(-2ib\mm_{1+}+b^2\right)\prod_{I=2,3}\gamma\left(-ib\mm_{I+}-ib\mm_{I-}\right)\\
&\gamma\left(ib\mm_{1+}-ib\mm_{1-}-b^2\right)\gamma\left(1+ib\mm_{2-}-ib\mm_{2+}+b^2\right)+
(1\leftrightarrow 3)+\\
&+e^{-\frac{16\pi^2}{g_2^2} \left(2ib\mm_{2+}-b^2\right)}\gamma\left(-2ib\mm_{2+}+b^2\right)\prod_{I=1,3}\gamma\left(-ib\mm_{I+}-ib\mm_{I-}\right)+\\
&+\prod_{I=1}^3\left[\gamma\left(2ib\mm_{I+}-b^2\right)\gamma\left(-ib\mm_{I+}-ib\mm_{I-}\right)\right]\gamma\left(1+ib\mm_{2-}-ib\mm_{2+}+b^2\right)\ .}
$Z_{\text{decoupled}}$ describes the decoupled fields $X$ and $\eta$ of table \ref{generalizedtable} while $Z_{\text{charged}}$ describes the partition function of the $U(1)^3$ GLSM with the charged fields $\psi$ of table \ref{generalizedtable}. The agreement with the relevant $S^2$ partition function is shown in appendix \ref{S2computations}.
\section*{Acknowledgments}

The author likes to thank Efrat Gerchkovitz and Zohar Komargodski for fruitful discussions, and Ofer Aharony for comments on a preliminary version of the paper.
The author is supported by the ERC STG grant 335182.
\appendix

\section{Computations of $S^2$ partition functions}
\label{S2computations}
In this appendix, we will use the results of \cite{Gomis:2014eya} for $S^2$ partition functions to show agreement between the expressions derived in section \ref{localization} and the corresponding $S^2$ partition functions. The $S^2$ partition functions we obtain are functions of $z_a=e^{2\pi it_a}$ where $t_a$ are the complex FI parameters, and of the dimensionless complex masses $m=lM+\frac{i}{2}R$. Here $M,\ R$ are the mass and R-charge of the chiral multiplet, and $l$ is the radius of the sphere.\footnote{More precisely, we compute partition functions on the squashed sphere given by $\frac{x_0^2}{r^2}+\frac{x_1^2+x_2^2}{l^2}=1$. However, the partition function doesn't depend on $r$ and is equivalent to the partition function computed on a round sphere of radius $l$.\cite{Gomis:2012wy}} The relation between the su(1$|$1) superalgebra of the two-dimensional theory and the su(1$|$1) superalgebra of the  four-dimensional theory preserved by the string enables us to make the identification \eql{complexmass}{m=bM_{4d}+\frac{i(1+b^2)}{2}R^{(R)}-\frac{ib^2}{2}R^{(J)}\ .} The dependence of the expressions derived in section \ref{localization} on $b$ allows us to distinguish between the two $R$-charges and find agreement for each one of them seperately. 
\subsection{$SU(N)^2$ quiver}
\subsubsection{No $\tilde{q}$ excitations}
In this section we will compute the $S^2$ partition function of $U(1)\times U(1)$ gauge theory with 4 chiral multiplets with charges $(\pm1,0),\ (0,\pm1)$ and masses $m_{0,N}^{\pm}$ and $2(N-1)$ chiral multiplets with charges $(\pm1,\mp1)$ and masses $m_{I}^{\pm},\ I=1,...,N-1$. We will show that this agrees with \eqref{sun2partition} under some 2d-4d map of parameters. Ignoring instantons, the partition function is
\eq{Z_{S^2}=&\int \frac{d^2\sigma_a}{(2\pi)^2}\prod_{a=1}^{2}(z_a\zb_a)^{i\sigma_a}\prod_{I=1}^{N-1}\left[\gamma(-i\sigma_1+i\sigma_2-im_I^+)\gamma(i\sigma_1-i\sigma_2-im_I^-)\right]\\
&\gamma(-i\sigma_1-im_0^+)\gamma(i\sigma_1-im_0^-)\gamma(-i\sigma_2-im_N^+)\gamma(i\sigma_2-im_N^-)\ .}
Here $z_a=e^{2\pi it_a}$ where $t_a$ are the complexified FI parameters. We can close the contours of integrations over $\sigma_a$ in the complex plane and obtain the partition function in the Higgs branch representation, given by
\eq{Z_{S^2}&=\sum_{j=1}^{N-1}|z_1|^{-2im_0^+}|z_2|^{-2i(m_J^-+m_0^+)}\prod_{I=1,\ I\neq J}^{N-1}\left[\gamma(-im_J^--im_I^+)\gamma(im_J^--im_I^-)\right]\\
&\gamma(-im_0^+-im_0^-)\gamma(im_0^++im_J^--im_N^+)\gamma(-im_0^+-im_J^--im_N^-)\gamma(-im_J^--im_J^+)\\
+&\sum_{j=1}^{N-1}|z_1|^{-2i(m_j^++m_N^+)}|z_2|^{-2im_N^+}\prod_{I=1,\ I\neq J}^{N-1}\left[\gamma(im_J^+-im_I^+)\gamma(-im_J^+-im_I^-)\right]\\
&\gamma(im_J^++im_N^+-im_0^+)\gamma(-im_J^+-im_N^+-im_0^-)\gamma(-im_N^+-im_N^-)\gamma(-im_J^+-im_J^-)\\
&+|z_1|^{-2im_0^+}|z_2|^{-2im_N^+}\prod_{I=1}^{N-1}\left[\gamma(im_0^+-im_N^+-im_I^+)\gamma(im_N^+-im_0^+-im_I^-)\right]\\&\gamma(-im_0^+-im_0^-)\gamma(-im_N^+-im_N^-)\ .
}

There is an agreement with \eqref{sun2partition} if
\eq{&m_0^++m_0^-=b\sum_{j=0}^{N-1}\mm_j\ ,\ m_N^++m_N^-=b\sum_{j=N+1}^{2N}\mm_j\ ,\ m_J^++m_N^++m_0^-=b\mm_0-b\mm_J-ib^2\ ,\\
&m_J^-+m_N^-+m_0^+=b\mm_{2N}-b\mm_{N+J}-ib^2\ ,\ m_N^+-m_0^+-m_J^-=b\sum_{j=N+1}^{2N-1}\mm_j+b\mm_{N+J}+ib^2\ ,\\& m_J^-+m_I^+=b(\mm_N-\mm_I-\mm_{N+J})-ib^2\ ,\  m_0^+-m_N^+-m_J^+=b\sum_{j=1}^{N-1}\mm_j+b\mm_{J}+ib^2\\
&m_J^-+m_I^+=b(\mm_N-\mm_I-\mm_{N+J})-ib^2\ ,\ m_J^--m_I^-=b(\mm_{N+I}-\mm_{N+J})\ ,\ m_J^+-m_I^+=b(\mm_{I}-\mm_{J})\ ,}
together with
\eq{t_a=\tau_a\ .}
This is solved by choosing up to gauge transformations
\eq{&m_0^-=b\mm_0\ ,\ m_N^-=b\mm_{2N}\ ,\ m_0^+=b\sum_{j=1}^{N-1}\mm_j\ ,\ m_N^+=b\sum_{j=N+1}^{2N-1}\mm_j\ ,\\& m_J^+=-b\mm_J-ib^2-b\sum_{j=N+1}^{2N-1}\mm_j\ ,\  m_J^-=-b\mm_{N+J}-ib^2-b\sum_{j=1}^{N-1}\mm_j\ .}

From the definition of $\mm$, we can write the explicit expressions
\eq{&m_0^-=b\mu'_0+\frac{i(1+b^2)}{2}\left(1-\frac{(2N-1)c_0}{C}\right)-\frac{ib^2}{2}\frac{2c_0}{C}\\& m_N^-=b\mu'_{2N}+\frac{i(1+b^2)}{2}\left(1-\frac{(2N-1)c_{2N}}{C}\right)-\frac{ib^2}{2}\frac{2c_{2N}}{C}\\& m_0^+=\sum_{j=1}^{N-1}\left(b\mu'_j+ibQ/2-\frac{ibQ(2N-1)c_j}{2C}-\frac{ib^2c_j}{C}\right)\\&m_N^+=\sum_{j=N+1}^{2N-1}\left(b\mu'_j+ibQ/2-\frac{ibQ(2N-1)c_j}{2C}-\frac{ib^2c_j}{C}\right)\\
& m_J^+=-b\mu'_J-\frac{ibQ}{2}+\frac{ibQ(2N-1)c_J}{2C}+\frac{ib^2c_J}{C}-ib^2-\sum_{j=N+1}^{2N-1}\left(b\mu'_j+\frac{ibQ}{2}-\frac{ibQ(2N-1)c_j}{2C}-\frac{ib^2c_j}{C}\right)\\& m_J^-=-b\mu'_{N+J}-\frac{ibQ}{2}+\frac{ibQ(2N-1)c_{N+J}}{2C}+\frac{ib^2c_{N+J}}{C}-ib^2-\sum_{j=1}^{N-1}\left(b\mu'_j+\frac{ibQ}{2}-\frac{ibQ(2N-1)c_j}{2C}-\frac{ib^2c_j}{C}\right)\ .}
Using the relation \eqref{complexmass}, we can extract the masses and R-charges of the different fields and check agreement with the table \ref{tablenotilde}.

\subsubsection{Including $\tilde{q}$ excitations}
In the previous section we studied only the case without $\tilde{q}$ excitations. We can have a weakly coupled theory with $\tilde{q}$ excitations. These are cases 2,3,4 from the list right after \eqref{weakcondition}. The worldsheet theory will be very similar to that of case (1), but with additional neutral fields coupled to the charged fields via a superpotential. The $S^2$ partition function is independent of the superpotential coefficients. This means that when adding these fields we just need to multiply the partition function by the partition function of the neutral fields.  In case (2), this factor is
\eq{Z_{\chi}=\gamma\left(1+ib\sum_{j=N+1}^{2N}\mm_j\right)\prod_{i=N+1}^{2N-1}\gamma\left(1+ib\mm_{2N}-ib\mm_{i}+b^2\right)\ .}
These are the neutral fields $\chi_{2N}$ and $\chi_i$ with $i=N+1,...,2N-1$ that are described in \ref{withqtilde} .
In addition, we replace $\eta_{2N,j,r}$ and $\eta_{2N,N,r}$ with $\tilde{\eta}_{2N,j,r}$ $j=N+1,...,2N-1$ $r=1,..., \frac{(c_i-c_{2N})K}{C}-K$ and $\tilde{\eta}_{2N,N,r}$ $r=1,...,-\frac{K}{C}\sum_{j=N+1}^{2N}c_j$ as described in \ref{withqtilde}. These changes of the partition function are exactly captured by the change in the $I_{2N}$ factor of equations \eqref{Id},\eqref{Ic} when moving from case (1) to case (2).

\subsection{Weakly coupled $SU(2)^n$ strings}
In this section we compute the $S^2$ partition function of a $U(1)^n$ GLSM with 4 chiral multiplets with charges $(\pm1,0,...),\ (...,0,\mp1)$ and masses $m_{1,n+1}^{\pm}$, and $2(n-1)$ chiral multiplets with charges $(0,,...,0,\overbrace{\mp1}^{I-1},\overbrace{\pm1}^{I},0,...,0)$ and masses $m_{I}^{\pm},\ I=2,...,n$, and 1 neutral field with mass $m_X$. Ignoring instantons, the partition function is
\eq{Z_{S^2}&=\int \frac{d^n\sigma_a}{(2\pi)^n}  \prod_{a=1}^{n}(z_a\zb_a)^{i\sigma_a}\frac{\Gamma\left(-im_X\right)}{\Gamma\left(1+im_X\right)}\prod_{I=2}^{n}\frac{\Gamma\left(-i\sigma_I+i\sigma_{I-1}-im_I^+\right)}{\Gamma\left(1+i\sigma_I-i\sigma_{I-1}+im_I^+\right)}\frac{\Gamma\left(i\sigma_I-i\sigma_{I-1}-im_I^-\right)}{\Gamma\left(1-i\sigma_I+i\sigma_{I-1}+im_I^-\right)}\\&\frac{\Gamma\left(-i\sigma_1-im_1^+\right)}{\Gamma\left(1+i\sigma_1+im_1^+\right)}\frac{\Gamma\left(i\sigma_1-im_1^-\right)}{\Gamma\left(1-i\sigma_1+im_1^-\right)}\frac{\Gamma\left(-i\sigma_n-im_{n+1}^-\right)}{\Gamma\left(1+i\sigma_n+im_{n+1}^-\right)}\frac{\Gamma\left(i\sigma_n-im_{n+1}^+\right)}{\Gamma\left(1-i\sigma_n+im_{n+1}^+\right)}\ .}
Again, upon closing the $\sigma_a$ integral in the complex plane, we get \eq{Z_{S^2}&=\sum_{l=1}^{n+1}\prod_{a=1}^{l-1}(z_a\zb_a)^{-i\sum_{b=1}^am_b^+}\prod_{a=l}^{n}(z_a\zb_a)^{-i\sum_{b=a+1}^{n+1}m_b^-}\gamma\left(-im_X\right)\prod_{I=1,I\neq l}^{n+1}\gamma\left(-im_I^+-im_I^-\right)\times\\&\gamma\left(i\sum_{J=l+1}^{n+1}m_J^--i\sum_{J=1}^{l}m_J^+\right)\gamma\left(i\sum_{J=1}^{l-1}m_J^+-i\sum_{J=l}^{n+1}m_J^-\right)\ .}
We find agreement with \eqref{ZS2ofsu2n} for
\eq{&m_X=-ib^2\ ,\ m_I^++m_I^-=2b\mm_I\ ,\ m_1^++m_1^-=b\mm_1+b\mm_0\ ,\ m_{n+1}^++m_{n+1}^-=b\mm_{n+1}+b\mm_{n+2}\\
	&\sum_{a=1}^{n+1}m_a^-=b(\mm_0-\mm_1)-ib^2\ ,\ \sum_{a=1}^{n+1}m_a^+=b(\mm_{n+2}-\mm_{n+1})-ib^2}
This is solved up to gauge transformations by
\eq{m_1^-=b\mm_0\ ,\ m_{l>1}^-=b\mm_l\ ,\ m_{n+1}^+=b\mm_{n+2}\ ,\ m_{l<n+1}^+=b\mm_l\ .}
Using the relation \eqref{complexmass}, we can extract the masses and R-charges of the different fields and check agreement with the table \ref{su2spectrum}.
\subsection{$\mathcal{S}$-dual strings}
\subsubsection{Linear quiver}
In this section we will compute the $S^2$ partition function for the worldsheet theories described in \ref{Sdual} and show agreement with the results obtained from $S^4_b$ partition function. Due to the weak$\to$ strong mapping of parameters $z_a\equiv e^{2\pi i t_a}=1-e^{2\pi i\tau_a}$ for some $a$, we want to expand the partition function around $1-z_a$. In \cite{Gerchkovitz:2017ljt} it was shown that the partition function of a U(1) GLSM with 2 chirals with charge $+1$ and masses $m_{1}^+,\ m_2^-$ and two chirals with charge $-1$ and masses $m_1^-,\ m_2^+$ at leading order in $1-z_1$ is
\eq{Z=&\frac{|1-z_1|^{2+2i(m_1^++m_2^-+m_1^-+m_2^+)}}{\gamma\left(2+i(m_1^++m_2^-+m_1^-+m_2^+)\right)}\\&+\frac{\gamma\left(-im_1^+-im_1^-\right)\gamma\left(-im_1^+-im_2^+\right)\gamma\left(-im_2^--im_1^-\right)\gamma\left(-im_2^--im_2^+\right)}{\gamma\left(-i(m_1^++m_2^-+m_1^-+m_2^+)\right)}\ .}
To compare with \ref{ZSontheedge}, we need to add two more fields with masses $m_{3}^\pm$ and gauge another U(1) under which $m_2^\pm$ have charges $\pm1$ while $m_3^\pm$ have charges $\mp1$.
The leading order partition function becomes
\eql{dualS2partition}{Z=&\int d\sigma (z_2\zb_2)^{i\sigma}\gamma(-im_3^++i\sigma)\gamma(-im_3^--i\sigma)\left[\frac{|1-z_1|^{2+2i(m_1^++m_2^-+m_1^-+m_2^+)}}{\gamma\left(2+i(m_1^++m_2^-+m_1^-+m_2^+)\right)}\right.\\&+\left.\frac{\gamma\left(-im_1^+-im_1^-\right)\gamma\left(-im_1^+-im_2^+-i\sigma\right)\gamma\left(-im_2^--im_1^-+i\sigma\right)\gamma\left(-im_2^--im_2^+\right)}{\gamma\left(-i(m_1^++m_2^-+m_1^-+m_2^+)\right)}\right]\\=&(z_2\zb_2)^{-im_3^-}\gamma(-im_3^+-im_3^-)\left[\frac{|1-z_1|^{2+2i(m_1^++m_2^-+m_1^-+m_2^+)}}{\gamma\left(2+i(m_1^++m_2^-+m_1^-+m_2^+)\right)}\right.\\&+\left.\frac{\gamma\left(-im_1^+-im_1^-\right)\gamma\left(-im_1^+-im_2^++im_3^-\right)\gamma\left(-im_2^--im_1^--im_3^-\right)\gamma\left(-im_2^--im_2^+\right)}{\gamma\left(-i(m_1^++m_2^-+m_1^-+m_2^+)\right)}\right]+\\
	+&(z_2\zb_2)^{-im_1^+-im_2^+}\gamma(-im_3^+-im_1^+-im_2^+)\gamma(-im_3^-+im_1^++im_2^+)\gamma\left(-im_1^+-im_1^-\right)\gamma\left(-im_2^--im_2^+\right)\ .}
There is an agreement with \ref{ZSontheedge} (up to an overall factor which is interpreted as counterterm) under the mapping
\eq{&m_3^++m_3^-=b\mathbb{M}_3+b\mathbb{M}_4\\
	&m_1^++m_1^-=b\mathbb{M}_1+b\mathbb{M}_0\\
	&m_2^++m_2^-=b\mathbb{M}_1-\mathbb{M}_0+i(1+b^2)\\
	&m_1^++m_2^+-m_3^-=2b(\mathbb{M}_1+\mathbb{M}_2)+ib^2\\&e^{2\pi it_1}=1-e^{2\pi i\tau_1}\ ,\ t_2=\tau_2\ .}
Up to gauge transformations, this can be solved by
\eq{m_3^+=b\mathbb{M}_4\ ,\ m_3^-=b\mathbb{M}_3\ ,\ m_1^\pm=\frac{b}{2}(\mathbb{M}_1+\mathbb{M}_0)\pm b\mathbb{M}_2\mp\frac{i(1+b^2)}{2}\ ,\ m_2^\pm=\frac{b}{2}(\mathbb{M}_1-\mathbb{M}_0)+\frac{i(1+b^2)}{2}\ ,}
in agreement with the spectrum described in table \ref{Sontheedge}.
\subsubsection{Generalized quiver}
In this section we want to write the $S^2$ partition function for a $U(1)^3$ GLSM with 8 chiral multiplets, whose masses and U(1) charges are \eq{(m_{1}^{\pm},\pm1,0,0)\ ,\ (m_{2}^{\pm},0,\pm1,0)\ ,\ (m_{3}^{\pm},0,0,\pm1)\ ,\ (m_{4}^{\pm},\pm1,\pm1,\mp1)\ ,}
expanded around $z_{1,3}=0\ ,\ z_2=1$. This partition function should be equivalent to \eqref{partitionfunctiongen} under some map of the masses. The easiest way to write the partition function will be to start from \eqref{dualS2partition}, with the renaming $z_1\to z_2\to z_3\ ,\ m_{1}^{\pm}\to m_{2}^{\pm}\to m_{4}^{\mp}\ ,\ m_3^{\pm}\to m_{3}^{\mp}$ 
\eq{Z=&(z_3\zb_3)^{-im_3^+}\gamma(-im_3^+-im_3^-)\left[\frac{|1-z_2|^{2+2i(m_2^++m_4^-+m_2^-+m_4^+)}}{\gamma\left(2+i(m_2^++m_4^-+m_2^-+m_4^+)\right)}\right.\\&+\left.\frac{\gamma\left(-im_2^+-im_2^-\right)\gamma\left(-im_2^+-im_4^-+im_3^+\right)\gamma\left(-im_4^+-im_2^--im_3^+\right)\gamma\left(-im_4^--im_4^+\right)}{\gamma\left(-i(m_2^++m_4^-+m_2^-+m_4^+)\right)}\right]+\\
	+&(z_3\zb_3)^{-im_2^+-im_4^-}\gamma(-im_3^--im_2^+-im_4^-)\gamma(-im_3^++im_2^++im_4^-)\gamma\left(-im_2^+-im_2^-\right)\gamma\left(-im_4^--im_4^+\right)\ .}
Now we will gauge another U(1) under which $m_{4}^{\pm}$ have charges $\pm1$ and add two chiral fields with masses $m_{1}^{\pm}$ and charges $(\pm1,0,0)$. The partition function becomes
\eq{Z=&\int d\sigma (z_1\zb_1)^{i\sigma}(z_3\zb_3)^{-im_3^+}\gamma(-im_1^+-i\sigma)\gamma(-im_1^-+i\sigma)\gamma(-im_3^+-im_3^-)\\&\left[\frac{|1-z_2|^{2+2i(m_2^++m_4^-+m_2^-+m_4^+)}}{\gamma\left(2+i(m_2^++m_4^-+m_2^-+m_4^+)\right)}\right.\\&+\left.\frac{\gamma\left(-im_2^+-im_2^-\right)\gamma\left(-im_2^+-im_4^-+im_3^++i\sigma\right)\gamma\left(-im_4^+-im_2^--im_3^+-i\sigma\right)\gamma\left(-im_4^--im_4^+\right)}{\gamma\left(-i(m_2^++m_4^-+m_2^-+m_4^+)\right)}\right]+\\
	&+\int d\sigma (z_1\zb_1)^{i\sigma}(z_3\zb_3)^{-im_2^+-im_4^-+i\sigma}\gamma(-im_1^+-i\sigma)\gamma(-im_1^-+i\sigma)\\&\gamma(-im_3^--im_2^+-im_4^-+i\sigma)\gamma(-im_3^++im_2^++im_4^--i\sigma)\gamma\left(-im_2^+-im_2^-\right)\gamma\left(-im_4^--im_4^+\right)\ .}
Closing contour for the $\sigma$ integral, we get
\eq{Z=&\prod_{a=1,3} |z_a|^{-2im_a^+}|1-z_2|^{2+2i(m_2^++m_4^-+m_2^-+m_4^+)}\frac{\gamma(-im_1^--im_1^+)\gamma(-im_3^+-im_3^-)}{\gamma\left(2+i(m_2^++m_4^-+m_2^-+m_4^+)\right)}\\+&\prod_{a=1,3}|z_a|^{-2im_a^+}\frac{\gamma\left(im_3^+-im_2^+-im_4^--im_1^+\right)\gamma\left(im_1^+-im_4^+-im_2^--im_3^+\right)\prod_{a=1}^4\gamma\left(-im_a^--im_a^+\right)}{\gamma\left(-i(m_2^++m_4^-+m_2^-+m_4^+)\right)}+\\
	+&|z_1|^{-2im_1^+}|z_3|^{-2i(m_2^++m_4^-+m_1^+)}\gamma(-im_1^--im_1^+)\gamma\left(-im_2^+-im_2^-\right)\gamma\left(-im_4^--im_4^+\right)\\&\gamma(-im_3^--im_2^+-im_4^--im_1^+)\gamma(-im_3^++im_2^++im_4^-+im_1^+)\\+&|z_1|^{-2i(m_4^++m_2^-+m_3^+)}|z_3|^{-2im_3^+}\gamma(-im_1^++im_2^-+im_3^++im_4^+)\gamma(-im_1^--im_2^--im_3^+-im_4^+)\\&\gamma\left(-im_2^+-im_2^-\right)\gamma(-im_3^+-im_3^-)\gamma\left(-im_4^--im_4^+\right)\ .}
We find agreement with \eqref{partitionfunctiongen} under the following map of parameters:
\eq{&z_{1,3}=q_{1,3}\ ,\ z_2=1-q_2\ ,\\
	&-im_2^+-im_2^--im_4^+-im_4^-=-2ib\mm_{2+}+1+b^2\\
	&-im_{I}^+-im_I^-=-ib\mm_{I+}-ib\mm_{I-}\ ,\ I=1,2,3\\
	&-im_4^+-im_4^-=ib\mm_{2-}-ib\mm_{2+}+b^2+1\\
	&-im_2^+-im_1^+-im_4^-+im_3^+=2ib\mm_{3+}\\
	&-im_2^--im_4^+-im_3^++im_1^+=2ib\mm_{1+}\\
	&-im_1^+-im_2^+-im_4^--im_3^-=ib\mm_{3+}-ib\mm_{3-}-b^2\\
	&-im_2^--im_4^+-im_3^+-im_1^-=ib\mm_{1+}-ib\mm_{1-}-b^2\ .}
Up to gauge transformations, this is solved by 
\eq{&m_{I}^\pm=b\mm_{I\pm}\ ,\ m_{4}^-=\frac{i(b^2+1)}{2}\ ,\ m_{4}^+=b\mm_{2+}-b\mm_{2-}+\frac{i(b^2+1)}{2}\  ,}
in agreement with the spectrum of table \ref{generalizedtable}.

\bibliography{Quiverbib}
	\end{document}